\documentclass[11pt,a4paper,twoside,american]{article}
\newif\ifdraft \global\drafttrue

\pdfoutput=1

\usepackage{filecontents,pgfplots,tikz}
\usetikzlibrary{automata, positioning}
\pgfplotsset{compat=1.14}
\usepackage[expansion=true]{microtype}

\usepackage[T1]{fontenc}
\usepackage[utf8]{inputenc}
\usepackage{times}
\usepackage{graphicx}
\usepackage{mathrsfs}
\usepackage{subfig}

\usepackage{hyperref}
\colorlet{darkblue}{blue!50!black}
\hypersetup{
    colorlinks,%
    citecolor=darkblue,%
    filecolor=red,%
    linkcolor=darkblue,%
    urlcolor=magenta,%
    pdfnewwindow=true,%
    pdfstartview={FitH}
}
\usepackage{fancyhdr}
\usepackage{latexsym}
\usepackage{amsmath}
\usepackage{amsthm}
\usepackage[margin=3.1cm]{geometry}

\usepackage{bbm}
\usepackage{bm}
\usepackage{amssymb}
\usepackage{amsfonts}
\usepackage{enumerate}
\usepackage{cancel}
\usepackage{enumitem}
\usepackage{xspace}
\setlist{noitemsep,topsep=0pt,parsep=5pt,partopsep=0pt}
\usepackage{stmaryrd}
\ifdraft
\fi

\newcounter{smallarabics}

\newcounter{smallroman}

\newcommand{\ben}{\begin{enumerate}[{\rm (1)}]}
\newcommand{\een}{\end{enumerate}}

\newtheorem{theorem}{Theorem}[section]
\newtheorem{proposition}[theorem]{Proposition}
\newtheorem{lemma}[theorem]{Lemma}

\newtheorem{definition}[theorem]{Definition}

\theoremstyle{definition}
\newtheorem{remark}[theorem]{Remark}
\newtheorem{example}[theorem]{Example}

\newcommand{\eee}{\mathrm{e}}

\newcommand{\R}{{\mathbb R}}

\newcommand{\IP}{{\mathbb P}}
\newcommand{\Q}{{\mathbb Q}}

\newcommand{\I}{{\mathbb I}}

\newcommand{\argdot}{{\bm \cdot }}

\def\rr{{\mathbb R}}
\def\zz{{\mathbb Z}}
\def\cc{{\mathbb C}}
\def\nn{{\mathbb N}}

\def\textsl{{}}

\def\c0inf{C_0^\infty}

\renewcommand{\proof}[1]{\noindent{\textbf{Proof{#1}.}}}

\def\cH{{\cal  H}}

\def\cA{{\cal A}}
\def\PP{\mathcal{P}}

\def\QQ{\mathbb{Q}}
\def\II{\mathbb{I}}
\def\LL{\mathbb{L}}

\def\cB{{\cal B}}

\def\fin{{\rm fin}}

\renewcommand{\d}{\mathrm{d}}

\newcommand{\ep}{\mathrm{ep}}

\def\qed{$\Box$\medskip}

\def\cP{{\cal P}}

\def\cC{{\cal C}}

\def\cF{{\cal F}}

\def\cN{{\cal N}}
\def\cM{{\cal M}}
\def\cA{{\cal A}}

\def\bar{\overline}
\def\ubar{\underline}

\def\12{\frac{1}{2}}

\def\supp{{\rm supp}}
\def\dd{{\mathrm d}}

\def\d{{\rm d}}

\def\cH{{\cal H}}

\def\ind{{\rm 1\mkern-4.25mu l}}

\def\ind{{\rm 1\mkern-4.25mu l}}
\def\Ent{{\rm Ent}}

\def\tr{{\rm tr}}

\def\P{\mathbb P}

\def\ie{{\sl i.e., }}

\def\wP{{\widehat{\mathbb P}}}

\newcommand{\aA}{{\cal A}}

\newcommand{\FF}{{\cal F}}

\newcommand{\MM}{{\cal M}}
\newcommand{\NN}{{\cal N}}

\newcommand{\N}{{\mathbb N}}

\numberwithin{equation}{section}

\begin{document}
\author{{Noé~Cuneo$^{1,3}$, Vojkan Jak\v{s}i\'c$^{1}$, Claude-Alain 
Pillet$^{2}$,
Armen~Shirikyan$^{1,3,4}$}
\\ \\ \\
$^1$Department of Mathematics and Statistics, McGill University\\ 
805 Sherbrooke Street West, Montreal, QC, H3A 2K6, Canada 
\\ \\
$^2$Aix Marseille Univ, Université de Toulon, CNRS, CPT, Marseille, France
\\ \\
$^3$Department of Mathematics, University of Cergy--Pontoise, CNRS UMR 8088\\
2 avenue Adolphe Chauvin, 95302 Cergy--Pontoise, France
\\ \\
$^4$Centre de Recherches Math\'ematiques, CNRS UMI 3457\\
Universit\'e de Montr\'eal, Montr\'eal,  QC, H3C 3J7, Canada}

\title{Large Deviations and Fluctuation Theorem for Selectively  Decoupled
Measures on Shift Spaces}
\date{}
\maketitle
\thispagestyle{empty}

\bigskip 
\smallskip 

{\small\textbf{Abstract.}  We establish the Level-1 and Level-3 Large
Deviation Principles (LDPs) for invariant measures on shift spaces over finite
alphabets under very general decoupling conditions for  which the thermodynamic
formalism does not apply. Such decoupling conditions arise naturally in
multifractal analysis, in Gibbs states with hard-core interactions, and in the
statistics of repeated quantum measurement processes. We also prove the LDP for
the entropy production of pairs of such measures  and derive the related
Fluctuation Relation. The proofs are based on Ruelle--Lanford functions, and the
exposition is essentially self-contained.

\bigskip


}
\tableofcontents

\setlength{\parskip}{4pt}

\section{Introduction}

This work concerns the Large Deviation Principle (LDP) for a class of invariant
probability measures on shift spaces over finite alphabets. We prove: 

\begin{enumerate}
\item	The LDP for averages  of continuous random variables ({\bf Level-1 LDP}).
\item The LDP for empirical measures ({\bf Level-3 LDP}).
\item The LDP for the  entropy  production of pairs of probability measures
({\bf Fluctuation Theorem}) together with the corresponding symmetry ({\bf
Fluctuation Relation}).
\end{enumerate}

The class of invariant probability measures we shall consider is characterized
by  certain decoupling properties  that are described in
Section~\ref{sec:assumptions}. 
As these decoupling assumptions do not imply any Gibbsian-type condition in general,
traditional methods of thermodynamic formalism (as, for example, in \cite{FO-1988,OP-1988,kifer-1990,EKW-1994,CJPS_phys})
do not apply.
The technical route that
proved effective  is based 
on Ruelle--Lanford functions. 

In this paper, we mean by Fluctuation Theorem (FT) the LDP 
for the entropy production observable, and 
Fluctuation Relation (FR) refers to the Gallavotti--Cohen symmetry \eqref{fr-i}
satisfied by the rate function governing the FT.
The FT will be established for general pairs of measures (subject to decoupling
assumptions), whereas the FR further requires the two measures to be related by
some form of involution (including, but not limited to, time reversal; see
Definition~\ref{def:familyinvolutions}). 

In the remaining part of this introduction we briefly discuss our main results, with special emphasis on Part 3 above, which is the original motivation for this work and its
most novel part. Part 3 extends and complements the results of \cite{BJPP-2017}, and the  reader may benefit from reading
introduction of \cite{BJPP-2017} in parallel with this one. We emphasize, however, that no knowledge of the works cited in this introduction is
required to understand our results and their proofs, as our exposition is
essentially self-contained starting from Section~\ref{sec-pre}.

Starting with the seminal work \cite{GC-1995a,GC-1995}, one traditional setting for  FT and FR is that of 
dynamical systems $(M, \varphi)$, where~$M$ is a compact metric space and $\varphi: M\rightarrow M$ a continuous map.
The history of the subject in this context has been reviewed in \cite{CJPS_phys}; see also \cite{MV-2003}. 
In \cite{BJPP-2017}  the metric space~$M$ was taken to be  ${\cal A}^\nn$, where
${\cal A}$ is a finite alphabet and~$\varphi$ is the left shift map. A
$\varphi$-invariant 
probability measure~$\P$ of interest arises through a repeated quantum
measurement process  generated by a quantum 
instrument on a finite-dimensional Hilbert space (we recall the precise setup in
Example~\ref{ex:QMP}). The time-reversed instrument and measurement process
yield another 
probability measure $\wP$, and the object of study is the entropic
distinguishability of the pair $(\P, \wP)$ that quantifies the 
emergence of the arrow of time in the repeated measurement process. Denoting
by~$\P_t $ and~$\wP_t$ the marginals of these 
measures on the first $t$ coordinates of $\cA^\nn$, the entropic
distinguishability is quantified by the sequence of entropy production
observables 
\[\sigma_t=\log\frac{\d \P_t}{\d\wP_t}, \qquad t \in \nn.\]
The statement of the FT is the LDP for the sequence of random variables
$(t^{-1}\sigma_t)_{t\geq 1}$ with respect to the measure $\P$. The main
application of the FT  concerns 
hypothesis testing of the pairs $(\P_t, \wP_t)$ as $t\rightarrow  \infty$. The
corresponding error exponents (Stein, Chernoff, Hoeffding) quantify 
the emergence of the arrow of time.  The proofs  in \cite{BJPP-2017}  follow a
strategy that goes back to \cite{LS-1999} and are 
centered around the so-called {\em entropic pressure} defined by 
\begin{equation}
e(\alpha)=\lim_{t\rightarrow \infty}\frac{1}{t}\log \left[\int \eee^{-\alpha
\sigma_t}\d \P_t\right], \qquad \alpha \in \rr.
\label{e-alpha}
\end{equation}
If the limit exists, and is finite and differentiable for all $\alpha \in \rr$,
then the FT follows from the G\"artner--Ellis theorem, with a rate function 
$I$ that satisfies the FR
\begin{equation}\label{fr-i}
I(-s)= I(s)+ s, \qquad s\in \rr.
\end{equation}
The difficulty with this strategy is that the measures $\P$ and $\wP$ that arise
through repeated quantum measurement processes often do not satisfy the usual
Gibbsian-type conditions
 that allow the application of the thermodynamic 
formalism and ensure the existence and regularity of the entropic pressure
defined in \eqref{e-alpha}; see Appendix \ref{ss:weakGibbsmeas} for further discussion of this point. In our setting the Gibbsian-type conditions are 
naturally replaced by a decoupling condition motivated by \cite[Proposition
2.8]{Feng2009}, which is generalized 
in Section~\ref{sec:assumptions} below under the name {\it selective lower
decoupling}. Under those decoupling conditions the measures $\P$ and $\wP$ can
exhibit a very singular behavior from the thermodynamic formalism point of view.
In \cite{BJPP-2017} a restricted form of selective symmetric decoupling (see
Section~\ref{sec:assumptions}) has been employed 
to develop a subadditive thermodynamic formalism that leads to the proof of the
existence and finiteness of the limit \eqref{e-alpha} for 
$\alpha \in [0,1]$ and the differentiability of $e$ on $(0,1)$. That sufficed
for the proof of the local LDP on the interval 
${\mathfrak J}=\,(e^\prime(0^+), e^\prime(1^-))$ (via the local G\"artner--Ellis
theorem), the validity of \eqref{fr-i} for $s\in {\mathfrak J}$, and the
development 
of hypothesis testing. It was however  clear that this route cannot be used for
the proof of the global LDP and FT since  the assumptions of \cite{BJPP-2017} 
allowed, for example, for situations where $e(\alpha)=+\infty$ for $\alpha
\not\in [0,1]$ and $|e^\prime(0^+)|=|e^\prime(1^-)|<\infty$; see the {\em
rotational instruments} in \cite{BCJPP-2017}.

The derivation of the LDPs in this work is very 
different from the one in \cite{BJPP-2017}, and is based on the  method of {\it
Ruelle--Lanford (RL) functions\/} that goes back to
\cite{ruelle_correlation_1965,lanford_entropy_1973}.\footnote{See the
introduction of \cite{lewis_entropy_1995} for a historical perspective.} The
method was then used in \cite{BAZA79}, and further developed~ in
\cite{lewis_large_1994,lewis_entropy_1995,lewis_thermodynamic_1995,pfister_thermodynamical_2002};
see also \cite{ogata_ruelle-lanford_2011}. The main ideas of the method are also
exposed in \cite[Section 4.1.2]{DZ2000}, although the terminology {\it
Ruelle--Lanford\/} does not appear there.
The method of RL functions provides a unified approach to the Level-1, Level-3,
and entropy production LDPs, and  no application of the contraction principle is
needed (in other words, the different levels are independent, although their
respective proofs have common threads).\footnote{Of course, the Level-1 LDP can
also be obtained from the Level-3 LDP by the contraction principle, and so can
the Level-2 LDP, which we do not discuss except in Remark~\ref{rem:level2}. We
only include an explicit derivation of the Level-1 LDP for pedagogical purposes,
as it is a simple illustration of the method. See also
Remark~\ref{rmq:contractionlvl3}.} We are not aware of any previous use of RL
functions in the study of entropy production.  It should be added that our application of the  RL function method
is specific to $M={\cal A}^\nn$ (with 
straightforward extension to $M={\cal A}^\zz$), and it is an interesting open question whether this  method can be applied to other classes of dynamical systems.

The paper is organized as follows. In Section~\ref{sec-pre} we describe our general
setting, state the assumptions and  main results, 
and discuss several examples. For reasons of space, the detailed discussion of examples
related to quantum measurement processes is postponed 
to \cite{BCJPP-2017}, which is a continuation of \cite{BJPP-2017} and this work.
The general construction leading to the proof 
of our main results together with a presentation of the method of RL functions
is given in Section~\ref{sec-LF}.  The proofs of the main results are presented
in Sections~\ref{sec-ldp1}, \ref{sec-ldpep}, and  \ref{sec-ldp3}. In the
appendix we describe further applications (in particular to weak Gibbs measures,
which do not fit directly into our assumptions) and develop a prototypical
example of hidden Markov chain where the present method applies but the usual
methods based on thermodynamic formalism and the methods of \cite{BJPP-2017} do not apply.

We finish with the following general remark. The RL functions method  turned out
to be  surprisingly effective for our purposes. Although this method is both
very powerful and natural, it appears to be a lesser used route to LDPs.
Even in the cases where  the respective 
LDPs  are well known, this  approach  gives a new perspective on the results and
their proofs. The 
assumptions under which the method is used here are different from the ones
existing in the literature, and we  hope that the  essentially self-contained 
presentation given in this paper will facilitate its future applications.

\subsection*{Acknowledgments} 
We are grateful to C.-E. Pfister for helpful discussions and advice. 
This research was supported by the {\it Agence Nationale de la Recherche\/}
through the grant NONSTOPS (ANR-17-CE40-0006-01, ANR-17-CE40-0006-02,
ANR-17-CE40-0006-03), the CNRS collaboration grant {\it Fluctuation theorems in
stochastic systems\/}, and the {\it Initiative d'excellence Paris-Seine\/}. NC
was supported by Swiss National Science Foundation Grant 165057. VJ acknowledges
the support of NSERC. The work of CAP has been carried out in the framework of
the Labex Archim\`ede (ANR-11-LABX-0033) and of the A*MIDEX project
(ANR-11-IDEX-0001-02), funded by the ``Investissements d'Avenir'' French
Government programme managed by the French National Research Agency (ANR). The
research of AS was carried out within the MME-DII Center of Excellence
(ANR-11-LABX-0023-01).

\section{Preliminaries and main results}
\label{sec-pre}

\subsection{Setup and notation}\label{ss:setupnot}

Let $\aA$ be a finite set and let\footnote{We adopt the convention that 
$\nn=\{1,2, 3,\ldots\}$ and $\nn_0 = \nn \cup \{0\}$.} $\Omega=\aA^\N$ 
be the set of sequences
$\omega=(\omega_j)_{j\in\nn}$ whose elements belong to~$\aA$.  We denote by
$\varphi:\Omega\to\Omega$ the left shift defined by
$\varphi(\omega)_j=\omega_{j+1}$ for $j\in\nn$.
We write also $\Omega_t={\cal A}^{\llbracket1,t\rrbracket}$, with $\llbracket1,t\rrbracket=[1,t]\cap\nn$, and 
call $w = (w_1, \dots, w_t)\in \Omega_t$ a {\em word} of length $|w|=t$.
Given a sequence $\omega \in \Omega$  and two integers $m\le n$, we set
$\omega_{[m,n]}=(\omega_m,\dots,\omega_n)$, and similarly for $w\in \Omega_t$ if
$m\le n \le t$. The set of words of finite length is denoted by   $\Omega_\fin =
\bigcup_{t\in \nn_0} \Omega_t$, with the convention that $\Omega_0 = \{\kappa\}$,
where $\kappa$ is the ``empty word'' ($|\kappa| = 0$). 
Given $u,v\in \Omega_\fin$,  $uv \in \Omega_\fin$ denotes the natural
concatenation of $u$ and $v$, which satisfies $|uv| = |u| + |v|$. For the empty
word $\kappa$ and any $w\in \Omega_\fin$,   $w\kappa = \kappa w = w$.

The set~$\Omega$  is endowed with  the product topology and the corresponding
Borel $\sigma$-algebra~$\cF$.
We denote by $C(\Omega)$ the Banach space of real-valued continuous
functions on $\Omega$ with the sup-norm.  The set $\cP(\Omega)$ of Borel probability measures on
$\Omega$ is endowed with the weak topology.\footnote{This topology is metrizable
and 
${\cal P}(\Omega)$ is a compact metric space.} We shall write~$\Q$ for generic
elements of~$\cP(\Omega)$ and use the symbol $\P$ for the  probability measure
that will be fixed throughout. $\QQ\in\cP(\Omega)$ is $\varphi$-invariant,
or invariant for short, whenever $\Q\circ\varphi^{-1}=\Q$. We denote by 
$\PP_\varphi(\Omega)$ the set of invariant elements of $\PP(\Omega)$.
For $\QQ \in\cP(\Omega)$ and $f\in L^1(\QQ)$ we write
\begin{equation*}
\langle f,\QQ \rangle = \int f \dd \QQ.
\end{equation*}
Given a word $w\in \Omega_t$ with $t\in \nn$, we introduce the cylinder set
$\cC_{w}=\{\omega \in \Omega: \omega_{[1,t]} = w\}$. We adopt the convention that 
$\cC_\kappa=\Omega$ for the empty word~$\kappa$, and we denote by $(\cF_t)_{t\in \nn_0}$ 
the filtration generated by the cylinder sets.\footnote{Notice that $\cF_t$ is the 
finite algebra generated by the elements of $\{\cC_w: w\in \Omega_t\}$.} 

For any $\QQ \in \cP(\Omega)$ and any~$t\in\nn$, $\QQ_t$ denotes  the
restriction of~$\QQ$ to $\cF_t$, which we identify with a function on $\Omega_t$
in the natural way:
\begin{equation*}
\QQ_t(w) = \QQ({\cC_w}) =: \QQ(w), \qquad w\in \Omega_t,
\end{equation*}
where the expression $\QQ(w)$ is used by a slight abuse of notation.
Consistently with the convention that $\cC_\kappa = \Omega$, we have
$\QQ(\kappa) = \QQ_0(\kappa) = 1$.

Throughout the paper, we fix an invariant probability measure $\P\in
\PP_\varphi(\Omega)$, which will be subject to some assumptions below. We write
\begin{equation}
\label{eq:OmegaPlus}
\Omega^+ = \supp\, \P = \{\omega \in \Omega: \P(\omega_{[1,t]}) > 0\text{ for all }t\in\nn\}
\end{equation}
and notice that~$\Omega^+$ is a {\em subshift,} \ie a closed subset of $\Omega$
satisfying $\varphi(\Omega^+) = \Omega^+$. For $t\in\nn$, let $\Omega_t^+ = \{w \in
\Omega_t : \P_t(w) > 0\}$ and set $\Omega_\fin^+ = \bigcup_{t\in \nn_0}\Omega_t^+ = \{w
\in \Omega_\fin : \P(w) > 0\}$. The set $\Omega_\fin^+$ is a {\it
language\/}\footnote{See for example \cite[Proposition 1.3.4]{lind1995introduction}. The notions of
subshift and language are not crucial in our study; they will only be used to
discuss how (weak) Gibbs measures on subshifts fit into our assumptions (see
Example~\ref{ex:WGSS} and Appendix~\ref{ss:weakGibbsmeas}).} in the sense that
for each $w \in \Omega_\fin^+$ the following holds: (1) each subword of $w$ is
also in $\Omega_\fin^+$ and (2) there exist non-empty words $u,v$ such that $uwv
\in \Omega_\fin^+$.

Finally, we use throughout the conventions that $\log 0 = -\infty$, and $0\log 0
= 0$.

\subsection{Assumptions}\label{sec:assumptions}

We now introduce a set of {\em decoupling} assumptions on $\P$ (which are not in
force throughout). Without further saying, we shall always assume that  the 
sequences $(\tau_t)_{t\in\nn}\subset\N_0$ and 
$(c_t)_{t\in\nn}\subset[0, \infty)$ which appear in these assumptions  
satisfy
\[
\lim_{t\to\infty} \frac {c_t}t =\lim_{t\to\infty} \frac {\tau_t}t = 0,
\]
which we write as of now $c_t = o(t)$ and $\tau_t = o(t)$.

The assumption that will play the central role in our work is
\newcommand{\hSLD}{\hyperlink{hyp.SLD}{\textbf{(SLD)}}\xspace}
\begin{description}
{\itshape
\item[\hypertarget{hyp.SLD}{Selective Lower Decoupling (SLD).}] 
For all $t\in\nn$, all $u \in \Omega_t$ and all $v\in \Omega_{\fin}$, $|v|\geq
1$, there exists $\xi \in \Omega_\fin$, $|\xi| \leq  \tau_t $, such that 
\begin{equation}\label{eq:sellowerdecoup}
\IP(u \xi v) \geq \eee^{- c_t} \IP(u) \IP(v).
\end{equation}
(Note that we take $|\xi|\leq  \tau_t$ and not $|\xi| =  \tau_t $; this is
crucial).
}
\end{description}

In order to refine some of the results (see Theorem~\ref{t1.9}), we will
sometimes also assume
\newcommand{\hUD}{\hyperlink{hyp.UD}{\textbf{(UD)}}\xspace}
\begin{description}
{\itshape
\item[\hypertarget{hyp.UD}{Upper Decoupling (UD).}] 
For all $t\in\nn$, all $u \in \Omega_t$ and all $v\in \Omega_{\fin}$, $|v|\geq
1$,
\begin{equation*}
\sup_{\xi \in \Omega_{ \tau_t}}\IP(u\xi v)\le \eee^{ c_t}\IP(u\bigr)\IP(v\bigr).
\end{equation*}
}
\end{description}

Some of our results involve a pair of measures $(\P, \widehat \P)$, where $\P$
is as above, and $\widehat \P\in \cP_\varphi(\Omega)$ is another invariant probability
measure. When we consider a pair $(\P, \widehat \P)$, we always assume the
following absolute continuity condition:
\begin{equation}\label{eq:abscontinuous}
\P_t  \ll \wP_t \quad\text {for all }t\in\nn.
\end{equation}
Interesting cases include when $\wP$ is the uniform measure\footnote{That is,
$\wP_t(w)=|{\Omega_t}|^{-1}$ for $w \in \Omega_t$.} on $\Omega$, 
and when $\wP$ is obtained from some transformation of $\P$ (see
Definition~\ref{def:familyinvolutions}). This leads to our final assumption that
concerns the pair $(\P, \widehat \P)$:
\newcommand{\hSSD}{\hyperlink{hyp.SSD}{\textbf{(SSD)}}\xspace}
\begin{description}
{\itshape
\item[\hypertarget{hyp.SSD}{Selective Symmetric Decoupling (SSD).}] 
For all $t\in\nn$, all $u \in \Omega_t$ and all $v\in \Omega_{\fin}$, $|v|\geq
1$, there exists $\xi \in \Omega_\fin$, $|\xi| \leq  \tau_t $, such that for
both  $\P^\sharp = \P$ and $\P^\sharp = \widehat \P$ we have
\begin{equation}\label{eq:timereversibledec}
 \eee^{- c_t} \IP^\sharp(u) \IP^\sharp(v) \leq \IP^\sharp(u \xi v)\leq  \eee^{
c_t}\IP^\sharp(u)\IP^\sharp(v).
\end{equation}
(Note that this is the same $\xi$ for both $\P$ and $\widehat \P$).
}
\end{description}

\begin{remark}\label{rem:weakerstrongerD} The same sequences $(c_t)_{t\in\nn}$ and 
$(\tau_t)_{t\in\nn}$ are used
in the different conditions above. This results in no loss of generality. This
is obvious for $c_t$, while for $\tau_t$ the argument is slightly more involved.
It is immediate that for~\hSLD and~\hSSD, $\tau_t$ can always be
replaced by some $\tau_t' \geq \tau_t$. The same is true of~\hUD, since we
have for all $\xi \in \Omega_{\tau'_t}$ that $\xi = \xi' b$ for some $\xi'\in
\Omega_{\tau_t}$ and $b \in \Omega_\fin$, and then 
\[\P(u\xi v) = \P(u \xi' bv) \leq \eee^{c_t}\P(u) \P(bv) \leq \eee^{c_t}\P(u)
\P(v).\]
\end{remark}

\begin{remark}\label{rem:otherassumptions} \hSLD is implied by the seemingly
weaker condition that 
\begin{equation*}
\sum_{\xi \in \Omega_\fin \,:\,|\xi|\leq  \tau_t }\IP(u \xi v) \geq \eee^{-
c'_t} \IP(u) \IP(v)
\end{equation*}
for some $c'_t = o(t)$. In this case~\eqref{eq:sellowerdecoup} is easily shown
to hold\footnote{We use that for any finite set $A$, $\sum_{x\in A} f(x) \leq
|A| \max_{x\in A} f(x)$.} with $c_t = c'_t +  \log(\sum_{i=0}^{ \tau_t }|\cA|^i
) = o(t)$.
Similarly, \hUD  implies the seemingly stronger assumption that
\begin{equation*}
\sum_{\xi \in \Omega_{ \tau_t }}\IP(u\xi v)\le \eee^{ c'_t}\IP(u)\IP(v)
\end{equation*}
if we choose $ c'_t =   c_t + \log(|\cA|^{ \tau_t }) = o(t)$.
\end{remark}

\begin{remark}\label{rem:hyposymmetricdec}
Unless $\tau_t \equiv 0$, \hSSD does not imply \hUD, since the upper bound in
\eqref{eq:timereversibledec} has to be satisfied only for the ``selected''
$\xi$. \hSSD does, however, imply \hSLD for both $\P$ and $\widehat \P$, with the
additional information that we can choose the {\em same} $\xi$ for both $\P$ and
$\widehat \P$. 
On the other hand, in order to have \hSSD, it is enough to have \hUD for both $\P$
and $\widehat \P$ as well as \hSLD for both $\P$ and $\widehat \P$ with the same
$\xi$ (in general $c_t$ and $\tau_t$ have to be increased, see
Lemma~\ref{lem:chgthyposym}).
\end{remark}

\begin{remark}\label{rem:ergo} The measure $\P$ is not assumed to be ergodic.
One can show, however, that it is ergodic if, for example, \hSLD holds with
$\sup_t \tau_t < \infty$ and $\sup_t c_t < \infty$ (see
Lemma~\ref{lem:ergodicQstar}).
\end{remark}

One special case of interest is when $\wP$ is related to $\P$ by a
transformation defined as follows. 
\begin{definition}\label{def:familyinvolutions} For each $t\in\nn$, let
$\theta_t: \Omega_t \to \Omega_t$ be an involution. Assume that the sequence
$\Theta = (\theta_t)_{t\in\nn}$ is such that one of the following holds for some
involution $u: \cA \to \cA$:
\begin{enumerate}
	\item  $\theta_t(w_1, w_2, \dots, w_t) = (u(w_1), u(w_2), \dots, u(w_t))$ for
each $t\in\nn$, $w\in \Omega_t$;
	\item  $\theta_t(w_1, w_2, \dots, w_t) = (u(w_t), u(w_{t-1}), \dots, u(w_1))$
for each $t\in\nn$, $w\in \Omega_t$.
 \end{enumerate}
For each $\Q\in \cP_\varphi(\Omega)$, we denote by $\Theta \Q$ the invariant
measure on $\Omega$ obtained by  extending the family\footnote{One easily shows
that both conditions imply that $\sum_{a\in \cA}\Q_{t+1}\circ \theta_{t+1}(wa) =
\sum_{a\in \cA}\Q_{t+1}\circ \theta_{t+1}(aw) = \Q_t\circ \theta_{t}(w)$, which,
by Kolmogorov's extension theorem, guarantees that such an invariant extension
exists.} of marginals $((\Theta \Q)_t)_{t\in\nn}$, where $(\Theta
\Q)_t=\Q_t\circ\theta_t$.
\end{definition}

We shall  see below that when $\widehat \P = \Theta \P$ for some $\Theta$ as
above, the  FT rate function satisfies the celebrated 
Fluctuation Relation \cite{ECM-1993,ES-1994,GC-1995,GC-1995a,gallavotti-1995}. 
\begin{remark}\label{rem:abscompattheta} By the absolute continuity assumption
\eqref{eq:abscontinuous}, in order for the choice $\wP = \Theta \P$ to be
allowed, $\Theta$ and $\P$ must be so that $\P_t \ll (\Theta \P)_t$ for all $t\in\nn$.
Since  $\theta_t$ is an involution (and hence a bijection), the support of
$\P_t$ and that of $(\Theta \P)_t$ (as subsets of $\Omega_t$) have the same
cardinality, and hence $\wP = \Theta\P$ implies that  $\P_t$ and $\wP_t$ are
equivalent for all $t$.
\end{remark}

We finish with several  comments on the relation between the decoupling
assumptions described in this section  and those to be found in the literature. 
\begin{itemize}
	\item Our decoupling assumptions are related to those in \cite[Definition
3.2]{pfister_thermodynamical_2002} (restricted to one-sided shift spaces). In view of
Remark~\ref{rem:weakerstrongerD}, the upper decoupling assumption is the same.
Our \hSLD condition is weaker than the lower decoupling condition in
\cite{pfister_thermodynamical_2002}, as we allow $|\xi| \leq \tau_t$ instead of
$|\xi| = \tau_t$. This weaker condition covers  some important  classes of
measures (see the examples below), which are not covered by any result in the
literature, as far as we are aware. The Ruelle--Lanford estimates, which are
done in the spirit of \cite{pfister_thermodynamical_2002}, are noticeably
complicated by the fact that we allow $|\xi| \leq \tau_t$ in the \hSLD condition.
\item The main feature of our \hSLD assumption, \ie allowing $|\xi| \leq \tau_t$,
is reminiscent of some variants of the specification property for subshifts of
$\Omega$, which allow for similar ``flexibility'' (see for example
\cite{pfister_billingsley_2003,PS-2005,thompson_irregular_2012,PS-2018}).
Specification properties are conditions on the structure of the subshift (viewed
as a metric space in itself), not on measures defined on it. We shall  discuss
Gibbs states and (weak) Gibbs measures whose supports satisfy such ``flexible''
forms of specification property in Examples~\ref{ex:WGSS}, \ref{eq:Gibbs1D} and
in Appendix~\ref{ss:weakGibbsmeas}.
\item  A property similar to \hSLD (with $\tau_t$ and $c_t$ independent of $t$)
was observed to hold for some products of matrices in \cite[Proposition
2.8]{Feng2009}, and some parts of our construction are similar to
\cite{Feng2009}. See Example~\ref{ex:matrixprod} below.
\item To the best of our knowledge, the only assumptions similar to \hSSD to be
found in the literature are in \cite{BJPP-2017}, with $c_t$ and $\tau_t$ not
allowed to depend on $t$ (see Assumptions (C) and (D) therein, and
Example~\ref{ex:QMP} below).
\end{itemize}

\subsection{Main results}

We endow $\R^d$ with the Euclidean structure and denote by $|\argdot|$ and
$(\argdot, \argdot)$ the corresponding  norm and inner product. Given a function
$f: \Omega \to \R^d$, we write $\|f\| = \sup_{\omega\in \Omega}|f(\omega)|$ and 
introduce
\begin{equation*}
 S_t f(\omega)=\sum_{s=0}^{t-1}f(\varphi^s(\omega)), \qquad t\in \nn. 
\end{equation*}
Let us recall that in the standard LDP terminology, a {\em rate function} is
always assumed to be lower semicontinuous, while a  {\em good rate function}
has, in addition, compact level sets. The next result follows from
Propositions~\ref{prop:expressureqfalpha} and~\ref{prop:fctcontinueadmissible}
below.

\begin{theorem}[Level-1 LDP]\label{thm:summarythmlev1} Assume \hSLD and let $f\in
C(\Omega, \R^d)$. 
\begin{enumerate}
	\item For all $\alpha\in \R^d$, the limit
\begin{equation*}
q_f(\alpha) :=  \lim_{t\to\infty}\frac 1 t \log \left\langle  \eee^{ ( \alpha ,
S_{t} f ) }, \P \right\rangle 
\end{equation*}
exists, is finite, and the mapping $\alpha \mapsto q_f(\alpha)$ is convex and
$\|f\|$-Lipschitz. 
\item The sequence of random variables $(\frac 1 t S_t f)_{t\in\nn}$ satisfies
the LDP with a good convex rate function $I_f$, in the sense that for every open
set $O \subset \R^d$ and every closed set $\Gamma\subset \R^d$,
\begin{align}
 \liminf_{t\to\infty} \frac 1 t \log \P\left(\frac 1 t S_t f \in O\right) & \geq
-\inf_{x\in  O} I_f(x),\label{eq:ldplev1up}\\
 \limsup_{t\to\infty} \frac 1 t \log \P\left(\frac 1 t S_t f\in \Gamma\right)
&\leq -\inf_{x\in  \Gamma} I_f(x).\label{eq:ldplev1lo}
\end{align}
Moreover, $I_f$ is the Fenchel--Legendre transform $q_f^*$ of $q_f$, \ie
\begin{equation*}
	I_f(x) = q_f^*(x) = \sup_{\alpha\in \R^d}((\alpha, x) - q_f(\alpha)), \qquad x
\in \rr^d.
\end{equation*}
\end{enumerate}
\end{theorem}

We define the {\it entropy production observable} over the time 
interval~$\llbracket1,t\rrbracket$
by
\begin{equation} \label{1.81}
\sigma_t=\log\frac{\dd\IP_t}{\dd\widehat\IP_t},
\end{equation}
which is $\cF_t$-measurable and well defined  $\P_t$-almost surely since $\P_t 
\ll \wP_t$.  The entropy production observable is also called
{\em log-likelihood ratio} in the context of hypothesis testing.

The next result follows from Propositions~\ref{prop:sigmapressure}
and \ref{prop:ldpsigmafrst}.

\begin{theorem}[LDP for entropy production]\label{main-2} Assume \hSSD.
\begin{enumerate}
\item For all $\alpha\in\rr$, the limit\footnote{Note that the sign of $\alpha$ in
\eqref{eq:defqalphaentrop} is different from that in \eqref{e-alpha}. See
Remark~\ref{rem:sigmaq}.}
\begin{equation}\label{eq:defqalphaentrop}
q(\alpha)
:=\lim_{t\to\infty}\frac1t\log\left\langle\eee^{\alpha \sigma_t},\P\right\rangle
\end{equation}
exists and defines a closed proper convex function\footnote{\ie it is convex, lower semicontinuous and not everywhere 
infinite.} on $\rr$
taking its values in $(-\infty, \infty]$. In particular $q(0) = 0$ and 
$q(-1)\leq 0$, so that $q$ is non-positive (and hence finite) on $[-1,0]$.
\item The sequence of random variables $(\frac 1 t \sigma_t)_{t\in\nn}$
satisfies the LDP with a convex rate function $I$ in the sense that
 for every open set $O \subset \R$ and every closed set $\Gamma\subset \R$,
\begin{align}
 \liminf_{t\to\infty} \frac 1 t \log \P\left(\frac 1 t \sigma_t \in O\right) &
\geq -\inf_{s\in  O} I(s),\label{eq:weakLDPsigmaup}\\
 \limsup_{t\to\infty} \frac 1 t \log \P\left(\frac 1 t \sigma_t \in
\Gamma\right) &\leq -\inf_{s\in  \Gamma} I(s).
\label{eq:weakLDPsigma}
\end{align}
Moreover,  $I$ is the Fenchel--Legendre transform $q^*$ of $q$, \ie
\begin{equation}\label{eq:legendreIentrop}
	I(s) = q^*(s) =   \sup_{\alpha\in \R}(\alpha s - q(\alpha)), \qquad s\in \rr.
\end{equation}
\item If $q$ is finite in a neighborhood of $0$, then $I$ is a good rate
function.
\item If $\wP = \Theta \P$ for some $\Theta$ as in
Definition~\ref{def:familyinvolutions},\footnote{The \hSSD assumption of the
theorem is still in force.} then $q$ satisfies the symmetry
\begin{equation}\label{eq:symmetryealpha}
q(-\alpha)=q(\alpha-1), \qquad \alpha \in \R,
\end{equation}
and $I$ satisfies the Fluctuation Relation (also known as the {\em
Gallavotti--Cohen} symmetry)
\begin{equation}\label{eq:GCsigma}
I(-s)=I(s)+s, \qquad s\in \R.
\end{equation}
\end{enumerate}
\end{theorem}

\begin{remark}\label{rem:sigmaq}
In the physics literature, it is more common to work with $e(\alpha) =
q(-\alpha)$ as in \eqref{e-alpha}. Then \eqref{eq:legendreIentrop} and
\eqref{eq:symmetryealpha} read respectively 
\[I(s)=-\inf_{\alpha\in \R}(\alpha s +e(\alpha)), \qquad s\in \rr,
\]
\begin{equation*}
e(\alpha)=e(1-\alpha), \qquad \alpha \in \R.
\end{equation*}
This is relevant for the applications to hypothesis testing that are discussed
in Section~\ref{ss:hypotesting}. 
\end{remark}

The following remark gives a sufficient condition for $q(\alpha)$ to be finite
for all $\alpha \in \rr$.
\begin{remark}\label{rem:finiteqalpha}
Assume that, in addition to \hSSD, we have for all $t\in\nn$, all $u \in \Omega_t$
and all $v\in \Omega_{\fin}$, $|v|\geq 1$, that both  $\P^\sharp = \P$ and
$\P^\sharp = \widehat \P$ satisfy
\begin{equation}\label{eq:strongLD}
\IP^\sharp(u  v) \geq  \eee^{- c_t}\IP^\sharp(u) \IP^\sharp(v).
\end{equation}	
Let moreover $c$ be the minimum of all the non-zero values achieved by $\P(a)$
and $\wP(a)$ for $a\in \cA$.
Then, for all $w\in \Omega_t^+$, we find $\P^\sharp(w) \geq
\eee^{-(t-1)c_1}\P^\sharp(w_1)\cdots \P^\sharp(w_t) \geq \eee^{-t(c_1 - \log
c)}$. As a consequence, we obtain that $\sup_{t\in\nn}\sup_{w\in \Omega_t^+}
|t^{-1} \sigma_t| < \infty$, which implies in particular that $q(\alpha)$ is
finite for all $\alpha \in \rr$.
\end{remark}

\begin{remark} If $\widehat \P$ is the uniform measure, then
 $\frac 1 t \sigma_t  = \frac 1 t \log \P_t + \log |\cA|$. Assuming \hSSD (note
that \eqref{eq:timereversibledec} trivially holds for~$\widehat\IP$), Parts 1-2
of the theorem above apply. We then have 
\begin{equation*}
	r(\alpha) := \lim_{t\to\infty}\frac 1 t \log \left\langle \eee^{\alpha \log
\P_t} , \P\right\rangle = q(\alpha) - \alpha \log |\cA|, \qquad \alpha\in \rr,
\end{equation*}
and  $r$ inherits the properties of existence, convexity and lower
semicontinuity of $q$. Moreover, $\frac 1 t \log \P_t$ satisfies the LDP with
convex rate function $J(s) = I(s + \log|\cA|)$, which can be identified with the
Fenchel--Legendre transform of $r$.  Part~3 extends in an obvious way to $J$ and
$r$.  In order to apply the discussion of Remark~\ref{rem:finiteqalpha}, it
suffices to verify \eqref{eq:strongLD} for $\P$, since   \eqref{eq:strongLD} is
trivially satisfied for the uniform measure  $\wP$. Note also that $r$ is
related to the R\'enyi entropy of $\P$, since
\begin{equation*}
r(\alpha ) 	= \lim_{t\to\infty}\frac 1 t \log \sum_{w\in \Omega_t^+}
(\P(w))^{1+\alpha}.
\end{equation*}
\end{remark}
\begin{remark}
By the Shannon--McMillan--Breiman (SMB) theorem,\footnote{For this result no
assumptions are needed; the SMB theorem holds 
for any $\P\in {\cal P}_\varphi(\Omega)$.} the limit
\[ H(\omega):=-\lim_{t\rightarrow \infty}\frac{1}{t}\log \P_t(\omega_{[1,t]})\]
exists $\P$-almost surely and in $L^1(\P)$, $H \circ \varphi=H$, and
$\int_\Omega H \d \P=h(\P)$, where $h(\P)$ is the Kolmogorov--Sinai 
entropy of $\P$. Thus, in the case when~$\wP$ is the uniform measure,
Theorem~\ref{main-2} provides the LDP 
counterpart to the SMB theorem, and establishes the result that was originally
intended for the fourth  instalment in the series of papers initiated 
by \cite{BJPP-2017}.
\end{remark}

We now turn to the Level-3 LDP. The sequence of {\em empirical  measures}
$(\mu_t)_{t\in\nn}$ is defined by 
\begin{equation*}
\mu_t(\omega) = \frac 1 t\sum_{s=0}^{t-1} \delta_{\varphi^s (\omega)} \in
\cP(\Omega), \qquad \omega \in \Omega,~t\in\nn.
\end{equation*}
We also recall that the relative entropy of two probability measures $\QQ$
and~$\QQ'$ on a measurable space~$(X, \FF)$ is given by 
\begin{equation} \label{1.9}
\Ent(\QQ'\,|\,\QQ)=
\left\{
\begin{aligned}
\int_X\log&\Bigl(\frac{\dd\QQ'}{\dd\QQ}\Bigr)\dd\QQ'&\quad &\mbox{if
$\QQ'\ll\QQ$},\\[4pt]
&+\infty&\quad&\mbox{otherwise}.  
\end{aligned}
\right.
\end{equation}
In what follows, we always assume that~$\PP(\Omega)$ is endowed with the weak
topology and the corresponding Borel $\sigma$-algebra. 
The next result follows from Propositions~\ref{prop:expressureqfalpha},
\ref{prop:level3ldp}, \ref{p1.6}, and \ref{p1.11}.
\begin{theorem}[Level-3 LDP]\label{t1.9} Assume \hSLD. 
\begin{enumerate}
\item For all $f\in C(\Omega, \R)$, the limit
\begin{equation}\label{eq:defQfl3}
Q(f) := \lim_{t\to\infty}\frac 1 t \log \left\langle \eee^{ S_t f },
\P\right\rangle
\end{equation}
exists and defines a convex, $1$-Lipschitz function on $C(\Omega, \R)$.
\item 	The sequence of random variables $(\mu_t)_{t\in\nn}$ satisfies the LDP on
the space $\cP(\Omega)$ with some good\footnote
{Notice that goodness follows from lower semicontinuity since ${\cal P}(\Omega)$
is  compact.} convex rate function $\I$, \ie for every open set $O \subset
\cP(\Omega)$, and every closed set $\Gamma\subset \cP(\Omega)$,
\begin{align*}
 \liminf_{t\to\infty} \frac 1 t \log \P\left(\mu_t \in O\right) & \geq
-\inf_{\QQ\in  O} \I(\QQ),\\
 \limsup_{t\to\infty} \frac 1 t \log \P\left(\mu_t \in \Gamma\right) &\leq
-\inf_{\QQ\in  \Gamma} \I(\QQ).
\end{align*}
Moreover, $\I$ is the restriction of the Fenchel--Legendre transform $Q^\ast$ of $Q$ to $\cP(\Omega)$, \ie
\begin{equation}\label{eq:defIIlev3}
\I(\QQ) = \sup_{f\in C(\Omega, \R)}\big(\langle f,\QQ\rangle -Q(f)\big), \qquad \QQ \in \cP(\Omega),
\end{equation}
and satisfies $\I(\QQ) = +\infty$ for $\QQ \in \cP(\Omega) \setminus
\cP_\varphi(\Omega)$. 
\item Assuming \hUD (in addition to \hSLD), we have for any 
$\Q\in\cP_\varphi(\Omega)$ that
\begin{equation} \label{eq:identificationIentrop}
{\II}(\Q)=\lim_{t\to\infty} \frac 1 t\Ent(\Q_t|\,\IP_t),
\end{equation}
and ${\II}$ is an affine function of $\Q\in\PP_\varphi(\Omega)$.
\item Assume again \hUD and \hSLD. Assume moreover that $\wP = \Theta \P$ for some
$\Theta$ as in Definition~\ref{def:familyinvolutions}. Then, for
any~$\Q\in\PP_\varphi(\Omega)$ such that $(\Theta \Q)_t$ and~$\Q_t$ are
equivalent for all~$t$,  ${\II}(\Q)<+\infty$, and  ${\II}(\Theta\Q)<+\infty$,
the following {\em Level-3 fluctuation relation} holds: 
\begin{equation} \label{lev-3}
{\II}(\Theta\Q)-{\II}(\Q)=\lim_{t\to\infty}\frac 1 t \langle\sigma_t,\Q\rangle.
\end{equation}
\end{enumerate}

\end{theorem}
\begin{remark}
Note that in general, the identification \eqref{eq:identificationIentrop} is not
possible for $\Q \in \cP(\Omega) \setminus \cP_\varphi(\Omega)$; for such
measures the left-hand side is infinite, but the right-hand side need not be.
\end{remark}
\begin{remark} The right-hand side of \eqref{lev-3} is interpreted as the {\rm
mean entropy production}  of the pair $(\P, \wP)$ w.r.t.~$\Q$. 
For a discussion of the Level-3 fluctuation relation (\ref{lev-3}), we refer the
reader to \cite{CJPS_phys}.
\end{remark}

\begin{remark}\label{rmq:contractionlvl3} As mentioned, the Level-1 LDP of
Theorem~\ref{thm:summarythmlev1}  can be retrieved from the Level-3
LDP by using the contraction principle. In our proofs, however, the Level-1 LDP
is established first, and then the Level-3 LDP is proved independently (although
the two proofs have many common points). A natural question is whether, as in
\cite{CJPS_phys}, the LDP for the entropy production can be retrieved by
 contraction from Level 3, and if then~\eqref{eq:GCsigma}
follows from~\eqref{lev-3}. We are not aware of a way of doing so at the level
of generality of \hSSD, as $\sigma_t$ may be highly ``non-additive'' (see
Example~\ref{ex:HiddenMarkov}). 
\end{remark}

\begin{remark}\label{rem:level2}Assuming \hSLD, the following {\em Level-2 LDP}
holds: the sequence
\[
\left(\frac 1 t\sum_{s=0}^{t-1}\delta_{\omega_s}\right)_{t\in\nn}\subset \cP(\cA)
\]
satisfies the LDP with respect to a good convex rate function
$I_2$, which can be expressed as
\begin{equation*}
I_2(\nu) =\sup_{f\in C(\cA)}  \left(\sum_{a\in \cA} f(a) \nu(a)	- Q(f)\right),
\qquad \nu\in \cP(\cA),
\end{equation*}
where $Q(f)$ is as in \eqref{eq:defQfl3} (viewing $f\in C(\cA)$ as a function on
$\Omega$ depending only on $\omega_1$). This result can be obtained in three
different ways: (1) by applying the Level-1 LDP to some well-chosen
$\rr^{\cA}$-valued function, (2) from the Level-3 LDP by the contraction
principle, or (3) independently of the others LDPs by applying the
Ruelle-Lanford functions method directly. We shall not discuss the Level-2 LDP
any further.
\end{remark}

\begin{remark}
Although the work \cite{BJPP-2017} was focused on the LDP for entropy 
production for pairs of probability measures $\P$ and $\Theta \P$ obtained from
repeated quantum measurement processes, with $\Theta$ as in Case~2 of
Definition~\ref{def:familyinvolutions}, the method of proof 
extends to the general setting of this paper and yields the following: (a) 
Theorems~\ref{thm:summarythmlev1} and \ref{t1.9}  hold\footnote{By adapting the
proof of \cite[Theorem 2.5]{BJPP-2017} to the case where $\sigma_n$ is replaced
by $S_n f$ for any function $f$ depending on finitely many variables, one
verifies the assumptions of \cite[Theorem 2.1]{kifer-1990}, which yield the
Level-3 LDP.} assuming \hUD  and \hSLD with $\tau_t$ and $c_t$ that do  not depend
on $t$; (b) a local version of Theorem~\ref{main-2} holds assuming \hSSD and \hUD
(for both $\P$ and $\wP$) with 
$c_t$ and $\tau_t$ that do not depend on $t$; (c) Theorem~\ref{main-2} holds
assuming \hSSD with $\tau_t \equiv 0$ and with $c_t$ that does not depend on~$t$.
In this context, see Example~\ref{ex:HiddenMarkov}. 
\end{remark}
\subsection{Hypothesis testing}\label{ss:hypotesting}

An important application of Theorem~\ref{main-2} concerns asymptotic hypothesis
testing of the pairs of measures $(\P_t, \wP_t)$ as 
$t\rightarrow \infty$. The discussion of this point  is nearly identical  to the
one presented in Section 2.9 of \cite{BJPP-2017} (see also
\cite{CJPS_phys,JNPPS}), and we shall only briefly comment on a few
changes that are needed due to the generality of our setting.  

Unless $\wP = \Theta \P$, the function $\alpha \mapsto e(\alpha)=q(-\alpha)$
that controls the
Chernoff and Hoeffding error exponents does not need to satisfy the symmetry 
$e(\alpha)=e(1-\alpha)$. In this case  the 
upper and lower Chernoff exponents $\bar c$ and $\ubar c$ satisfy\footnote{If
the symmetry holds, as in \cite{BJPP-2017}, then obviously 
$\bar c=\ubar c=e(1/2)$.} 
\[ \bar c=\ubar c =\min_{\alpha \in [0,1]}e(\alpha).\]
The formula for the Hoeffding error exponents (Theorem 2.13 in \cite{BJPP-2017})
remains unchanged. We also remark that although the 
analysis of the Chernoff and Hoeffding error exponents presented in 
\cite{JOPS-2012} (and used in \cite{BJPP-2017}) required the function
$e$ to be differentiable on the interval $(0,1)$, this assumption was used only through the
application of the induced local LDP for 
the entropy production.\footnote{The local LDP followed by an application of the
G\"artner--Ellis theorem.} In case $e$ exists but is not necessarily differentiable,
and if the required LDP with a convex rate function is established by other means, as is the  case in
Theorem~\ref{main-2}, then the analysis of  \cite{JOPS-2012} carries through without
changes; see \cite{JNPPS} for details.  

The interpretation of all  three types of exponents (Stein, Chernoff, Hoeffding)
in terms of hypothesis testing and support 
separation of the pairs $(\P_t, \wP_t)$ as $t\rightarrow \infty$ as presented in
\cite{BJPP-2017} remains unchanged. Obviously, 
the support separation is linked to the emergence of the arrow of  time only in
Case~2 of Definition~\ref{def:familyinvolutions}.

\subsection{Examples}

We start with five examples where our results apply, but for which the
conclusions are well known and have been reached in the literature by other
means. We believe however that 
in all these  cases the Ruelle--Lanford functions method presented here offers a
different perspective on the resulting LDPs.

\begin{example} {\bf Bernoulli measures}. Let $P$ be a
probability measure on $\cA$, and define\footnote{See
\cite[Section~A.4]{lanford_entropy_1973} for a pedagogical exposition of this
case.} the measure $\P$ by $\P_t(w) = \prod_{i=1}^t P(w_i)$, $w\in \Omega_t$.
Then obviously \hUD and \hSLD are satisfied with $\tau_t \equiv 0$ and $c_t \equiv
0$. If $\widehat \P$ is defined similarly for some probability measure
$\widehat P$ on $\cA$, then \hSSD also holds with the same sequences $(\tau_t)_{t\in\nn}$ and $(c_t)_{t\in\nn}$, and all our results apply provided that $P\ll\widehat P$.
\end{example}

\begin{example}\label{ex:markovirred} {\bf Irreducible Markov processes.} Let
$\P\in \cP_\varphi(\Omega)$ be a Markov process. Then \hUD holds. Assume furthermore
that it is irreducible (\ie that for all $a,b\in \cA$, there exists $\xi^{(a,b)}\in
\Omega_\fin$ such that $\P(a\xi^{(a,b)} b) > 0$). Then \hSLD holds. If, in
addition, $\widehat \P\in \cP_\varphi(\Omega)$ is another Markov process such
that $\P_2 \ll \widehat \P_2$ (hence $\widehat \P$ is also irreducible), then
\hSSD holds. See Lemma~\ref{lem:irreducibleMarkov} for the proof of these  claims.
Note that no aperiodicity condition is required; if the Markov process~$\P$ is
irreducible  and {\em aperiodic}, then \eqref{eq:sellowerdecoup} also holds with
the condition $|\xi| \leq \tau_t$ strengthened to $|\xi| = \tau_t$. 
\end{example}

The next example consists of (weak) Gibbs measures, which have been studied
extensively, and for which the LDPs and FR have been obtained via the thermodynamic
formalism (see for example
\cite{young-1990,varandas_nonuniform_2012,varandas_weak_2015,PS-2018,CJPS_phys}).

\begin{example}\label{ex:WGSS}{\bf  Gibbs and weak Gibbs measures on subshifts}.
Assume that the subshift $\Omega^+$ satisfies the following {\em weak
specification property}:\footnote{A typical example would be a subshift of
$\{0,1,2\}^\nn$ where the only restriction is that for each $t\in\nn$, every
occurrence of the word $01^t0$ must be followed by the word $2^{\lfloor
\sqrt{t+2} \rfloor }$. Then the weak specification property is satisfied with
$\tau_t = \lfloor \sqrt {t} \rfloor $.} for  all $u,v \in \Omega_\fin^+$, there
exists $\xi \in \Omega_\fin^+$, $|\xi| \leq \tau_{|u|}$ such that $u\xi v \in
\Omega_\fin^+$. Assume moreover that $\P$ is a Gibbs measure for some potential
$f\in C(\Omega^+)$, \ie  that for some $p\in \rr$, $d\geq 0$ and all $\omega \in
\Omega^+$,
\begin{equation}\label{eq:strongGibbsshiftNre}
 \eee^{-d + S_tf(\omega)- tp} \leq \P_t(\omega_1, \dots, \omega_t)\leq  \eee^{d
+ S_tf(\omega)- tp}.
\end{equation}
Then it is easy to realize that \hUD and \hSLD are satisfied with $\tau_t$ as in the
above specification property and $c_t = 3d + \tau_t \|f-p\|$. Moreover, \hSSD is
satisfied if one of the following conditions holds: (a) $\widehat \P = \Theta
\P$ with $\Theta$  as in Definition~\ref{def:familyinvolutions} and 
$\theta_t(\Omega_t^+) = \Omega_t^+$; (b) $\widehat \P$ is also a  Gibbs measure
(\ie satisfies \eqref{eq:strongGibbsshiftNre} for some $\hat f$ and $\hat p$,
and all $\omega$ in the support $\Omega^+$ of $\P$). More generally, we say that
$\P$ is a {\em weak Gibbs measure} if \eqref{eq:strongGibbsshiftNre} holds with
 $d$ replaced by $d_t = o(t)$. In this case, the decoupling assumptions above do
not hold in general.\footnote{As discussed in Appendix~\ref{ss:weakGibbsmeas},
our decoupling assumptions are not comparable with the weak Gibbs property.}
However, we show in Appendix~\ref{ss:weakGibbsmeas} that our results can easily 
be adapted to this case. 
\end{example}

An interesting special case of Example~\ref{ex:WGSS} is:
\begin{example}{\bf $\beta$-shifts.} Consider the
$\beta$-shift for some $\beta > 1$ (see \cite{PS-2005} and references therein). 
The weak specification property described in Example~\ref{ex:WGSS} is satisfied
for Lebesgue-almost all $\beta > 1$ (see the discussion after Corollary 5.1 in
\cite{PS-2005}; the quantity defined in equation (5.9) therein plays the role of
$\tau_t$), and hence for such $\beta$'s our results apply to any (weak) Gibbs
measure.  
\end{example}

We next turn to Gibbs states. Such measures satisfy at the same time our
decoupling assumptions and the weak Gibbs condition.

\begin{example}\label{eq:Gibbs1D}{\bf Gibbs states in 1D}. Let $\P^*$ be an
invariant Gibbs state (in the Dobrushin--Lanford--Ruelle sense, see for example
\cite{ruelle2004,van_enter_regularity_1993,EKW-1994,lewis_entropy_1995}) for
some absolutely summable interaction  on the full two-sided shift $\cA^{\zz}$.
Then the marginal $\P$ of $\P^*$ on the one-sided shift $\cA^{\nn}$ satisfies \hUD
and \hSLD with $\tau_t \equiv 0$ \cite[Lemma 2.9]{lewis_entropy_1995}. If one
considers also  hard-core interactions, \ie if $\P^*$ is an invariant Gibbs
state on a subshift $M$ of $\cA^{\zz}$, then the proof can be adapted provided
$M$ satisfies the following condition:\footnote{This condition is slightly more
``flexible'' than Condition~D in \cite[Section 4.1]{ruelle2004} in the sense
that the position where~$\omega_{[1,t]}$ appears in $\eta'_{[1, t+\tau_t]}$ may
depend on $\omega$ (for fixed $t$).} for all $\eta, \omega\in M$, and all $t\in\nn$,
there exists $\eta' \in M$ such that $\omega_{[1,t]}$ appears in $\eta'_{[1,
t+\tau_t]}$, and such that $\eta_i = \eta'_i$ for all $i \in \zz\setminus\llbracket1,
t+\tau_t\rrbracket$.  The discussion of \hSSD in this setup is similar to the (weak) Gibbs
case discussed in Appendix~\ref{ss:weakGibbsmeas}.
\end{example}

The LDPs and FR for Gibbs states have also been obtained using the thermodynamic
formalism \cite{comets_86,FO-1988,OP-1988,Olla_88,EKW-1994,CJPS_phys}, and the proofs
therein do not require $\P$ to be $\varphi$-invariant. The condition on $M$
spelled out in Example~\ref{eq:Gibbs1D} seems to be slightly more general than
those found in the literature on Gibbs states.

We now turn to examples which genuinely require the full generality of our
assumptions.

\begin{example}\label{ex:HiddenMarkov}{\bf A class of hidden Markov chains.} In
Appendix~\ref{ss:hiddenMarkChain} we describe a prototypical pair of hidden
Markov chains, which satisfies \hSSD with $\tau_t \equiv 1$ and $\sup_t c_t <
\infty$, and for which the function $q$ defined in
\eqref{eq:defqalphaentrop} displays different types of singularities. Depending
on the parameters of the model, one can have that:
\begin{itemize}
\item  $q$ is finite but not differentiable everywhere on $\rr$;
\item  there exists $\alpha_* \geq 0$ such that $q$ is finite (and even
analytic) on $(-\infty, \alpha_*)$ and infinite on $(\alpha_*, \infty)$, with
either
\begin{itemize}
\item[$\argdot$]  $\lim_{\alpha \uparrow \alpha_*}q(\alpha) = q(\alpha_*) = + \infty$;
\item[$\argdot$] $q(\alpha_*) < \infty$ and $q'(\alpha_*^-)=+ \infty$;\footnote{Here and
below $q'(\alpha_*^-)$ denotes the left-derivative of $q$ at $\alpha_*$.} or
\item[$\argdot$] $q(\alpha_*) < \infty$ and $q'(\alpha_*^-)< \infty$.
\end{itemize}
\end{itemize}
This leads to situations where neither of \cite{CJPS_phys,BJPP-2017} apply, or to give the global LDP in Theorem
\ref{main-2}. This example illustrates how ``non-additive'' (or
``non-extensive'' in physical terms) $\sigma_t$ can be under our assumptions, in
the sense that the sequence $(t^{-1} \sigma_t(\omega))_{t\in\nn}$ may be
unbounded for some $\omega \in \Omega$. A closely related, and physically
relevant, example of {\em rotational quantum instrument} will be discussed in
\cite{BCJPP-2017}.
\end{example}

The following example arises naturally in multifractal analysis, see
\cite{olivier2006weak,olivier2006infinite}.

\begin{example}{\bf Matrix product probability measures.}\label{ex:matrixprod}
Let $M: {\cal A}\rightarrow {\mathbb M}_{N}(\rr)$ be a map taking  values in the
algebra of real $N\times N$ matrices that satisfies the following 
assumptions: 
\begin{itemize}
\item[\textbf{(A1)}]The entries of $M(a)$ are non-negative for all $a\in {\cal
A}$.
\item[\textbf{(A2)}]The matrix $S=\sum_{a\in {\cal A}}M(a)$ and its transpose~$S^T$ 
satisfy
 \[Sv=\lambda v, \qquad S^T w=\lambda w\]
 for some $\lambda>0$ and vectors  $v, w \in \rr^N$ 
 with strictly positive entries. 
\end{itemize}
For each $t\in\nn$ we define a probability measure $\P_t$ on $\Omega_t$  by
\[
\P_t(\omega_1, \dots, \omega_t)=\frac{1}{\lambda^t} \left( w, M(\omega_1)\cdots
M(\omega_t)v\right).
\]
One easily verifies that there exists a unique $\P\in {\cal P}_\varphi(\Omega)$
whose family of marginals is given by $(\P_t)_{t\in\nn}$. We shall 
call such $\P$ the matrix product measure associated with  the triple $(M, v,
w)$. We have: 
\begin{itemize}
\item \hUD holds for $\P$ with $\tau_t \equiv 0$, and $\sup_{t}c_t < \infty$. 
\item If the entries of $M(a)$ are strictly positive for all 
$a\in {\cal A}$, then  \hSLD holds with $\tau_t\equiv0$ and  $\sup_tc_t<\infty$. 
\item  If for some $a\in {\cal A}$ all the entries of $M(a)$ are strictly
positive, then \hSLD holds 
with $\tau_t\equiv1$ and $\sup_tc_t<\infty$ (by taking $\xi=a$ in
\eqref{eq:sellowerdecoup}). 

\item Suppose that  for all $a\in {\cal A}$ some entries  of $M(a)$ are
vanishing. If the matrix $S$ is irreducible, \ie for some $r\in \nn$ the matrix
$(I +S)^r$ 
has strictly positive entries (here~$I$ denotes the identity matrix), 
then \hSLD holds with $\tau_t \equiv r$ and $\sup_tc_t < \infty$. Note that if $S$
is irreducible, 
then \textbf{(A2)} automatically holds and $\P$ is ergodic.  It is easy to construct
examples 
of $M$ for which \textbf{(A1)} and \textbf{(A2)} hold, $S$ is not irreducible, and \hSLD fails. 

\item Let  $\widehat M: {\cal A}\rightarrow {\mathbb M}_{N}(\rr)$  be another
map satisfying \textbf{(A1)} and \textbf{(A2)}, and let $\wP$ be the induced 
probability measure. If the matrix $\sum_{a\in {\cal A}}M(a)\otimes \widehat
M(a)$ acting on $\rr^N \otimes \rr^N$ is irreducible, then \hSSD holds 
for the pair $(\P, \wP)$ with $\tau_t$ and $c_t$ that do not depend on $t$ (by
an adaptation of the proof of Proposition~2.6 in \cite{BJPP-2017}; see also
\cite{BCJPP-2017}). 

\item If $\Theta$ is as in Definition~\ref{def:familyinvolutions} and
$\wP=\Theta\P$, then 
in Case 1 (of Definition~\ref{def:familyinvolutions}), the measure $\wP$ is the
matrix product measure associated with  $(\widehat M,  v, w)$, where $\widehat
M(a)=M(u(a))$, and in Case 2 the measure
$\wP$ corresponds to $(\widehat M,  w, v)$ (note the order of $w$ and $v$),
where $\widehat M(a)=M^T(u(a))$.
\item  If \hSSD holds, then the quantity $q(\alpha)$ in \eqref{eq:defqalphaentrop} (which exists as a limit by Theorem~\ref{main-2}) is finite for all
$\alpha$. Indeed, since all the non-zero entries of the matrices at hand are
bounded below by some constant $c>0$, the integrand in
\eqref{eq:defqalphaentrop} increases at most exponentially in $t$ on the support
of $\P$.
\end{itemize}

For reasons of space we postpone the detailed discussion of various concrete
examples of matrix product probability measures to  \cite{BCJPP-2017}.
 \end{example}

As a final example, we recall here the setup of the quantum instruments studied
\cite{BJPP-2017}, as these were our initial motivation. We note that any matrix
product measure can also be obtained by a well-chosen positive instrument (see
\cite{BCJPP-2017}).

\begin{example}{\bf Positive  instruments.}\label{ex:QMP} Let $\cH$ be a
finite-dimensional complex Hilbert space and denote by $\cC=\cB(\cH)$ the 
$\ast$-algebra of all linear maps $A:\cH\rightarrow \cH$ equipped with the 
inner product $(A, B)=\tr(A^\ast B)$. Let $\Phi:{\cal A}\rightarrow\cB(\cC)$ be a map 
satisfying the following assumptions: 
\begin{itemize}
	\item[\textbf{(B1)}]The map $\Phi(a)$  is positive\footnote{$\Psi\in \cB(\cC)$ is
positive if $\Psi[X]\geq 0$ for any $X\geq0$.} for all $a\in {\cal A}$.
	\item[\textbf{(B2)}]The map  $S=\sum_{a\in {\cal A}}\Phi(a)$ and its adjoint~$S^*$
satisfy
 \[S[\nu] =\lambda \nu, \qquad S^\ast[\rho]=\lambda \rho\]
 for some $\lambda>0$ and strictly positive $\nu, \rho\in \cC$.
\end{itemize}

For each $t\in\nn$ we define a probability measure $\P_t$ on $\Omega_t$  by
\[
\P_t(\omega_1, \dots, \omega_t)=\frac{1}{\lambda^t\tr(\rho \nu)} \tr(
\rho(\Phi(\omega_1)\circ\cdots \circ \Phi(\omega_t))[\nu]).
\]

One easily verifies that there exists a unique  $\P\in {\cal P}_\varphi(\Omega)$
whose family of marginals is given by $(\P_t)_{t\in\nn}$. We shall 
call such $\P$ the positive instrument  process  associated with  the positive
instrument  $(\Phi, \nu, \rho)$. If $\Phi(a)$ is completely
positive\footnote{$\Psi\in\cB(\cC)$ is completely positive if for all $k\in\nn$ the map $\rm{id}_k \otimes \Psi \in \cB(\cB(\cc^{k})\otimes\cC)$ is
positive, where $\rm{id}_k$ is the identity map on $\cB(\cc^k)$.} for all $a\in {\cal A}$,  $\nu$ is the identity map, and $\lambda=1$,
then $(\Phi, \rho)$ is called a {\em quantum instrument} and $\P$ describes 
the statistics of the repeated quantum measurement process generated by $(\Phi,
\rho)$; see \cite{BJPP-2017} for additional information and references 
regarding quantum instruments and induced processes.
We have:
\begin{itemize}

\item \hUD holds for $\P$ with $\sup_tc_t< \infty$, \cite[Lemma 3.4]{BJPP-2017}.

\item If  $\Phi(a)$ is positivity improving\footnote{$\Psi\in\cB(\cC)$ is
positivity improving  if $\Psi[X]> 0$ for any non-zero $X\geq0$.} for all 
$a\in {\cal A}$, then \hSLD holds  with $\tau_t \equiv 0$, $\sup_tc_t<\infty$.

\item If $\Phi(a)$ is positivity improving for some $a\in {\cal A}$, then
\hSLD holds 
with $\tau_t \equiv 1$, (by taking $\xi=a$ in \eqref{eq:sellowerdecoup}), and
$\sup_tc_t<\infty$. 

\item Suppose that none of the $\Phi(a)$'s is positivity improving. If the map
$S$ is irreducible, 
\ie  for some $r\in \nn$ the map $(\imath +S)^r$ is positivity
improving\footnote{See \cite[Section 2.1]{EHKSP87}.} ($\imath$
denotes the identity map on $\cB(\cC)$), then \hSLD holds with $\tau_t \equiv r$ and
$\sup_tc_t<\infty$. We remark that if $S$ is irreducible, 
then \textbf{(B2)} automatically holds and $\P$ is ergodic.  

\item Let  $\widehat \Phi: {\cal A}\rightarrow \cB(\cC)$  be another map
satisfying \textbf{(B1)} and \textbf{(B2)}, and let $\wP$ be the induced 
positive instrument process. If the map  $\sum_{a\in {\cal A}}\Phi(a)\otimes
\widehat \Phi(a)$ acting on $\cC\otimes \cC$ is irreducible, then \hSSD
holds 
for the pair $(\P, \wP)$ with $\tau_t$ and $c_t$ that do not depend on $t$; see
\cite[Proposition~2.6]{BJPP-2017}.\footnote{It is easy to realize that
Assumption (C) in \cite{BCJPP-2017} together with \hUD imply \hSSD, see also
\cite{BCJPP-2017}.}

\item If $\Theta$ is as in Definition~\ref{def:familyinvolutions} and $\wP=\Theta\P$, then 
in Case~1 the measure $\wP$ is the positive instrument  process associated with 
$(\widehat \Phi,  \nu, \rho)$, where $\widehat \Phi(a)=\Phi(u(a))$. In Case 2,  
$\wP$ is the positive instrument process associated with $(\widehat \Phi,  \rho,
\nu)$, where $\widehat \Phi(a)=\Phi^\ast(u(a))$.

\item Unlike for matrix product measures, we do not have in general that
$q(\alpha)<\infty$ for all $\alpha\in \rr$. See the {\em rotational instruments}
in \cite{BCJPP-2017}.
\end{itemize}

Again, for details and discussion of concrete examples we refer the reader to
\cite{BCJPP-2017}.

\end{example}

\section{General constructions and abstract LDP}
\label{sec-LF}

We start with some further notation and conventions that will be used throughout
the paper.
A function~$f$ on $\Omega$ is $\FF_t$-measurable if and only if $f(\omega)$ depends only
on $\omega_1, \dots, \omega_t$. We identify the space
of $\FF_t$-measurable functions and the space\footnote{Since $\Omega_t$ is
endowed with the discrete topology, all functions on $\Omega_t$ are continuous.}
$C(\Omega_t)$ in the obvious way, and for such functions we write $f(\omega)$
and $f(\omega_1, \dots, \omega_t)$ interchangeably. The space $C_{\fin}(\Omega)$
consisting of all functions which are  $\FF_t$-measurable for some $t$ is dense
in $C(\Omega)$.

By this identification, any function $f\in C(\Omega_t)$ is associated with  a
$\cF_t$-measurable random variable on $(\Omega, \cF, \P)$. Conversely, any
$\cF_t$-measurable real-valued random variable $f$ on $(\Omega, \cF, \P)$ is
associated with a function  $f\in C(\Omega_t^+)$, and can be extended to a
function in $C(\Omega_t)$ by defining $f(w)$ arbitrarily for $w\in \Omega_t
\setminus \Omega_t^+$. We note that with  this convention, $\langle f,\P \rangle =
\sum_{w\in \Omega_t^+} f(w) \P(w)$.
These considerations extend to $\R^d$-valued functions, and  the corresponding
spaces are denoted by $C(\Omega, \R^d)$ and $C_\fin(\Omega, \R^d)$.

Following these conventions, the quantity $\sigma_t$ defined in \eqref{1.81} can
be expressed as
\begin{equation}\label{eq:identificationrhot}
	\sigma_t(w)=\log\frac{\IP_t(w)}{\widehat\IP_t(w)}, \qquad w\in\Omega_t^+,
\end{equation}
which is well defined since $\P_t \ll \widehat \P_t$.

\subsection{Construction of a map $\psi_{n,t}$}\label{subs:constrPsint}

For any pair $(t,n)\in\nn^2$ with $t\geq n$, we 
define\footnote{$\lfloor x\rfloor=\max\{n\in\zz:n\leq x\}$ denotes the floor function.}
\begin{equation}\label{eq:defNTprime}
	N = N(t,n) = 2 \left\lfloor \frac t {2(n+\tau_n)} \right\rfloor \qquad
\text{and} \qquad t' = t'(t,n) = Nn ,
\end{equation}
where $(\tau_n)_{n\in\nn}$ is the integer sequence introduced in 
Section~\ref{sec:assumptions}. 
Observe that~$N$ is even. An important inequality following from the above definition
is
\begin{equation*}
\frac {t}{1+\tau_n/n} - 2n\leq t'(t,n) \leq t	,
\end{equation*}
which implies that
\begin{equation}\label{eq:boundintgpart}
	\lim_{n\to\infty} \limsup_{t\to\infty} \left|\frac{t'(t,n)}{t}-1\right| \leq
\lim_{n\to\infty} \limsup_{t\to\infty} \left(\frac {2n} t + \frac {\tau_n}n
\frac 1 {1+\tau_n/n}\right)=0 .
\end{equation}

For each $n\in\nn$, we define the decoupled measure $\P^{(n)} = (\P_n)^{\times \nn}$
(which is $\varphi^n$-invariant, but not $\varphi$-invariant). For $t=mn+j$ with 
$0\leq j<n$, the marginal $\P^{(n)}_t$ is given by
\begin{equation*}
\P^{(n)}_t (w) = \left(\prod_{k=0}^{m-1}\P_n(w_{[kn+1,(k+1)n]})\right)
\P_j(w_{[mn+1,mn+j]}) ,
\end{equation*}
where the last term is $1$ if $j=0$. We also define
\begin{equation}\label{eq:defLambdatprime}
\Lambda_{t'} = (\Omega_n^+)^N = \{w\in \Omega_{t'}: \P_n(w_{[kn+1,(k+1)n]}) >
0,~ k=0,1, \dots, N-1\},
\end{equation}
which is the support of $\P_{t'}^{(n)}$. Note that obviously $\Omega_{t'}^+
\subset \Lambda_{t'}$.

The main result of this subsection is the following proposition that provides a
way to compare the two discrete probability spaces $(\Omega_{t'},
\P_{t'}^{(n)})$ and $(\Omega_{t}, \P_{t})$. 
\begin{proposition}\label{prop:constructionpsi}Assume \hSLD. For any pair
$(t,n)\in\nn^2$ with $t\geq n$, and with $N$ and $t'$ defined by~\eqref{eq:defNTprime},
there exists a map $\psi_{n,t} : \Omega_{t'} \to \Omega_t$ such that the following 
holds.
\begin{enumerate}
	\item There exists $g(n,t)\geq 0$ such that 
\begin{gather}
\P_{t'}^{(n)}\circ \psi_{n,t}^{-1} \leq  \eee^{g(n,t)} \, \P_{t},
\label{2.3}\\
\lim_{n\to\infty} \limsup_{t\to\infty} \frac 1 t g(n,t)  = 0	.
\label{eq:boundgnT}
\end{gather}
\item Assume furthermore \hSSD. Then $\psi_{n,t}$ can be chosen so that, in
addition to the above,
\begin{equation}\label{eq:sigmapsicompatible}
\lim_{n\to\infty }\limsup_{t\to\infty}\frac 1 t
\sup_{w\in\Lambda_{t'}}\left|\sigma_{t}(\psi_{n,t}(w)) - \sum_{k=0}^{N-1}
\sigma_n(w_{[kn+1, (k+1)n]}) \right|  = 0.
\end{equation}
\end{enumerate}
\end{proposition}
\begin{remark}
For further reference, we make the immediate observation that \eqref{2.3} is
equivalent to the fact that for all $A \subset \Omega_{t'}$,
\begin{equation}\label{eq:Pndecouplsets}
\P_{t'}^{(n)} (A) \leq  \eee^{g(n,t)}  \P_{t}(\psi_{n,t}(A)),
\end{equation}
and to the fact that for each function $h: \Omega_t \to [0, \infty)$ we have 
\begin{equation}\label{eq:decouplexpectation}
\sum_{w \in \Omega_{t'}} h(\psi_{n,t}(w)) \P_{t'}^{(n)}(w) \leq \eee^{g(n,t)}
\sum_{w \in \Omega_{t}} h(w) \P_t(w).
\end{equation}
\end{remark}
\begin{remark} It follows from~\eqref{eq:Pndecouplsets}
that $\psi_{n,t}(\Lambda_{t'}) \subset \Omega_t^+$. In particular, all the
quantities in \eqref{eq:sigmapsicompatible} are well defined (see
\eqref{eq:identificationrhot}). Note that the map $\psi_{n,t}$ is in general
neither injective nor surjective. There will be two contributions in $g(n,t)$:
one coming from the ratio $\P_t(\psi_{n,t}(w))/\P_{t'}^{(n)}(w)$, and one coming
from the maximal number of points $w\in \Omega_{t'}$ which share the same image
$\psi_{n,t}(w)\in \Omega_t$.
\end{remark}
\begin{remark}
The structure of $\psi_{n,t}$ here is very similar to a construction used in
Section~2 of \cite{Feng2009} in the context of products of matrices.
\end{remark}

We start with two technical lemmas.
\begin{lemma}\label{lem:extendword}
There exists a constant $C$ such that the following holds. For all $t, k\in \nn$
and all $v\in \Omega_t$, there exists $b\in \Omega_k$ such that
	\begin{equation}\label{eq:Pnothatcomplete}
	 \P(bv) \geq  \P(v)\eee^{-Ck}.
	\end{equation}
Assuming \hSSD, $b$ can be chosen so that, in addition to the above,
 	\begin{equation}\label{eq:sigmacomplete}
		 \widehat\P(bv) \geq  \widehat\P(v)\eee^{-Ck}.
	\end{equation}
\end{lemma}
\proof{} The first statement holds with $C = \log |\cA|$, since for all $v\in
\Omega_t$ we have 
\[\P(v) = \sum_{b\in \Omega_k}\P(bv) \leq |\cA|^k\max_{b\in \Omega_k}\P(bv).\]

The second statement is less trivial because \eqref{eq:Pnothatcomplete} and
\eqref{eq:sigmacomplete}  have to hold for the same $b$. Fix a symbol $a\in \cA$ such
that $\P_1(a)\widehat \P_1(a) > 0$ (which is possible by the absolute continuity
condition \eqref{eq:abscontinuous}). We claim that the result holds with $C =
c_1 -\log (\P_1(a)\wedge  \widehat \P_1(a))$. Assume  first that $k=1$.
Then for all $v \in \Omega_t$, there exists $\xi\in\Omega_\textrm{fin}$ such that
$|\xi| \leq \tau_1$ and
\[\P^\sharp(a\xi v) \geq \eee^{-c_1} \P^\sharp(a) \P^\sharp(v) \geq   \eee^{-C}
\P^\sharp(v)\]
 for both  $\P^\sharp = \P$ and $\P^\sharp = \widehat \P$. Now, let $b$ be the last letter in the word $a\xi$. We then have 
 \[\P^\sharp(b v) \geq \P^\sharp(a\xi v) \geq  \eee^{-C} \P^\sharp(v),\]
  which shows that both \eqref{eq:Pnothatcomplete} and \eqref{eq:sigmacomplete}
hold in the case $k=1$. The general statement follows by induction on $k$.
\hfill \qed

The following lemma is immediate.
\begin{lemma}\label{lem:mapping2probaspace}
Let $(X_1, P_1)$ and $(X_2, P_2)$ be two discrete probability spaces (each with
its discrete $\sigma$-algebra).  Let $\psi:X_1 \to X_2$ be a mapping which is at
most $r$-to-one, and assume that $P_1(\omega)\leq c P_2(\psi(\omega)) $ for some
$c>0$ and all $\omega \in X_1$.  Then 
\begin{equation*}
	P_1\circ \psi^{-1} \leq c r P_2.
\end{equation*}
\end{lemma}

We can now prove the main result of this subsection.

\proof{ of Proposition~\ref{prop:constructionpsi}}
The map $\psi_{n,t} : \Omega_{t'} \to \Omega_t$ is constructed  as follows.  For
$w\in \Omega_{t'}$, we write $w = w^{1}w^{2} \dots w^{N}$ with $w^i \in
\Omega_n$, and define
\begin{equation}\label{eq:DefPsi}
		\psi_{n,t}(w) =b w^1 \xi^1 w^2 \xi^2 \dots  w^{N-1}\xi^{N-1} w^N 
\end{equation}
for some $\xi^i, b \in \Omega_\fin$ to be chosen below that will satisfy 
$|\xi^i  |\leq \tau_n$ and 
\[|b| = \delta := t - t'- \sum_{i=1}^{N-1}|\xi^i|\]
(which may be zero), so that $|\psi_{n,t}(w)|=t$. 
Observe that
\begin{equation}\label{eq:bornesb}
t - t' \geq \delta  \geq t - N(n+\tau_n) \geq 0.
\end{equation}
Using \hSLD, we first choose $\xi^{N-1}$ such that $|\xi^{N-1}|\leq \tau_n$ and
\begin{equation*}
	\P\left(w^{N-1}\xi^{N-1}w^N\right) \geq
\eee^{-c_n}\P\left(w^{N-1}\right)\P\left(w^N\right).
\end{equation*}
Next, we choose $\xi^{N-2}$ such that $|\xi^{N-2}|\leq \tau_n$ and
\begin{align*}
	\P\left(w^{N-2}\xi^{N-2}w^{N-1}\xi^{N-1}w^N\right) &\geq
\eee^{-c_n}\P\left(w^{N-2}\right)\P\left(w^{N-1}\xi^{N-1}w^N\right)\\
& \geq
\eee^{-2c_n}\P\left(w^{N-2}\right)\P\left(w^{N-1}\right)\P\left(w^N\right).
\end{align*}
Continuing this process, we choose $\xi^{N-3}, \dots, \xi^1$ such that
$|\xi^i|\leq \tau_n$ and
\begin{equation*}
	\P\left( w^1 \xi^1 w^2 \xi^2 \dots  w^{N-1}\xi^{N-1} w^N\right)\geq
\eee^{-(N-1)c_n}\P\left(w^1\right) \P\left(w^2\right) \cdots \P\left(w^N\right)
= \eee^{-(N-1)c_n}\P^{(n)}(w).
\end{equation*}
Finally, if $\delta \geq 1$, we choose $b\in \Omega_{\delta}$ so that 
\eqref{eq:Pnothatcomplete} holds with $v= w^1 \xi^1 w^2 \xi^2 \dots 
w^{N-1}\xi^{N-1} w^N$ and $k=\delta$, so that
\begin{equation*}
	\P(\psi_{n,t}(w))\geq \eee^{-(N-1)c_n-C\delta}\P^{(n)}(w).
\end{equation*}
If $\delta=  0$, we choose $b$ as the empty word, and the above also holds.
Next, \eqref{eq:bornesb} implies that
\[(N-1)c_n+ C\delta  \leq (N-1) c_n + (t-t')C =:g_1(n,t),\]
and so
\begin{equation}\label{eq:Ppsintprop31}
	\P(\psi_{n,t}(w))\geq \eee^{-g_1(n,t)}\P^{(n)}(w).
\end{equation}

The mapping $\psi_{n,t}$ is not injective. In order to retrieve $w\in
\Omega_{t'}$ from $\psi_{n,t}(w)$, it suffices to know the length of $\xi^1,
\dots, \xi^{N-1}$, 
and there are at most $(\tau_n+1)^{N-1}$ possibilities. Thus, $\psi_{n,t}$ is at
most  $(\tau_n+1)^{N-1}$-to-one. By Lemma~\ref{lem:mapping2probaspace}, we
obtain \eqref{2.3} with
\begin{equation*}
g(n,t) =  g_1(n,t)	+(N-1)\log(\tau_n+1) \leq  g_1(n,t)	+N \tau_n .
\end{equation*}
To finish the proof  of Part 1,  observe that since $t\geq t'=nN$, we have
\begin{equation*}
\frac {g(n,t)} t  \leq 	\frac {c_n}n + C\left(1-\frac{t'(t,n)}{t}\right) 
+\frac{\tau_n}n,
\end{equation*}
which by \eqref{eq:boundintgpart}, shows that \eqref{eq:boundgnT} also holds.

To prove Part 2 of the proposition, assume \hSSD and let  $w\in \Lambda_{t'}$. We
then proceed exactly as above. By \hSSD one can choose $\xi^1, \dots, \xi^{N-1}$
such that 
\begin{align*}
\eee^{-(N-1)c_n}\P^\sharp\left(w^1\right)\P^\sharp\left(w^2\right)\cdots
\P^\sharp\left(w^N\right)
& \leq 	\P^\sharp\left( w^1 \xi^1 w^2 \xi^2 \dots  w^{N-1}\xi^{N-1} w^N\right)
\\
& \leq \eee^{(N-1)c_n}\P^\sharp\left(w^1\right)\P^\sharp\left(w^2\right)\cdots
\P^\sharp\left(w^N\right)
\end{align*}
for both  $\P^\sharp = \P$ and $\P^\sharp = \widehat \P$. Note that all
quantities here are positive, since $w \in \Lambda_{t'}$ implies, by definition,
that all the $w^i$ are in the support of $\P_{n}$, and hence in that of
$\wP_{n}$ by \eqref{eq:abscontinuous}.

Defining $\delta$ as above and choosing $b\in \Omega_\delta$ as in
Lemma~\ref{lem:extendword} (with $b=\kappa$ if $\delta =0$), we obtain
that~$\psi_{n,t}(w)$ defined by~\eqref{eq:DefPsi} satisfies
\begin{equation*}
\eee^{-(N-1)c_n-C\delta }\P^\sharp\left(w^1\right)\cdots
\P^\sharp\left(w^N\right) \leq 	\P^\sharp(\psi_{n,t}(w)) \leq
\eee^{(N-1)c_n}\P^\sharp\left(w^1\right)\cdots \P^\sharp\left(w^N\right).
\end{equation*}
Recalling definition of $\sigma_t$ and $\sigma_n$ and using the inequality
\[(N-1)c_n \leq (N-1)c_n + C\delta \leq g_1(n,t) \leq g(n,t),\]
we finally obtain 
\begin{equation*}
\left|\sigma_{t}(\psi_{n,t}(w)) - \sum_{k=1}^{N} \sigma_n(w^k) \right| \leq
2g(n,t),
\end{equation*}
which implies \eqref{eq:sigmapsicompatible}. This completes the  proof of
Proposition~\ref{prop:constructionpsi}.\hfill\qed

\subsection{Ruelle--Lanford functions}

Let $X$ be a locally convex Hausdorff topological vector space endowed with its Borel 
$\sigma$-algebra. Let $\cN_0$ be a neighborhood basis of $0\in X$, so 
that~$\NN_x=\NN_0+x$ is a neighborhood basis of $x\in X$.

Given a sequence $(z_t)_{t\in\nn}$ of $X$-valued random variables on $(\Omega,
\cF, \P)$, 
we define the following two non-decreasing set functions on the Borel sets of $X$:
\begin{equation}
	\begin{split}\label{eq:defuolines}
		\underline s (A) &= 	\liminf_{t\to\infty} \frac 1 t \log \P\left(\frac 1 t z_t
\in A\right),\\
\overline s (A) &= 	\limsup_{t\to\infty} \frac 1 t \log \P\left(\frac 1 t z_t
\in A\right).
	\end{split}
\end{equation}

\begin{definition}\label{def:RLfct}
Assume that for all $x\in X$, we have
	\begin{equation}\label{eq:definitionsx}
	 \inf_{G\in \cN_x} \underline s(G) =  \inf_{G\in \cN_x} \overline s(G).
	\end{equation}
Then, the function $s: X \to [-\infty, 0]$, whose value $s(x)$ is defined by the two expressions in
\eqref{eq:definitionsx} for each $x\in X$, is called the {\em Ruelle--Lanford function} of the sequence $(z_t)_{t\in\nn}$.\footnote{The two infima in \eqref{eq:definitionsx} are independent of the choice
of the neighborhood basis $\NN_x$ of $x$, and hence so is $s(x)$.}
\end{definition}

When defined, the function $s$ is upper semicontinuous. Indeed,
for all $x\in X$ and $\varepsilon>0$, there exists  $G \in \cN_x$ such that
$\overline s(G) \leq s(x) + \varepsilon$, and for each $x'\in G$ there exists
$G'\in\NN_{x'}$ such that $G'\subset G$. It follows that
$s(x') \leq  \overline s(G')\leq  \overline s(G) \leq s(x) + \varepsilon$.

We now give sufficient conditions for the Ruelle--Lanford function to exist.

\begin{definition}
We say that the sequence $(z_t)_{t\in\nn}$ is {\em admissible} if for all
$x_1, x_2 \in X$
and for every neighborhood $G$ of $x:=\frac 12 x_1 + \frac 12 x_2$, there exist
$G_1\in\NN_{x_1}$ and $G_2\in\NN_{x_2}$ such that
	\begin{equation}\label{eq:concavityS}
	\underline s(G) \geq \frac 12 \overline s(G_1) +\frac 12 \overline s(G_2) .
	\end{equation}
\end{definition}

\begin{proposition}\label{prop:abstractLDP}
Let the sequence $(z_t)_{t\in\nn}$ be admissible. Then \eqref{eq:definitionsx} holds 
for all $x\in X$, so that the Ruelle--Lanford function $s$ is well defined. Moreover, 
$(\frac1t z_t)_{t\in\nn}$ satisfies the {\em weak} LDP with convex rate function $-s$,
in the sense that for every open set $O \subset X$,
\begin{equation}\label{eq:lowerLDPabstrait}
\underline s(O) \geq \sup_{x\in O}s(x)
\end{equation}
and that for every {\em compact} set $\Gamma \subset X$,
\begin{equation}\label{eq:upperldp}
\overline s(\Gamma) \leq \sup_{x\in \Gamma}s(x).
\end{equation}
If, in addition, the laws of $(\frac 1 t z_t)_{t\in\nn}$ form an exponentially
tight family\footnote{This means that for any $\epsilon>0$ there exists a compact set 
$K_\epsilon$ such that $\bar s(K_\epsilon^c)\leq-1/\epsilon$.}, then $-s$ is a good 
rate function, and $(\frac 1 t z_t)_{t\in\nn}$
satisfies the LDP, \ie \eqref{eq:upperldp} holds for any closed set
$\Gamma\subset X$.
\end{proposition}
\proof{} For the reader's convenience, we include a complete proof, although
this is a classical result (see
\cite[Proposition~3.5]{pfister_thermodynamical_2002} or \cite[Lemmas 4.1.11 and
4.1.21]{DZ2000}).
First, the special case $x = x_1 = x_2$ in \eqref{eq:concavityS} immediately
implies that the two infima in \eqref{eq:definitionsx} are equal, so that $s$ is
well defined.
Next, if $x = \frac 12 x_1 + \frac 12 x_2$, then~\eqref{eq:concavityS} yields
$s(x) \geq \frac 1 2(s(x_1) + s(x_2))$. Since $s$ is upper semicontinuous, this
inequality implies that~$s$ is concave (by a bisection method).

We now turn to the LDP.  The lower bound~\eqref{eq:lowerLDPabstrait} is immediate.
Indeed, for any open set $O \subset X$ and every $x \in O$ we have $x\in G\subset O$
for some $G\in\NN_x$, so that $\underline s(O)\geq\underline s(G)\geq s(x)$. Since this 
holds for all $x\in O$, we obtain~\eqref{eq:lowerLDPabstrait}. 

The upper bound \eqref{eq:upperldp} is more involved. Let $\Gamma\subset X$ be closed, 
and let $\varepsilon > 0$. 
It suffices to prove~\eqref{eq:upperldp} in the following two cases.
\begin{itemize}
	\item Case 1: the laws of $(\frac 1 t z_t)_{t\in\nn}$ are
exponentially tight. Then, there exists a compact set $K$ such that $\overline
s(K^c) \leq -1/\varepsilon$. We let then $G_0 = \Gamma \cap K^c$. Thus, for each
$x \in G_0$, we have $s(x) \leq - 1/ \varepsilon$. 
	\item Case 2: $\Gamma$ is compact. Then we let $G_0 = \emptyset$.
\end{itemize}

Observe that in both cases $\Gamma \setminus G_0$ is compact.
For each $x \in \Gamma\setminus G_0$, there exists $G(x)\in\NN_x$
such that $ \overline s(G(x)) \leq s(x) + \varepsilon \leq \sup_{y\in
\Gamma}s(y)+ \varepsilon $. Now,  $\{G(x): x \in \Gamma\setminus G_0\}$ is an open
cover of $\Gamma \setminus G_0$, and by compactness one can extract a finite
subcover $\{G_i : i=1, \dots, n\}$. Since $\Gamma \subset \bigcup_{i=0}^n G_i$,
one has
\begin{equation*}
	\overline s(\Gamma)   \leq 	\limsup_{t\to\infty} \frac 1 t \log
\left(\sum_{i=0}^n\P\left(\frac{z_t}t \in G_i  \right)\right)
 \leq  \max_{i=0, \dots, n} \overline s(G_i) \leq \max(-1/\varepsilon, \sup_{x
\in \Gamma}s(x) + \varepsilon ).
\end{equation*}
Sending $\varepsilon \to 0$ completes the proof of \eqref{eq:upperldp}.

Finally, we show that exponential tightness implies the goodness of the rate
function $I:=-s$ (see for example
\cite[Lemma 1.2.18]{DZ2000}). Let $a \in \R$, and let $L_a = \{x\in X: I(x) \leq
a\} $ be the corresponding level set (which is closed by lower semicontinuity of
$I$). Assuming exponential tightness, there is a compact set $K$ such that
$\overline s(K^c) < -a$, and applying \eqref{eq:lowerLDPabstrait} to $O = K^c$
yields $\inf_{x\in K^c}I(x) > a$. Thus, $L_a \subset K$, and hence $L_a$ is
compact.\hfill \qed

\subsection{Compatible observables}

In this subsection we assume \hSLD, so that the map $\psi_{n,t}$ is well defined and 
Part~1 of Proposition~\ref{prop:constructionpsi} holds. Moreover, $N =
N(t,n)$ and $t' = t'(t,n)$ are as in \eqref{eq:defNTprime} and $\Lambda_{t'}$ is
as in \eqref{eq:defLambdatprime}. Finally, for $x\in X=\rr^d$, we choose
the neighborhood basis
$\NN_x=\{B(x,\varepsilon)\}_{\varepsilon>0}$, where $B(x,\varepsilon)$ denotes 
the open ball of radius $\varepsilon$ in $X$, centered at $x$.

\begin{definition}\label{def:psicompat}Let $X = \mathbb R^d$ and let
$(z_t)_{t\in\nn}$ be  a sequence of $X$-valued random variables on $(\Omega,
\cF, \P)$. We say that $(z_t)_{t\in\nn}$ is $\psi$-compatible if $z_t$ is
$\cF_t$-measurable for each $t$, and the quantity
\begin{equation*}
h(n,t) := \frac 1 t \sup_{w \in \Lambda_{t'}}\left|  z_t(\psi_{n,t}(w)) -
\sum_{k=0}^{N-1} z_n(w_{[kn+1, (k+1)n]})\right| 
\end{equation*}
satisfies 
\begin{equation}\label{eq:boundhnT}
\lim_{n\to\infty} \limsup_{t\to\infty} h(n,t)  = 0.
\end{equation}
\end{definition}

\begin{proposition}\label{prop:admissibleZtcompatible}
Let $(z_t)_{t\in\nn}$ be a sequence of $\psi$-compatible $\rr^d$-valued random variables
on $(\Omega,\cF, \P)$. Then the following holds.
\begin{enumerate}
\item For all $x= \frac 12 x_1 + \frac 12 x_2 \in \rr^d$ and  $0<\varepsilon <
\varepsilon'$, we have 
\begin{equation*}
	\underline s(B(x, \varepsilon')) \geq \frac 12 \overline s(B(x_1, \varepsilon))
+\frac 12 \overline s(B(x_2, \varepsilon)).
\end{equation*}
In particular, $(z_t)_{t\in\nn}$ is admissible and the conclusions of
Proposition~\ref{prop:abstractLDP} hold.
\item There exists a sequence $(\gamma_t)_{t\in\nn}$ with $\gamma_t \to 0$ such
that for all $\varepsilon > 0$, all $t\in\nn$ and all $x \in
\rr^d$,\footnote{Note that the bound is uniform in both $t$ and $x$. This will
be crucial in the proof of Proposition~\ref{prop:ldpsigmafrst} below. }
\begin{equation}\label{eq:unifcontrolnx}
 \frac {1}{t} \log \P\left(\frac {z_t} t \in B(x, \varepsilon)\right)  \leq
\gamma_t + \sup_{y\in B(x, \varepsilon + (1+|x|)\gamma_t)} s(y),
\end{equation}
where $s$ is as in Definition~\ref{def:RLfct} and Proposition~\ref{prop:abstractLDP}.
\end{enumerate}
\end{proposition}
\proof{}
We shall prove that there exists a sequence $(\gamma_t)_{t\in\nn}$ with
$\gamma_t \to 0$ such that for all $\varepsilon > 0$, all $t\in\nn$ and all $x=
\frac 12 x_1 + \frac 12 x_2 \in \rr^d$, we have
\begin{equation}\label{eq:toprovesunderl}
 \frac {1}{2t} \log \P\left(\frac {z_t} t \in B(x_1, \varepsilon)\right) + \frac
{1}{2t}\log \P\left(\frac {z_t} t \in B(x_2, \varepsilon)\right) \leq \underline
s(B(x, \varepsilon + (1+|x|)\gamma_t))+ \gamma_t.
\end{equation}
This relation yields both Part 1 and Part 2 of the proposition. Part 1 follows
by taking the limit $t\to\infty$ and using that, for  fixed $x$, we have 
\[(1+|x|)\gamma_t \leq \varepsilon'-\varepsilon\]
for $t$ large enough. For Part 2, we take $x_1 = x_2= x$ in
\eqref{eq:toprovesunderl}, and by  \eqref{eq:upperldp} we obtain
\eqref{eq:unifcontrolnx} with the ball  $B(x, \varepsilon + (1+|x|)\gamma_t)$
replaced by its closure in the right-hand side. Replacing $\gamma_t$ with
$\gamma_t+ t^{-1}$, we then obtain \eqref{eq:unifcontrolnx}. 

To prove \eqref{eq:toprovesunderl}, let $x = \frac 12 x_1 + \frac 12 x_2 \in
\R^d$ and $\varepsilon > 0$. In the following, $j(k) = 1$ if $k$ is odd,
$j(k)=2$ if $k$ is even, and we write $I_k = [kn+1, (k+1)n]$. By assumption, for
any $w\in\Lambda_{t'}$, 
\begin{align*}
	\left|\frac 1 t   z_t(\psi_{n,t}(w)) - x\right| & \leq \left|\frac 1 t
\sum_{k=0}^{N-1}z_n(w_{I_k})-x\right| +  h(n,t)\\
& \leq     \left|  \frac n t \sum_{k=0}^{N-1}\left(
\frac{z_n(w_{I_k})}n-x\right)\right| + h(n,t)+ |x|\left(1-\frac {t'}t\right)\\
& =     \left|  \frac n t \sum_{k=0}^{N-1}\left(
\frac{z_n(w_{I_k})}n-x_{j(k)}\right)\right| + h(n,t)+ |x|\left(1-\frac
{t'}t\right),
\end{align*}
where the last equality holds because $N$ is even and $x = \frac 12 x_1 + \frac
12 x_2$. Using further that $\frac n t \leq \frac 1 N$ leads to
\begin{equation*}
	\left|\frac 1 t   z_t(\psi_{n,t}(w)) - x\right|  \leq   \frac 1 N
\sum_{k=0}^{N-1}\left|  \frac{z_n(w_{I_k})}n-x_{j(k)}\right| + h(n,t)+
|x|\left(1-\frac {t'}t\right).
\end{equation*}
Let $u_n = n^{-1} +  \limsup_{t\to\infty}\max\left(h(n,t),1-\frac {t'}t\right)$.
By \eqref{eq:boundintgpart} and \eqref{eq:boundhnT} we have $u_n\to 0$. In
addition, for each fixed $n$, there exists $t_0(n)$ such that for all $t\geq
t_0(n)$,
\begin{equation*}
	\left|\frac 1 t   z_t(\psi_{n,t}(w)) - x\right|  \leq   \frac 1 N
\sum_{k=0}^{N-1}\left|  \frac{z_n(w_{I_k})}n-x_{j(k)}\right| +(1+|x|)u_n,
\end{equation*}
and hence
\begin{equation*}
\psi_{n,t}\left(\bigcap_{k=0}^{N-1}\left\{w \in \Lambda_{t'}:  \frac 1 n 
z_n(w_{I_k})\in B(x_{j(k)}, \varepsilon)\right\}\right)  \subset \left\{w \in
\Omega_t^+: \frac 1 t  z_t(w)\in 
B'\right\},
\end{equation*}
where $B ' = B(x, \varepsilon + (1+|x|)u_n)$.
Using \eqref{eq:Pndecouplsets}, for all $t\geq
t_0(n)$ we derive
\begin{align*}
\P\left(\frac { z_t} t \in B'\right) & \geq
\P_{t}\left(\psi_{n,t}\left(\bigcap_{k=0}^{N-1}\left\{w \in \Lambda_{t'}: \frac
1 n   z_n(w_{I_k})\in  B(x_{j(k)}, \varepsilon)\right\}\right)\right) \\
&  \geq \eee^{-g(n,t)}\P^{(n)}_{t'}\left(\bigcap_{k=0}^{N-1}\left\{w \in
\Lambda_{t'}: \frac 1 n   z_n(w_{I_k})\in  B(x_{j(k)},
\varepsilon)\right\}\right) \\
&  =  \eee^{-g(n,t)}\left(\P_n\left( \frac { z_n} n\in B(x_{1},
\varepsilon)\right)\right)^{\frac {N}2}  \left(\P_n\left( \frac { z_n} n\in
B(x_{2}, \varepsilon)\right)\right)^{\frac {N}2}.
\end{align*}

Using  also that $\frac {N}t \leq \frac 1n$, and sending $t\to \infty$, we
obtain
\begin{equation*}
\underline s (B(x, \varepsilon + (1+|x|)u_n)) \geq \frac {1}{2n} \log
\P\left(\frac {z_n} n \in B(x_{1}, \varepsilon)\right) + \frac {1}{2n}\log
\P\left(\frac {z_n} n \in B(x_{2}, \varepsilon)\right) - u_n',
\end{equation*}
where $u_n '=  \limsup_{t\to\infty}\frac {g(n,t)}t$. By \eqref{eq:boundgnT} we
have $u_n' \to 0$. Defining $\gamma_n = \max(u_n, u_n')$  completes the proof 
of \eqref{eq:toprovesunderl}.\hfill\qed

\begin{lemma}\label{lem:expression}
Let $(z_t)_{t\in\nn}$ be a sequence of $\psi$-compatible $\rr^d$-valued random variables
on $(\Omega,\cF, \P)$. Then, for all $\alpha \in \R^d$, the limit
\begin{equation*}
q(\alpha) := \lim_{t\to\infty}\frac 1 t \log \left \langle  \eee^{ ( \alpha ,
z_t ) },\P\right\rangle
\end{equation*}
exists and takes value in $(-\infty, \infty]$. Moreover, the function 
$\rr^d\ni\alpha\mapsto q(\alpha)$ is convex and lower semicontinuous.
\end{lemma}

\proof{} Let  $h(n,t)$ be as in Definition~\ref{def:psicompat}, and consider
\begin{equation*}
A_t(\alpha) =  \left \langle \eee^{(\alpha  , z_t) } ,\P\right\rangle.
\end{equation*}

For each finite $t$, the map $\alpha \mapsto A_t(\alpha)$ is continuous. Recall
that by definition of $\psi$-compatibility, $z_n$ is $\cF_n$-measurable. Thus,
by \eqref{eq:decouplexpectation},
\begin{align*}
( A_n(\alpha))^{N} & = \sum_{w \in \Lambda_{t'}} 
\exp\biggl\{\sum_{k=0}^{N-1}
\bigl(\alpha, z_n(w_{[kn+1, (k+1)n]})\bigr)\biggr\}
\P^{(n)}_{t'}(w)\\
& \leq \eee^{|\alpha|th(n,t)} \sum_{w \in \Lambda_{t'}} \eee^{( \alpha, 
z_t(\psi_{n,t}(w)))}\P^{(n)}_{t'}(w)\\
& \leq \eee^{|\alpha|th(n,t) + g(n,t)} \sum_{w \in \Omega_{t}^+} \eee^{( \alpha,
 z_t(w))}\P_{t}(w)\\
& =  \eee^{|\alpha|th(n,t) + g(n,t)}   A_t(\alpha).
\end{align*}
It follows that 
\begin{equation*}
	\frac 1 t \log A_t(\alpha)  \geq \frac {N}t \log A_n(\alpha)- |\alpha|h(n,t) -
\frac {g(n,t)}t
 =   \frac 1 n\frac {t'}t \log A_n(\alpha)-  |\alpha|h(n,t) -\frac {g(n,t)}t.
\end{equation*}
By \eqref{eq:boundintgpart}, \eqref{eq:boundgnT} and \eqref{eq:boundhnT}, there
exists $\delta_n\to 0$ such that
\begin{equation}\label{eq:liminf1TAtf}
\liminf_{t\to\infty} 	\frac 1 t \log A_t(\alpha) \geq  \frac 1 n \left(1+
\delta_n\right) \log A_n(\alpha) - (1+|\alpha|)\delta_n~.
\end{equation}
Taking now the $\limsup$ as $n\to\infty$ yields
\begin{equation*}
\liminf_{t\to\infty} 	\frac 1 t \log A_t(\alpha) \geq \limsup_{n\to\infty} \frac
1 n  \log A_n(\alpha) ,
\end{equation*}
and so $q(\alpha)$ exists. Combining this with~\eqref{eq:liminf1TAtf}, we derive
\begin{equation*}
q(\alpha) =  \sup_{n\in\nn} \left( \frac 1 n \left(1+ \delta_n\right) \log
A_n(\alpha) - (1+|\alpha|)\delta_n\right).
\end{equation*}
It follows that $q(\alpha) > -\infty$ for all $\alpha\in \rr$, and since the
right-hand-side is a supremum over a family of continuous functions (with
respect to $\alpha$), we also derive that $q$ is lower semicontinuous. Finally,
it follows from H\"older's inequality that the functions $\alpha \mapsto \frac 1
t \log A_t(\alpha)$ are convex and, hence, so is the limit $q$.~\hfill\qed

\section{Level-1 LDP}
\label{sec-ldp1}

In this section we assume again \hSLD. Thus, Part 1 of
Proposition~\ref{prop:constructionpsi} holds, and again  $N = N(t,n)$ and $t' =
t'(t,n)$ are as in \eqref{eq:defNTprime}.

\begin{lemma}\label{lem:finvarcompatible}
Let $f\in C(\Omega_r,\R^d)$ for some $r\in \nn$ and set $z_t = S_{t-r+1} f$. Then, $(z_t)_{t\in\nn}$ is 
$\psi$-compatible. (We take the convention that $S_{j} f = 0$ if $j\leq0$). 
\end{lemma}
\proof{}
Clearly $z_t$ is $\cF_t$-measurable. Recall that by its definition
\eqref{eq:DefPsi}, $\psi_{n,t}$ is expressed as
\begin{equation*}
		\psi_{n,t}(w) = bw_{[1, n]} \xi^1 w_{[n+1, 2n]} \xi^2 \dots  \xi^{N-1}
w_{[(N-1)n+1,Nn]}, \qquad w\in \Omega_{t'}.
\end{equation*}
For $n\geq r$ and $t$ large enough, we have
\begin{align*}
h(n,t) & =  \frac 1t\sup_{w\in \Lambda_{t'}}
\left| z_t(\psi_{n,t}(w)) - \sum_{k=0}^{N-1} z_n(w_{[kn+1, (k+1)n]})\right| \\
&= \frac 1t\sup_{w\in \Lambda_{t'}}\left| \sum_{s=0}^{t-r}
f\bigl(\varphi^s(\psi_{n,t}(w)\bigr) 
- \sum_{k=0}^{N-1}\sum_{s=0}^{n-r} 
f\bigl(\varphi^s(w_{[kn+1, (k+1)n]})\bigr)\right| \\
&\leq \frac{\|f\|}{t}(t-r +1 - N(n-r+1))  \leq 	\|f\|\frac{(t- t' + Nr)}{t} 
\leq\|f\|\left(1-\frac {t'} t + \frac r n\right),
\end{align*}
where the first inequality follows from the observation that all the terms of
the iterated sum are also present in the first sum. By~\eqref{eq:boundintgpart},
it follows that~\eqref{eq:boundhnT} holds, and so  $(z_t)_{t\in\nn}$ is
$\psi$-compatible, as claimed.\hfill \qed 

\begin{proposition}\label{prop:expressureqfalpha}
For all $f\in C(\Omega, \R^d)$ and all $\alpha\in \R^d$, the limit
\begin{equation*}
q_f(\alpha) :=\lim_{t\to\infty}\frac 1 t \log \left\langle \eee^{ ( \alpha ,
S_{t} f ) },\P \right \rangle
\end{equation*}
exists and is finite. Moreover, the map $(f, \alpha) \mapsto q_f(\alpha)$ is convex in 
both arguments, $|\alpha|$-Lipschitz with respect to~$f$, and $\|f\|$-Lipschitz with
respect to $\alpha$.
\end{proposition}
\proof{} For each $t$, the function 
\[(\alpha, f)\mapsto \frac 1 t \log  \left\langle \eee^{ ( \alpha , S_{t} f )
},\P \right \rangle\]
 has the convexity and Lipschitz properties stated in the proposition (convexity
follows again from H\"older's inequality). By Lemmas~\ref{lem:expression}
and~\ref{lem:finvarcompatible}, for every $r\in \nn$ and $f\in C(\Omega_r, \R^d)$, the limit 
 \[\lim_{t\to\infty}\frac 1 t \log  \left\langle \eee^{ ( \alpha , S_{t} f )
},\P \right \rangle= \lim_{t\to\infty}\frac 1 t \log  \left\langle \eee^{ (
\alpha , S_{t-r+1} f ) },\P \right \rangle \]
  exists and is finite for all $\alpha\in\R^d$. Thus, $q_f(\alpha)$ exists for all $f\in C_\fin(\Omega,
\R^d)$ and $\alpha \in {\rr^d}$. Since $C_\fin(\Omega, \R^d)$ is dense in
$C(\Omega, \R^d)$, the $|\alpha|$-Lipschitz continuity in~$f$ implies that the
limit also exists for all $f\in C(\Omega, \R^d)$. The convexity and Lipschitz
properties are preserved in the limit.
\hfill\qed 

\begin{proposition}\label{prop:fctcontinueadmissible}
Let $f \in C(\Omega, \R^d)$ and set $z_t = S_t f$. Then, $(z_t)_{t\in\nn}$ is
admissible and the laws of $(\frac 1 t z_t)_{t\in\nn}$ are
exponentially tight. Thus $(\frac 1 t z_t)_{t\in\nn}$ satisfies the LDP (see
\eqref{eq:ldplev1up} and \eqref{eq:ldplev1lo}) with good convex rate function
$I_f$, where $I_f$ is the Fenchel--Legendre transform of $q_f$.
\end{proposition}
\proof{} We first prove the admissibility claim, with $X = \R^d$. Let $x = \frac
12 x_1 + \frac 12 x_2 \in \R^d$ and let $\varepsilon > 0$.
Since $f$ is continuous, for $\delta = \varepsilon/6$ there exists an integer
$r\geq 1$ and an $\cF_r$-measurable function $\tilde f$ such that $\|f-\tilde
f\| \leq \delta$. Define now $\widetilde z_t = S_{t-r+1}\widetilde f$, which is
$\cF_t$-measurable.
We have
\begin{equation}\label{eq:Zgrand2delta}
\left\|\frac 1 t z_t - \frac 1t\widetilde z_t\right\|
=\left\|\frac 1 t S_t f - \frac 1t S_{t-r+1}\widetilde f\right\| \leq \delta +
\frac {r-1}t \|f\|  .
\end{equation}
By Lemma~\ref{lem:finvarcompatible}, $(\widetilde z_t)_{t\in\nn}$ is
$\psi$-compatible. Denote by $\overline s$ and $\underline s$ the functions
defined in \eqref{eq:defuolines}, and let $\tilde{\overline s}$ and
$\underline{\tilde s}$ be the corresponding functions defined for $\tilde z_t$.
As a consequence of Proposition~\ref{prop:admissibleZtcompatible}, we have
\begin{equation}\label{eq:ineqballse}
	\tilde{\overline s}( B(x, \varepsilon-2\delta))\geq \frac {\underline{\tilde s}
(B(x_1, \varepsilon-3\delta))}2  + \frac {\underline{\tilde s}(B(x_2,
\varepsilon-3\delta))}2  .
\end{equation}
Using \eqref{eq:Zgrand2delta}, we obtain that for $t$ large enough, 
\[\P\left(\frac 1 t  z_t \in B(x, \varepsilon)\right) \geq \P\left(\frac 1 t
\widetilde z_t \in B(x, \varepsilon-2\delta)\right),\]
\[\P\left(\frac 1 t \widetilde z_t \in B(x_i, \varepsilon-3\delta)\right) \geq
\P\left(\frac 1 t  z_t \in B(x_i, \varepsilon-5\delta)\right).\]
By combining this with \eqref{eq:ineqballse}, we obtain that
\begin{equation*}
s( B(x, \varepsilon)) \geq \frac {\underline{ s} (B(x_1, \varepsilon-5\delta))}2
 + \frac {\underline{ s}(B(x_2, \varepsilon-5\delta))}2.
\end{equation*}
Hence, $(z_t)_{t\in\nn}$ is admissible. Since $\|\frac 1 t z_t\| \leq \|f\|$ for
all $t$, the laws of $(\frac 1 t z_t)_{t\in\nn}$ are exponentially
tight. Thus, by Proposition~\ref{prop:abstractLDP}, the LDP holds with good
convex rate function $I_f$ (which is equal to $-s$ in the notation of
Definition~\ref{def:RLfct}). We now denote by $I_f^*$ the Fenchel--Legendre
transform of $I_f$, and by $q_f^*$ the Fenchel--Legendre transform of $q_f$.

In order to identify $I_f$ and $q_f^*$ (see \cite[Theorem 4.5.10]{DZ2000} for a
similar argument), we use Varadhan's integral theorem and the convexity of
$I_f$. For $\alpha, u\in \R^d$, let $\phi_\alpha(u) = (\alpha, u)$ (which is
continuous as a function of $u$ for fixed $\alpha$), and let $P_t$ be the
distribution of $\frac 1 t z_t$. We have
\begin{equation*}
q_f(\alpha) = \lim_{t\to\infty} \frac 1 t \log \int_{\R^d} \eee^{t
\phi_\alpha(u)} \dd P_t(u).
\end{equation*}
Since for any $\gamma > 1$ we have $q_f(\gamma \alpha) < \infty$ by
Proposition~\ref{prop:expressureqfalpha}, the conditions of Varadhan's theorem
(see \cite[Theorem 4.3.1]{DZ2000} or \cite[Theorem 2.1.10]{DS1989}) are met, and
we obtain  $q_f(\alpha) = I_f^*(\alpha)$. Since this is true for all $\alpha \in
\R^d$ and since $I_f$ is convex and continuous (in particular, lower
semicontinuous), we find $I_f = q_f^*$, which completes the proof.\hfill \qed

\section{LDP for entropy production}
\label{sec-ldpep}

In this section  we assume \hSSD, so that $\psi_{n,t}$ is well defined and Parts~1
and~2 of Proposition~\ref{prop:constructionpsi} hold. In particular,
\eqref{eq:sigmapsicompatible} shows that  $(\frac 1 t \sigma_t)_{t\in\nn}$ is
$\psi$-compatible and hence admissible by
Proposition~\ref{prop:admissibleZtcompatible}.

\begin{proposition}\label{prop:sigmapressure}The limit
\begin{equation*}
	q(\alpha) : = \lim_{t\to\infty}\frac 1 t \log \left\langle \eee^{\alpha
\sigma_t} , \P\right\rangle 
\end{equation*}
exists for all $\alpha \in \rr$ and takes value in $(-\infty, \infty]$. The
function $q$ is lower semicontinuous and convex. We have $q(0)=0$ and $q(-1)
\leq  0$, so that $q$ is non-positive (and hence finite) on $[-1,0]$.
\end{proposition}
\proof{} Since $(\frac 1 t \sigma_t)_{t\in\nn}$ is admissible, we find by
Lemma~\ref{lem:expression} (with $d = 1$)  that $q$ exists, takes value in
$(-\infty, \infty]$, and is lower semicontinuous and convex. We have obviously
$q(0) = 0$. Moreover, 
\begin{equation*}
\frac 1 t \log  \left\langle \eee^{-\sigma_t},\P\right\rangle = \frac 1 t \log
\sum_{w \in \Omega_t^+}  \frac{\widehat\P_t(w)}{ \P_t(w)}\P_t(w)  = \frac 1
t\log \widehat \P_t(\Omega_{t}^+) \leq  0, 
\end{equation*}
and so  $q(-1) \leq 0$. By convexity, $q$ is non-positive (and hence finite) on
$[-1,0]$. This completes the proof. \hfill\qed

In the sequel, we denote by $I^*$ and $q^*$ the Fenchel--Legendre transforms of
$I$ and $q$.

\begin{proposition}\label{prop:ldpsigmafrst}The sequence $(\frac 1 t
\sigma_t)_{t\in\nn}$ satisfies the LDP (see \eqref{eq:weakLDPsigmaup} and
\eqref{eq:weakLDPsigma}) with a convex rate function $I$ given by $I(s) =
q^*(s)$ for all $s\in \rr$. Moreover, if $q(\alpha)<\infty$ for all $\alpha$ in
a neighborhood of 0, then $I$ is a good rate function.  
\end{proposition}
\proof{} Since $(\frac 1 t \sigma_t)_{t\in\nn}$ is admissible, it satisfies by
Proposition~\ref{prop:abstractLDP} the weak LDP with convex rate function $I$.
To strengthen the result to the LDP (\ie to show that \eqref{eq:weakLDPsigma} is
true also for unbounded $\Gamma$), we separate the following two cases (recall
that $q$ is finite  and non-positive on $[-1,0]$, and that $q(-1) \leq 0 =
q(0)$).
\begin{itemize}
	\item If $q(\alpha) < \infty$ in a neighborhood of the origin, a standard
application of Chebychev's inequality shows that the laws of $(\frac 1
t \sigma_t)_{t\in\nn}$ are exponentially tight,  so that the weak LDP is in fact
the LDP, and $I$ is a good rate function.
\item If $q(\alpha) = \infty$ for all $\alpha> 0$, then we have $
\lim_{x\to+\infty}q^*(x) = 0$. The identification $I = q^*$, which we prove
below, implies that $\lim_{x\to+\infty}I(x) = 0$ (in particular $I$ is not a
good rate function). We now show that the LDP still holds. If  $\Gamma$ is a
closed set such that $\sup \Gamma = +\infty$, then $\inf_{x\in \Gamma} I(x) =
0$, and hence \eqref{eq:weakLDPsigma} is trivial. Assume on the contrary that
$\Gamma$ is a closed set such that $\sup \Gamma <\infty$ (but possibly $\inf
\Gamma = -\infty$). Then, since $q(-1) < \infty$, Chebychev's inequality
provides the necessary exponential tightness on the negative half-line in order
to show  \eqref{eq:weakLDPsigma} (by a minor and standard adaptation of the
argument in the proof of Proposition~\ref{prop:abstractLDP}).
\end{itemize}

 We now turn to the comparison of $I$ and $q^*$.
If $q(\alpha)<\infty$ for all $\alpha\in \R$, we can proceed exactly as in
Proposition~\ref{prop:fctcontinueadmissible}, by using Varadhan's theorem to
obtain that $q = I^*$, and then the convexity of $I$ to obtain that $I= q^*$.
However, in the general case, more specific estimates are required in order to
show that $q = I^*$. We split the proof of this identity into three steps. Steps
1 and 3 are almost identical to the proof of  Varadhan's theorem (see
\cite[Theorem 4.3.1]{DZ2000} or \cite[Theorem 2.1.10]{DS1989}), although our
assumptions are slightly different. Step 2, however, is quite specific to our
setup (see Remark~\ref{rem:afterp44} below).

{\em Step 1: $q \geq I^*$}. We denote by $P_t$ the law of $t^{-1}\sigma_t$. 
 For any $x$, $\alpha \in \R$ and $\varepsilon > 0$, we have
\begin{align*}
	q(\alpha) &\geq \liminf_{t\to\infty} \frac 1 t  \log \int_{|x-y|<\varepsilon} 
\eee^{t\alpha y}P_t(dy)\\
&  \geq (\alpha x - |\alpha|\varepsilon) + \liminf_{t\to\infty} \frac 1 t\log
P_t((x-\varepsilon,x+\varepsilon))\\[5pt]
& \geq (\alpha x - |\alpha|\varepsilon) - \inf_{|x-y|<\varepsilon} I(y)
\geq \alpha x - |\alpha|\varepsilon-I(x).
\end{align*}
Letting $\varepsilon \to 0$, we get $q(\alpha) \geq \alpha x - I(x)$. Since
$\alpha$ and $x$ are arbitrary, we obtain $q \geq I^*$.

{\em Step 2: Tail estimates.}
Let
\begin{equation*}
\alpha_\pm=\lim_{x\to\pm\infty}\frac {I(x)}{x}.
\end{equation*}
By convexity these limits exist, and since $I$ is non-negative we have
$\pm\alpha_\pm\in[0, \infty]$. Moreover, since $0
\geq  q(-1) \geq I^*(-1) = \sup_{x\in \rr} (-x-I(x))$, we actually have
$\alpha_- \in [-\infty, -1]$, and in particular $\alpha_- < \alpha_+$. Let
$\alpha \in (\alpha_-, \alpha_+)$, and set $\delta = \frac 12 \min(1,
|\alpha-\alpha_-|, |\alpha-\alpha_+|)$. Then there exists $c > 0$ such that
\begin{equation*}
I(x) \geq \alpha x + \delta |x| - c \quad \text{for all }x\in \rr.
\end{equation*}
Using this and \eqref{eq:unifcontrolnx} we find
\begin{align*}
\frac 1 t \log P_t\bigl((k-1,k+1)\bigr)& \leq  \gamma_t - \inf_{y\in B(k, 1 +
(1+|k|)\gamma_t)} I(y)\\
& \leq - \alpha k - \delta |k| + c' + c''|k|\gamma_t,
\end{align*}
where $\gamma_t \to 0$, and the constants $c'$, $c''$ are independent of $t$ and
$k$. It follows that, for all $t$ large enough,
\begin{equation*}
P_t \bigl((k-1,k+1)\bigr)\leq 
\exp\Bigl(\bigl(- \alpha k - \tfrac\delta 2 |k| + c'\bigr)t\Bigr),
\end{equation*}
whence there exists $C>0$, depending only on $\alpha$, such that for all $K>0$,
\begin{equation}
\label{eq:Rk}
R_K :=\limsup_{t\to\infty} \frac 1 t	\log \int_{|x| > K} \eee^{\alpha t
x}P_t(dx)\leq   -K\frac{\delta}2 + C.
\end{equation}

{\em Step 3: $q \leq I^*$}.
If $\alpha \notin [\alpha_-, \alpha_+]$, we clearly have $I^*(\alpha) = + \infty
\geq q(\alpha)$. It therefore remains to show that $q(\alpha) \leq I^*(\alpha)$
for all $\alpha \in [\alpha_-, \alpha_+]$. Since both $I^*$ and $q$ are convex,
lower semicontinuous functions, it is enough to consider $\alpha \in (\alpha_-,
\alpha_+)$. We now fix $\alpha \in (\alpha_-, \alpha_+)$, $\varepsilon > 0$ and
$K>0$. For all $x\in [-K, K]$, there exists an open neighborhood $G_x$ such that
\[\inf_{y\in \overline{G}_x} I(y) \geq (I(x)-\varepsilon)\wedge
\varepsilon^{-1}, \qquad \sup_{y\in G_x} \alpha y \leq \alpha x+\varepsilon.\]
 We extract a finite subcover $\{G_{x_1}, \dots, G_{x_n}\}$ of $[-K,K]$ and
write
\begin{align*}
\limsup_{t\to\infty} \frac 1 t	\log \int_{G_{x_i}} \eee^{\alpha t y}P_t(dy) 
&\leq \limsup_{t\to\infty} \frac 1 t \log \left( \eee^{\alpha t x_i +
\varepsilon t}P_t(G_{x_i})\right)\\
& \leq \alpha x_i +\varepsilon - (I(x_i)-\varepsilon)\wedge \varepsilon^{-1} \\
& = \max\left(\alpha x_i - I(x_i)+ 2\varepsilon, \alpha x_i + \varepsilon -
\varepsilon^{-1} \right) \\
& \leq \max\left(I^*(\alpha)+ 2\varepsilon, |\alpha|K + \varepsilon -
\varepsilon^{-1} \right)  .
\end{align*}
It follows that
\begin{align*}
q(\alpha) &\leq \limsup_{t\to\infty} \frac 1 t	\log \left(\int_{|x|>K}
\eee^{\alpha t x}P_t(dx) + \sum_{i=1}^n \int_{G_{x_i}} \eee^{\alpha t x}P_t(dx)
\right) \\
&   \leq \max\left\{R_K,I^*(\alpha)+ 2\varepsilon, |\alpha|K + \varepsilon -
\varepsilon^{-1} \right\}. 
\end{align*}
Sending $\varepsilon \to 0$ shows that $q(\alpha) \leq  \max(R_K,I^*(\alpha))$.
Finally, sending $K\rightarrow \infty$ and using   \eqref{eq:Rk} yields
$q(\alpha) \leq I^*(\alpha)$, which completes the proof.\hfill \qed 

\begin{remark}\label{rem:afterp44}
The tail estimates in Step 2 above are equivalent to the statement that $
	q(\alpha)  < \infty$ for all $\alpha \in (\alpha_-, \alpha_+)$, which is
obviously a necessary condition in order to have $q = I^*$ . 
The uniform bound \eqref{eq:unifcontrolnx} is crucial in Step 2. An instructive
example of what can go wrong without it is given by the family of distributions
\[
\frac {dP_t} {dx} = (1-\eee^{-t^2}) \sqrt{t/\pi} \eee^{-tx^2}  +\frac 12
\eee^{-t^2} (\delta_t(x) +\delta_{-t}(x)), \qquad t \in \nn,
\]
which satisfies  the LDP with rate
function $I(x) = x^2$. Here  $\alpha_\pm = \pm \infty$, while\footnote{The
quantity $\ind_{|\alpha| > 1}$ is equal to $1$ if $|\alpha| > 1$ and to $0$
otherwise.} 
 \[q(\alpha) = \lim_{t\to\infty} \frac 1 t \log \int \eee^{t\alpha x} P_t(dx)=
\alpha^2/4  +\infty \ind_{|\alpha| > 1}.\]
 One now easily checks that $q$ and $I^*$ coincide only on $[-1,1]$, and that 
$I$ and $q^*$ coincide only on $[-1/2,1/2]$.
\end{remark}

\begin{lemma}
If $\wP = \Theta \P$ with $\Theta$ as in Definition~\ref{def:familyinvolutions},
then $q$ satisfies
\begin{equation}\label{eq:symmetryealphaproof}
q(-\alpha)=q(\alpha-1), \qquad \alpha \in \R,
\end{equation}
and $I$ satisfies the Gallavotti--Cohen symmetry
\begin{equation}\label{eq:GCsigmaproof}
I(-s)=I(s)+s, \qquad s\in \R.
\end{equation}
\end{lemma}
\proof{}
Recalling that $\theta_t = \theta_t^{-1}$ and that $\theta_t$ leaves
$\Omega_t^+$ invariant (see Remark~\ref{rem:abscompattheta}), we find
\begin{align*}
\left\langle \eee^{-\alpha\sigma_t},\P\right\rangle  &= \sum_{w\in \Omega_t^+}
\P_t^{1-\alpha}(w)\wP_t^{\alpha}(w)= \sum_{w\in \Omega_t^+}
\P_t^{1-\alpha}(\theta_t(w))\wP_t^{\alpha}(\theta_t(w))\\
&= \sum_{w\in \Omega_t^+} \wP_t^{1-\alpha}(w)\P_t^{\alpha}(w) = \left\langle
\eee^{(\alpha-1)\sigma_t},\P\right\rangle,
\end{align*}
which yields \eqref{eq:symmetryealphaproof}.
Although one can derive \eqref{eq:GCsigmaproof} from
\eqref{eq:symmetryealphaproof} and the identity $I = q^*$, we provide here a
direct derivation based on the LDP and the following {\em transient fluctuation
relation} (see \cite{CJPS_phys,JPRB-2011} and references therein): using that $\sigma_t
\circ \theta_t = -\sigma_t$, we find
\begin{align*}
\P_t\left(\frac 1 t\sigma_t = s\right) &= \sum_{w \in \Omega_t^+: \sigma_t(w) =
ts}\P_t(w) =  \sum_{w \in \Omega_t^+: \sigma_t(w) = ts}\eee^{ts}\wP_t(w)\\
& = \sum_{w \in \Omega_t^+: \sigma_t(\theta_t(w)) =
ts}\eee^{ts}\wP_t(\theta_t(w)) =\sum_{w \in \Omega_t^+: \sigma_t(w) =
-ts}\eee^{ts}\P_t(w) \\
&= \eee^{ts} \P_t\left(\frac 1 t\sigma_t = -s\right).
\end{align*}
From this we obtain that, for all $\varepsilon > 0$ and $s\in \R$,
\begin{equation*}
\left|\liminf_{t\to\infty} \frac 1 t \log \P_t\left(\frac 1 t \sigma_t \in
B(s,\varepsilon)\right)-\liminf_{t\to\infty} \frac 1 t \log \P_t\left(\frac 1 t
\sigma_t \in B(-s,\varepsilon)\right) - s\right| \leq\varepsilon.
\end{equation*}
By the construction of the rate function $I$, sending $\varepsilon \to 0$ gives
$|{-}I(s) + I(-s) - s| =0$, which is~ \eqref{eq:GCsigmaproof}.\hfill\qed

\section{Level-3 LDP}
\label{sec-ldp3}

\subsection{Main result}

In this section, we assume \hSLD and prove Theorem~\ref{t1.9}. For technical
reasons (in fact, in order to invert a Fenchel--Legendre transform in the proof
of Proposition~\ref{prop:level3ldp} below), we consider a slightly more general
situation, viewing $\mu_t(\omega) := \frac 1 t\sum_{s=0}^{t-1}
\delta_{\varphi^s(\omega)}$ as an element of the space $ X = \MM(\Omega)$
of finite signed Borel measures on~$\Omega$. We endow $\MM(\Omega)$ with the 
{weak-$\star$} topology with respect to the natural pairing\footnote{We shall
reserve the symbols $\P, \Q$ for the elements of $\PP(\Omega)$ and denote by
$\mu, \nu$ the elements of $\MM(\Omega)$.}
\begin{equation*}
\langle f,\nu \rangle = \int f \dd \nu, \qquad \nu \in \cM(\Omega),\quad  f\in
C(\Omega).
\end{equation*}
Recall that $C(\Omega)$ is endowed with the topology of uniform convergence.
With these topologies, the spaces $\cM(\Omega)$  and  $C(\Omega)$ are the
continuous dual of each other (with the natural identification). The induced
topology on ${\cal P}(\Omega)$ is the weak topology that we have considered so
far.

We shall  show that for every open set $O \subset \cM(\Omega)$ and every closed
set $\Gamma\subset \cM(\Omega)$,
\begin{align}
 \liminf_{t\to\infty} \frac 1 t \log \P\left(\mu_t \in O\right) & \geq
-\inf_{\nu\in  O} \I(\nu),\label{eq:lowerldplev3nu}\\
 \limsup_{t\to\infty} \frac 1 t \log \P\left(\mu_t \in \Gamma\right) &\leq
-\inf_{\nu\in  \Gamma} \I(\nu)\label{eq:upperldplev3nu},
\end{align}
where $\II$ is given by \eqref{eq:defIIlev3} on $\cP(\Omega)$, and where
$\I(\nu) = + \infty$ on $\cM(\Omega) \setminus \cP(\Omega)$. Since
$\mu_t(\omega)\in\cP(\Omega)$ for all~$\omega$, this will immediately  imply the
LDP on $\cP(\Omega)$ in Theorem~\ref{t1.9}.

For $f\in C(\Omega)$, let 
\begin{equation}\label{eq:deflimQf}
Q(f) = \lim_{t\to\infty}\frac 1 t \log  \left \langle  \eee^{ S_t f }, \P
\right\rangle .
\end{equation}
By Proposition~\ref{prop:expressureqfalpha} (in the special case $d = 1, \alpha
= 1$), 
the limit \eqref{eq:deflimQf} exists and is finite, and the function $Q$ is
convex and $1$-Lipschitz .

\begin{proposition}\label{prop:level3ldp}
	The sequence $(\mu_t)_{t\in\nn}$ satisfies the LDP with respect to the
{weak-$\star$} topology on~$\cM(\Omega)$ for some good rate function~$\I$ (see
\eqref{eq:lowerldplev3nu} and~\eqref{eq:upperldplev3nu}). Moreover, $\I$ is the
Fenchel--Legendre transform of $Q$, \ie for all $\nu\in \cM(\Omega)$,
\begin{equation}\label{eq:IIasltf}
\I(\nu) = \sup_{f\in C(\Omega, \R)}\big(\langle f,\nu \rangle -Q(f)\big).
\end{equation}
Finally, $\I(\nu) = +\infty$ for all $\nu \in \cM(\Omega) \setminus
\cP_\varphi(\Omega)$. 
\end{proposition}

\proof{} We set $z_t = t \mu_t$, and we define $\overline s $ and $\underline s$
as in \eqref{eq:defuolines}. We first show that the sequence $(z_t)_{t\in\nn}$ is 
admissible. A neighborhood basis of $\nu\in\cM(\Omega)$ is given by 
\[
\cN_\nu=\left\{G(\nu,f,\varepsilon)
:=\{\mu\in\cM(\Omega):|\langle f,\mu-\nu\rangle|<\varepsilon\}:
\varepsilon>0,f\in C(\Omega,\rr^d),d\geq1\right\}.
\]
We immediately have
\[
\frac1tz_t(\omega)=
\mu_t(\omega) \in G(\nu, f, \varepsilon) \iff \frac 1 t S_t f(\omega) \in
B(\langle  f, \nu \rangle, \varepsilon).
\]
Fix now $\nu = \frac 12 \nu_1 + \frac 12 \nu_2 \in \cM(\Omega)$, and consider a
neighborhood $G(\nu, f, \varepsilon)$ of $\nu$. Let $x = \frac 12 x_1 + \frac 12
x_2$ with $x_i = \langle  f, \nu_i \rangle$. Since $(S_t f)_{t\in\nn}$ is
admissible by Proposition~\ref{prop:fctcontinueadmissible}, there exists
$\varepsilon'>0$ such that 
\begin{align*}
\underline s(G(\nu, f, \varepsilon)) &= 
\liminf_{t\to\infty} \frac 1 t \log \P\left(\frac 1 t  S_t f \in B(x,
\varepsilon)\right)\\
 &\geq  \limsup_{t\to\infty} \frac 1 {2t} \log \P\left(\frac 1 t  S_tf \in
B(x_1, \varepsilon')\right)
 +\limsup_{t\to\infty} \frac 1 {2t} \log \P\left(\frac 1 t S_tf \in B(x_2,
\varepsilon')\right)\\
& = \frac 12 \overline s(G(\nu_1, f, \varepsilon'))+\frac 12 \overline
s(G(\nu_2, f, \varepsilon')).
\end{align*}
This implies that $(z_t)_{t\in\nn}$ is admissible. Moreover, since $\mu_t$
belongs to the compact subset $\cP(\Omega)$ for all $t$, the laws of
$(\frac1t z_t)_{t\in\nn}$ trivially form an exponentially tight family, so that by
Proposition~\ref{prop:abstractLDP}, $\mu_t$ satisfies the LDP with good convex
rate function~$\I$ defined by 
\[\I(\nu) = -\inf_{G\in \cN_\nu} \underline s(G) = -\inf_{G\in \cN_\nu}
\overline s(G).\]

We now show that  $\I(\nu) = +\infty$ when $\nu \notin \cP_\varphi(\Omega)$.
Since $\mu_t \in \cP(\Omega)$, and since $\cP(\Omega)$ is closed, one
immediately obtains $\I(\nu) = +\infty$ if $\nu \notin \cP(\Omega)$. Now, if
$\nu \in \cP(\Omega) \setminus \cP_\varphi(\Omega)$, one can find a function
$g\in C(\Omega)$ such that $f:=g-g\circ\varphi$ satisfies $\langle f, \nu\rangle=1$. 
Then, for all $\mu\in G(\nu,f,1/2)$, we have 
$\langle f,\mu\rangle > 1/2$.
However, by construction,
\[\langle  f, \mu_t(\omega)\rangle = \frac 1 t(g(\omega)-g\circ
\varphi^t(\omega)) \leq \frac 2 t\|g\|,\]
which is eventually $< 1/2$. Thus, $\P(\mu_t\in G(\nu,f,1/2)) = 0$ for $t$ large
enough, and
\[\I(\nu) \geq -\liminf_{t\to\infty}\frac 1 t \log \P (\mu_t \in G(\nu,f,1/2))  =
+\infty.\]

Following the same ideas as in Proposition~\ref{prop:fctcontinueadmissible} (see
also \cite[Theorem 4.5.10]{DZ2000} and \cite{pfister_thermodynamical_2002}), we
now identify $\I$ and $Q^*$ using Varadhan's integral theorem and the convexity
of $\I$. For fixed $f\in C(\Omega)$, let $\phi_f= \langle f, \argdot\rangle$,
which is a continuous function on $\cM(\Omega)$. Denoting by $P_t$ be the law
of $\mu_t$, we have
\begin{equation*}
Q(f) = \lim_{t\to\infty} \frac 1 t \log \int_{\cM(\Omega)} \eee^{t \phi_f(\nu)}
\dd P_t(\nu).
\end{equation*}
Since for any $\gamma > 1$ we have $Q(\gamma f) < \infty$ (or more simply, using
that $P_t$ is supported on the compact set $\cP(\Omega)$), we can apply
Varadhan's theorem, and obtain\footnote{Recall that $C(\Omega)$ is the dual of
$\cM(\Omega)$ with the weak-$\star$ topology, so $\I^*$ is naturally defined on
$C(\Omega)$.}
\[Q(f) = \I^*(f)= \sup_{\nu \in \cM(\Omega)}(\langle f, \nu \rangle - \I(\nu)).\]
 Since this is true for all $f \in C(\Omega)$, and since $\I$ is convex and
lower semicontinuous, we find that $\I = Q^*$, which is \eqref{eq:IIasltf} (see 
\cite[Theorem 3.10]{bronsted64} or  \cite[Lemma 4.5.8]{DZ2000} for variants of
the duality principle between convex conjugate functions that apply in the
present setup). \hfill \qed

\subsection{Alternative expression for the rate function}

Assuming also \hUD, we now derive an alternative expression for the rate
function ${\II}$ of Proposition~\ref{prop:level3ldp}. This new expression
will imply, in particular, that~${\II}$ is affine on $\cP_\varphi(\Omega)$.

Given $\Q\in\PP(\Omega)$ and $t\in\nn$, consider the relative entropy (recall
\eqref{1.9})
\begin{equation*} 
\Ent(\Q_t|\IP_t)=\varsigma_t(\Q)-h_t(\Q),
\end{equation*}
where we set, with the usual convention that $0\log 0 = 0$,
\begin{align*}
\varsigma_t(\Q)&=\left\{
\begin{aligned}
-\sum_{w\in\Omega_t}&\Q_t(w)\log\IP_t(w) &\quad &\mbox{if $\Q_t\ll\P_t$},\\
&+\infty&\quad&\mbox{otherwise}, 
\end{aligned}
\right.
\\
h_t(\Q)&=-\sum_{w\in\Omega_t}
\Q_t(w)\log\Q_t(w).
\end{align*}

For $\Q \in \cP_\varphi(\Omega)$, we have
\begin{equation*}
\lim_{t\to\infty}\frac 1 th_t(\Q) = h(\Q),
\end{equation*}
where $h(\Q)$ is the Kolmogorov--Sinai entropy of $\Q$ with respect to
$\varphi$.  The limit exists, is finite, and the mapping $h: \cP_\varphi(\Omega)
\to [0, \infty)$ is upper semicontinuous and affine.\footnote{See
Corollary~4.3.14, Corollary~4.3.17 and the remark following it in the
book~\cite{KH1995}.} For completeness, a proof of these elementary properties of
the Kolmogorov--Sinai entropy is provided in Lemma~\ref{lem:propsKS}.

We first need a technical lemma.

\begin{lemma}\label{lem:decobservabletaun} Assume \hUD. Let $f, g$ be two
non-negative random variables on $(\Omega, \cF, \P)$ such that  $f$ is
$\cF_n$-measurable and $g$ is $\cF_r$-measurable with $n,r\in \nn$. Then
\begin{equation*}
\left \langle f \,  (g \circ \varphi^{n+\tau_n}), \P \right\rangle \leq
\eee^{c_n+\tau_n\log |\cA|}\left \langle f, \P \right\rangle\left \langle g, \P
\right\rangle.
\end{equation*}
\end{lemma}

\proof{} Recalling the conventions in the beginning of Section~\ref{sec-LF}, we
obtain by \hUD
\begin{align*}
\left \langle f \,   (g \circ \varphi^{n+\tau_n}), \P \right\rangle 
&= \sum_{u\xi v\in \Omega_{n+\tau_n+r}^+}f(u)g(v) \P(u\xi v)\\
&\le  \sum_{u\xi v\in \Omega_{n+\tau_n+r}^+}f(u)g(v) \eee^{c_n}\P(u)\P(v)\\
& \le \sum_{\xi\in \Omega_{\tau_n}} \eee^{c_n} \left \langle f, \P
\right\rangle\left \langle g, \P \right\rangle, 
\end{align*}
which implies the claim, since $|\Omega_{\tau_n}| = |\cA|^{\tau_n}$. (The factor
involving $\cA$ would not be needed with the alternative \hUD assumption mentioned
in Remark~\ref{rem:otherassumptions}.)
\hfill\qed

\begin{proposition} \label{p1.6}
Suppose that \hUD is satisfied.\footnote{Although \hSLD is a standing assumption in
this section, observe that this proposition does not rely on it provided that we
{\em define} $\I$ by \eqref{eq:IIasltf}. This proposition does not rely on the
validity of the LDP for $\mu_t$ either.} Then the limit
\begin{equation} \label{1.12}
\varsigma(\Q):=\lim_{t\to\infty}\frac 1 t\varsigma_t(\Q)
\end{equation}
exists for any measure~$\Q\in\PP_\varphi(\Omega)$, and the mapping $\varsigma:
\cP_\varphi(\Omega) \to [0, \infty]$ is lower semicontinuous. Moreover, for any
$\Q\in \cP_\varphi(\Omega)$, we have
\begin{equation} \label{1.13}
{\II}(\Q)=\varsigma(\Q)-h(\Q)=\lim_{t\to\infty} \frac 1 t\Ent(\Q_t|\,\IP_t),
\end{equation}
and ${\II}$ is an affine function of $\Q\in\PP_\varphi(\Omega)$. 
\end{proposition}

\proof{} We first prove the existence and lower semicontinuity of the
limit~\eqref{1.12}
by using  a classical subadditivity argument. For each pair of integers $t, n$
with $t\geq n+\tau_n$, we let $M = \lfloor t/(n+\tau_n)\rfloor$. By using \hUD
$M-1$ times, we find for all $w \in \Omega_t^+$
\begin{equation*}
	\log \P_{t}(w)\leq \log\P_{(n+\tau_n)M}(w_{[1,(n+\tau_n)M]}) 
\leq    \sum_{k=0}^{M-1}\log\P_n(w_{[k(n+\tau_n)+1, k(n+\tau_n)+n]}) + (M-1)c_n,
\end{equation*}
where both sides may be $-\infty$. Integrating this inequality with respect
to~$-t^{-1}\Q$ and using the translation invariance of~$\Q$ yields
\begin{equation*}
\frac 1 t \varsigma_{t}(\Q)\geq  \frac M t\varsigma_{n}(\Q)- \frac
{(M-1)c_n}{t} \geq   \left (\frac 1{n+\tau_n} - \frac 1
t\right)\varsigma_{n}(\Q)- \frac {c_n}{n+\tau_n},
\end{equation*}
where we have also used that $\varsigma_n(\Q) \geq 0$.
Sending now $t\to \infty$ yields
\begin{equation*}
\liminf_{t\to\infty} \frac 1 t \varsigma_{t}(\Q)\geq  \frac
1{1+\tau_n/n}\left(\frac 1{n}\varsigma_{n}(\Q)- \frac {c_n}{n}\right).
\end{equation*}
Taking the $\limsup$ as $n\to\infty$ shows that the limit $\varsigma(\Q)$ exists
(it can be infinite). We then find that 
\begin{equation*}
\varsigma(\Q)  = \sup_{n\in\nn}\frac 1{1+\tau_n/n}\left(\frac
1{n}\varsigma_{n}(\Q)- \frac {c_n}{n}\right),
\end{equation*}
and thus, since $\Q \mapsto \varsigma_n(\Q)$ is continuous for all $n$, we
obtain that $\varsigma$ is lower semicontinuous.

\smallskip
We now fix $\Q\in \cP_\varphi(\Omega)$ and establish \eqref{1.13}. The second
equality in \eqref{1.13} follows from the definitions, and we need to  show that
$\I(\Q) = \LL(\Q) := \lim_{t\to\infty} \frac 1 t\Ent(\Q_t|\,\IP_t)$.

We first deal with the special case where there exists $t_0\in\nn$ such that
$\Q_{t_0}$ is not absolutely continuous with respect to $\IP_{t_0}$. Then
$\Ent(\Q_t | \,\IP_t) = \infty$ for all $t\geq t_0$, and thus $\LL(\Q) =
\infty$. Let us choose $w\in\Omega_{t_0}$ such that 
$\Q(w) > 0 = \IP(w)$. 
For all $n\in \nn$, let $f_n(\omega) = n\ind_{\omega_{[1,t_0]} = w}$.  
Observe that $\langle f_n,\Q\rangle = n \Q(w)$ and that $Q(f_n) = 0$,
since~$S_{t}f_n$ vanishes on the support of $\IP$. Thus, using
\eqref{eq:IIasltf}, we see that
$$
{\II}(\Q) \geq \langle f_n,\Q\rangle- Q(f_n) = n \Q(w)
\quad\text{for all }n\in\nn,
$$
so that ${\II}(\Q) = \infty$.

Suppose now that $\Q_t \ll \IP_t$ for all $t\in\nn$. We shall first prove that 
$\LL(\Q) \geq {\II}(\Q)$. Let~$f\in C_{\mathrm{fin}}(\Omega)$ be
$\cF_r$-measurable for some $r\in\nn$, and let
\begin{equation*}
A_t=\left\langle \eee^{S_t f},\IP\right\rangle. 
\end{equation*}
By Jensen's inequality and the invariance of $\Q$, we have
\begin{align*}
\log A_t &=\log\,\bigl\langle \eee^{S_t f},\IP \bigr\rangle
=\log\,\int_{\Sigma_{t+r-1}}
\eee^{S_t f}\frac{\dd\IP_{t+r-1}}{\dd\Q_{t+r-1}}\,\dd\Q_{t+r-1} \\[2pt]
&\geq\int_{\Sigma_{t+r-1}}
\Bigl(S_t f-\log\frac{\dd\Q_{t+r-1}}{\dd\IP_{t+r-1}}\Bigr)\,\dd\Q_{t+r-1}\\[2pt]
&=t \langle f,\Q\rangle- \Ent({\Q}_{t+r-1}|{\IP}_{t+r-1}),
\end{align*}
where $\Sigma_t$ is the support of~$\Q_t$. Dividing by $t$ and sending $t\to
\infty$ shows  that $\LL({\mathbb Q})\geq  \langle f,\Q\rangle-Q(f)$. Since
$f\in C_\fin(\Omega)$ is arbitrary, $C_{\mathrm{fin}}(\Omega)$ is dense in
$C(\Omega)$, and  $Q$ is Lipschitz, we find
\begin{equation*}
	\LL(\Q) \geq \sup_{f\in C_\fin(\Omega)}( \langle f,\Q\rangle-Q(f))=\sup_{f\in
C(\Omega)}( \langle f,\Q\rangle-Q(f)) = \I(\Q).
\end{equation*}

It remains to prove that  $\LL(\Q) \leq {\II}(\Q)$. Fix two integers $n,M\geq 1$
and let $t = n'M$ where $n' = \tau_n+n$. Consider the $\cF_n$-measurable
function
$f=\frac  1{n'}\log \frac{\Q_{n}}{\IP_{n}}$. This function is well defined on
the support of~$\Q_{n}$ (and, hence, on the support of~$\P_{n}$), and we define
it by~$-\infty$ on the complement. Note that 
\begin{equation} \label{1.32}
\Ent(\Q_{n}|\,\IP_{n})=n'\langle f,\Q\rangle.
\end{equation}

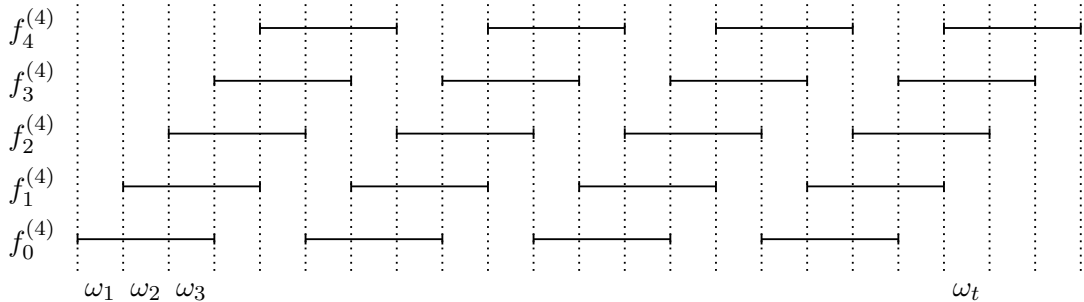
\begin{figure}[ht]
    \centering
  \begin{tikzpicture}[line width = 0.7pt]

    \draw[-] (0.0,0.0)--(1.8,0.0) ;
    \draw[-] (0.0,-0.07)--(0.0,0.07) ;
    \draw[-] (1.8,-0.07)--(1.8,0.07) ;
    
    \draw[-] (0.6,0.7)--(2.4,0.7) ;
    \draw[-] (0.6,0.63)--(0.6,0.77) ;
    \draw[-] (2.4,0.63)--(2.4,0.77) ;
    
    \draw[-] (1.2,1.4)--(3.0,1.4) ;
    \draw[-] (1.2,1.33)--(1.2,1.47) ;
    \draw[-] (3.0,1.33)--(3.0,1.47) ;
    
    \draw[-] (1.8,2.1)--(3.6,2.1) ;
    \draw[-] (1.8,2.03)--(1.8,2.17) ;
    \draw[-] (3.6,2.03)--(3.6,2.17) ;
    
    \draw[-] (2.4,2.8)--(4.2,2.8) ;
    \draw[-] (2.4,2.73)--(2.4,2.87) ;
    \draw[-] (4.2,2.73)--(4.2,2.87) ;
    
    \draw[-] (3.0,0.0)--(4.8,0.0) ;
    \draw[-] (3.0,-0.07)--(3.0,0.07) ;
    \draw[-] (4.8,-0.07)--(4.8,0.07) ;
    
    \draw[-] (3.6,0.7)--(5.4,0.7) ;
    \draw[-] (3.6,0.63)--(3.6,0.77) ;
    \draw[-] (5.4,0.63)--(5.4,0.77) ;
    
    \draw[-] (4.2,1.4)--(6.0,1.4) ;
    \draw[-] (4.2,1.33)--(4.2,1.47) ;
    \draw[-] (6.0,1.33)--(6.0,1.47) ;
    
    \draw[-] (4.8,2.1)--(6.6,2.1) ;
    \draw[-] (4.8,2.03)--(4.8,2.17) ;
    \draw[-] (6.6,2.03)--(6.6,2.17) ;
    
    \draw[-] (5.4,2.8)--(7.2,2.8) ;
    \draw[-] (5.4,2.73)--(5.4,2.87) ;
    \draw[-] (7.2,2.73)--(7.2,2.87) ;
    
    \draw[-] (6.0,0.0)--(7.8,0.0) ;
    \draw[-] (6.0,-0.07)--(6.0,0.07) ;
    \draw[-] (7.8,-0.07)--(7.8,0.07) ;
    
    \draw[-] (6.6,0.7)--(8.4,0.7) ;
    \draw[-] (6.6,0.63)--(6.6,0.77) ;
    \draw[-] (8.4,0.63)--(8.4,0.77) ;
    
    \draw[-] (7.2,1.4)--(9.0,1.4) ;
    \draw[-] (7.2,1.33)--(7.2,1.47) ;
    \draw[-] (9.0,1.33)--(9.0,1.47) ;
    
    \draw[-] (7.8,2.1)--(9.6,2.1) ;
    \draw[-] (7.8,2.03)--(7.8,2.17) ;
    \draw[-] (9.6,2.03)--(9.6,2.17) ;
    
    \draw[-] (8.4,2.8)--(10.2,2.8) ;
    \draw[-] (8.4,2.73)--(8.4,2.87) ;
    \draw[-] (10.2,2.73)--(10.2,2.87) ;
    
    \draw[-] (9.0,0.0)--(10.8,0.0) ;
    \draw[-] (9.0,-0.07)--(9.0,0.07) ;
    \draw[-] (10.8,-0.07)--(10.8,0.07) ;
    
    \draw[-] (9.6,0.7)--(11.4,0.7) ;
    \draw[-] (9.6,0.63)--(9.6,0.77) ;
    \draw[-] (11.4,0.63)--(11.4,0.77) ;
    
    \draw[-] (10.2,1.4)--(12.0,1.4) ;
    \draw[-] (10.2,1.33)--(10.2,1.47) ;
    \draw[-] (12.0,1.33)--(12.0,1.47) ;
    
    \draw[-] (10.8,2.1)--(12.6,2.1) ;
    \draw[-] (10.8,2.03)--(10.8,2.17) ;
    \draw[-] (12.6,2.03)--(12.6,2.17) ;
    
    \draw[-] (11.4,2.8)--(13.2,2.8) ;
    \draw[-] (11.4,2.73)--(11.4,2.87) ;
    \draw[-] (13.2,2.73)--(13.2,2.87) ;
    
    \draw[dotted] (0.0,-0.42)--(0.0,3.15) ;
    \draw[dotted] (0.6,-0.42)--(0.6,3.15) ;
    \draw[dotted] (1.2,-0.42)--(1.2,3.15) ;
    \draw[dotted] (1.8,-0.42)--(1.8,3.15) ;
    \draw[dotted] (2.4,-0.42)--(2.4,3.15) ;
    \draw[dotted] (3.0,-0.42)--(3.0,3.15) ;
    \draw[dotted] (3.6,-0.42)--(3.6,3.15) ;
    \draw[dotted] (4.2,-0.42)--(4.2,3.15) ;
    \draw[dotted] (4.8,-0.42)--(4.8,3.15) ;
    \draw[dotted] (5.4,-0.42)--(5.4,3.15) ;
    \draw[dotted] (6.0,-0.42)--(6.0,3.15) ;
    \draw[dotted] (6.6,-0.42)--(6.6,3.15) ;
    \draw[dotted] (7.2,-0.42)--(7.2,3.15) ;
    \draw[dotted] (7.8,-0.42)--(7.8,3.15) ;
    \draw[dotted] (8.4,-0.42)--(8.4,3.15) ;
    \draw[dotted] (9.0,-0.42)--(9.0,3.15) ;
    \draw[dotted] (9.6,-0.42)--(9.6,3.15) ;
    \draw[dotted] (10.2,-0.42)--(10.2,3.15) ;
    \draw[dotted] (10.8,-0.42)--(10.8,3.15) ;
    \draw[dotted] (11.4,-0.42)--(11.4,3.15) ;
    \draw[dotted] (12.0,-0.42)--(12.0,3.15) ;
    \draw[dotted] (12.6,-0.42)--(12.6,3.15) ;
    \draw[dotted] (13.2,-0.42)--(13.2,3.15) ;
    \draw (0.3,-0.7) node [] {$\omega_{1}$} ;
    \draw (0.9,-0.7) node [] {$\omega_{2}$} ;
    \draw (1.5,-0.7) node [] {$\omega_{3}$} ;
    \draw (11.7,-0.7) node [] {$\omega_{t}$} ;
    \draw (-0.6,0.0) node [] {$f_{0}^{(4)}$} ;
    \draw (-0.6,0.7) node [] {$f_{1}^{(4)}$} ;
    \draw (-0.6,1.4) node [] {$f_{2}^{(4)}$} ;
    \draw (-0.6,2.1) node [] {$f_{3}^{(4)}$} ;
    \draw (-0.6,2.8) node [] {$f_{4}^{(4)}$} ;

\end{tikzpicture}
\caption{Illustration of \eqref{eq:decoupV} in the case $n=3$, $\tau_n=2$,
$M=4$}
\label{fig:decoupV}
\end{figure}

We have (see Figure~\ref{fig:decoupV}) the decomposition
\begin{equation}\label{eq:decoupV}
S_{t} f=\sum_{s=0}^{n'-1}f_s^{(M)}, \quad 
f_s^{(M)}(\omega)
=\sum_{k=0}^{M-1}f(\omega_{[kn'+s+1,kn'+s+n]}). 
\end{equation}
Using H\"older's inequality and translation invariance leads to
\begin{equation*} 
\bigl\langle \eee^{S_{t}f},\IP\bigr\rangle \le \prod_{s=0}^{n'-1}\bigl\langle
\eee^{n'f_s^{(M)}},\IP\bigr\rangle^{1/n'} = \bigl\langle
\eee^{n'f_0^{(M)}},\IP\bigr \rangle .
\end{equation*}
Using then Lemma~\ref{lem:decobservabletaun} recursively $M-1$ times, we obtain
\begin{equation*} 
\bigl\langle \eee^{S_{t}f},\IP\bigr\rangle\leq \eee^{(M-1)d_n}\bigl\langle
\eee^{n'f},\IP\bigr \rangle^M\!\!\!= \eee^{(M-1)d_n}\left\langle
\frac{\Q_{n}}{\IP_{n}},\IP\right\rangle^M = \eee^{(M-1)d_n}
\left(\Q_n(\Omega_n^+)\right)^M\leq \eee^{(M-1)d_n},
\end{equation*} 
where  $d_n = c_n + \tau_n \log |\cA| = o(n)$. Thus,
\begin{equation*}
	\frac 1 t \log A_t \leq  \frac {(M-1)d_n}{t} \leq \frac {d_n}{n'} ,
\end{equation*}
whence $Q(f) \leq  \frac {d_n}{n'}$. Combining this with~\eqref{1.32}, we derive
\begin{equation*}
\frac 1 n \Ent(\Q_{n}|\,\IP_{n})=\frac {n'}{n}\langle f,\Q\rangle
\leq \frac {n'}{n}\left(\langle f,\Q\rangle 
-Q(f) \right)+\frac {d_n}{n}\leq   \frac {n'}{n}\I(\Q)+\frac {d_n}{n}.
\end{equation*}
Passing to the limit as $n\to \infty$ shows that $\LL(\Q) \leq {\II}(\Q)$, and
\eqref{1.13} follows. Finally, since both $\Q \mapsto h(\Q)$  and  $\Q\mapsto
\varsigma(\Q)$ are affine, we obtain from \eqref{1.13} that so is
$\I$.\hfill\qed

\medskip

\subsection{Level-3 fluctuation relation}

\begin{proposition} \label{p1.11}
Assume \hUD\footnote{The same remark as in Proposition~\ref{p1.6} applies:
although it is a standing assumption in this section, \hSLD is not necessary in
this proposition if we simply define $\I$ by \eqref{eq:IIasltf}.} and that $\wP
= \Theta \P$ with $\Theta$ as in Definition~\ref{def:familyinvolutions}. Then,
for any~$\Q\in\PP_\varphi(\Omega)$ such that $(\Theta \Q)_t$ and~$\Q_t$ are
equivalent for all~$t$,  ${\II}(\Q)<+\infty$, and  ${\II}(\Theta\Q)<+\infty$, we
have
\begin{equation} \label{1.82}
{\II}(\Theta\Q)={\II}(\Q)+\ep(\Q),
\end{equation}
where we set $\ep(\Q)=\varsigma(\Theta\Q)-\varsigma(\Q)$. Moreover,
\begin{equation} \label{1.83}
\ep(\Q)=\lim_{t\to\infty}\frac 1 t \langle\sigma_t,\Q\rangle. 
\end{equation}
\end{proposition}

\proof{}  \hUD and  Proposition~\ref{p1.6} imply 
\begin{equation} \label{1.84}
{\II}(\Q)=\varsigma(\Q)-h(\Q), \quad 
{\II}(\Theta\Q)=\varsigma(\Theta\Q)-h(\Theta\Q).
\end{equation}
Since $\theta_t$ is a bijection, we see that
\begin{equation*}
h_t(\Q)=-\sum_{w\in\Omega_t}
\Q_t(w)\log\Q_t(w)=-\sum_{w\in\Omega_t}
\Q_t\bigl(\theta_t(w)\bigr)
\log\Q_t\bigl(\theta_t(w)\bigr)=h_t(\Theta\Q).   
\end{equation*}
It follows that $h(\Theta\Q)=h(\Q)$. Combining this with~\eqref{1.84}, we arrive
at~\eqref{1.82}. We now prove \eqref{1.83}. As was already observed in the proof
of Proposition~\ref{p1.6}, the conditions ${\II}(\Q)<+\infty$, and 
${\II}(\Theta\Q)<+\infty$ imply that $\QQ_t \ll \P_t$ and $(\Theta \QQ)_t \ll
\P_t$ for all $t$. We remark that
\begin{align*}
\langle\sigma_t,\Q\rangle
&= -\left\langle\log\frac{\Q_t}{\IP_t},\Q_t\right\rangle +
\left\langle\log\frac{\Q_t}{\wP_t},\Q_t\right\rangle =
-\left\langle\log\frac{\Q_t}{\IP_t},\Q_t\right\rangle +
\left\langle\log\frac{(\Theta \Q)_t}{\P_t},(\Theta \Q)_t\right\rangle\\
&=-\Ent(\Q_t|\IP_t)+\Ent((\Theta\Q)_t|\IP_t). 	
\end{align*}

Dividing this relation by~$t$, passing to the limit as $t\to\infty$, and
using~\eqref{1.13}, we obtain 
$$
\lim_{t\to\infty}\frac 1 t \langle\sigma_t,\Q\rangle={\II}(\Theta\Q)-{\II}(\Q). 
$$
Comparing this with~\eqref{1.82}, we arrive at the required
relation~\eqref{1.83}. 
\hfill\qed

\appendix 
\section{Appendix}

\subsection{Technical results}

We first prove two lemmas justifying Remarks~\ref{rem:hyposymmetricdec}
and~\ref{rem:ergo}.

\begin{lemma}\label{lem:chgthyposym}Assume that \hUD holds for both $\P$ and
$\widehat \P$,
 and that \hSLD holds for both $\P$ and $\widehat \P$ with the same $\xi$, in the sense that
for all $t\in\nn$, all $u \in \Omega_t$ and all $v\in \Omega_{\fin}$, $|v| \geq
1$, there exists $|\xi| \leq  \tau_t $ such that for both  $\P^\sharp = \P$ and
$\P^\sharp = \widehat \P$,
\begin{equation*}
 \eee^{- c_t} \IP^\sharp(u) \IP^\sharp(v) \leq \IP^\sharp(u \xi v).
\end{equation*}
Then \hSSD holds (for some larger $\tau_t $ and $c_t$).
\end{lemma}
\proof{} Let $u \in \Omega_t$ and $v \in \Omega_\fin$. Then, by
Lemma~\ref{lem:extendword}, there exists $b\in \Omega_{\tau_t }$ such that
$\P^\sharp (v) \geq \P^\sharp (bv) \geq \P^\sharp (v) \eee^{-C \tau_t }$. By 
assumption there is $|\xi| \leq \tau_t $ such that
\begin{equation*}
\P^\sharp(u\xi b v) \geq \eee^{-c_t}\P^\sharp(u)\P^\sharp(bv) \geq
\eee^{-c_t-C\tau_t }\P^\sharp(u)\P^\sharp(v).
\end{equation*}
Let then $\xi' = \xi b$. By \hUD, we have
\begin{equation*}
\P^\sharp(u\xi' v) \leq
\eee^{c_t}\P^\sharp(u)\P^\sharp(\xi'_{[\tau_t+1,|\xi'|]}v) \leq
\eee^{c_t}\P^\sharp(u)\P^\sharp(v).
\end{equation*}
(Note that $\xi'_{[\tau_t+1,|\xi'|]}$ may be the empty word.)
Thus,  
\begin{equation*}
\eee^{-c_t-C\tau_t }\P^\sharp(u)\P^\sharp(v)\leq \P^\sharp(u\xi' v) \leq
\eee^{c_t}\P^\sharp(u)\P^\sharp(v).
\end{equation*}
Since $|\xi'| \leq 2\tau_t$, \hSSD holds with $\tau_t $ and $c_t$ replaced with
the sequences $2\tau_t $ and $c_t + C\tau_t $, which are also $o(t)$.\hfill \qed

Turning to Remark~\ref{rem:ergo}, we now give a sufficient condition for $\P$ to
be ergodic (which is fulfilled, in particular, if \hSLD holds with $\sup_t \tau_t
< \infty$ and $\sup_t c_t < \infty$).

\begin{lemma}\label{lem:ergodicQstar} Assume the following form of lower
decoupling: there exist $c>0$ and $k\in \N_0$ such that for all $t\in\nn$, all
$u \in \Omega_t$ and all $v\in \Omega_{\fin}$, $|v|\geq 1$,
\begin{equation*}
\sum_{\substack{\xi\in \Omega_\fin \\ \tau_t - k \leq |\xi| \leq \tau_t}}\IP(u
\xi v) \geq  \eee^{- c} \IP(u) \IP(v) .
\end{equation*}
Then $\P$ is ergodic. 
\end{lemma}
\proof{} Consider first two cylinder sets $\cC_1$ and~$\cC_2$ given by $\cC_i =
\{\omega \in \Omega ~|~\omega_{[1,r]} \in C_i\}$ for some $r\in\nn$ and sets
$C_i \subset \Omega_r$, $i=1,2$. Observe that, by assumption,
\begin{align*}
\sum_{j = \tau_t-k}^{\tau_t}\P(\cC_1 \cap \varphi^{-r-j} \cC_2) &= \sum_{j =
\tau_t-k}^{\tau_t}\sum_{u \in C_1}\sum_{v \in C_2}\sum_{\xi \in \Omega_{j}} \P(u
\xi v)\\
& \geq \eee^{-c}\sum_{u \in C_1}\sum_{v \in C_2}\P(u) \P(v)
= \eee^{-c}\,\P(\cC_1)\,\P(\cC_2).
\end{align*} 
Thus, there exists $t \in [r+\tau_t-k, r+\tau_t]$ such that
\begin{equation*}
\P(\cC_1 \cap \varphi^{-t} \cC_2) \geq C\,\P(\cC_1)\,\P(\cC_2),
\end{equation*}
where $C = \eee^{-c}/(k+1) > 0$. Since any Borel set in~$\Omega$ can be
approximated by cylinder sets (and the constant $C$ is independent of the choice
of $\cC_i$), it follows that~$\P$ is ergodic. The details are as follows. Assume
$B \subset \Omega$ is an invariant Borel set (\ie $\IP(B \triangle \varphi^{-1}
B)=0$),\footnote{The symmetric difference $A \triangle B$ of two sets $A$ and
$B$ is defined as $(A\setminus B) \cup (B\setminus A)$.} and let $\varepsilon >
0$. We can find two cylinder sets $\cC_1, \cC_2$ that approximate~$B$ and~$B^{c}
$, in the sense that $\P(B^c \triangle \cC_1) + \P(B \triangle \cC_2) \leq
\varepsilon$. Then,
\begin{align*}
0 &= \sup_{t\in \nn_0}\P(B^c\cap  \varphi^{-t} B) \geq \sup_{t \in \nn_0} \P(\cC_1 \cap
\varphi^{-t} \cC_2) - \varepsilon \\
& \geq C\, \P(\cC_1)\,\P(\cC_2) -\varepsilon 
\geq  C (\P(B^c)-\varepsilon)(\P(B)-\varepsilon) -\varepsilon\\
& \geq C\,\P(B^c)\P(B)-(2C+1)\varepsilon .
\end{align*}
Since $\varepsilon$ was arbitrary, we have $\P(B^c)\P(B) = 0$, so that $\P(B)
\in \{0, 1\}$. This completes the proof that $\P$ is ergodic.\hfill\qed

The next lemma proves the properties of irreducible Markov processes mentioned
in Example~\ref{ex:markovirred}.

\begin{lemma}\label{lem:irreducibleMarkov}Let $\P\in \cP_\varphi(\Omega)$ be a
Markov process. Then \hUD holds. Assume furthermore that it is irreducible (\ie that for
all $a,b\in \cA$, there exists $\xi^{(a,b)}\in \Omega_\fin$ such that
$\P(a\xi^{(a,b)} b) > 0$). Then \hSLD holds. If, in addition, $\widehat \P\in
\cP_\varphi(\Omega)$ is another Markov process such that $\P_2 \ll \widehat
\P_2$, then \hSSD holds.
\end{lemma}
\proof{} Since $\P$ is Markov and shift-invariant, we have
\begin{equation*}
	\P(w) = \P_1(w_1) P(w_1;w_2) P(w_2;w_3) \cdots P(w_{t-1};w_t), \qquad w\in
\Omega_t,
\end{equation*}
for some transition matrix $(P(a;b))_{a,b\in \cA}$.

{\em Upper Decoupling.} We show that \hUD holds with $\tau_t \equiv 0$ and 
\begin{equation*}
c_t \equiv - \min_{a \in \cA : \P_1(a) > 0}\log\P_1(a). 
\end{equation*}
Indeed, given $u\in \Omega_t$ and $v\in \Omega_\fin$ such that $\P_1(v_1) > 0$,
we have
\begin{equation*}
{\P(u v)} = \frac{P(u_t; v_1)}{\P_1(v_1)} \P(u)\P(v) \leq \eee^{c_t}\P(u)\P(v).
\end{equation*}
If $\P_1(v_1) = 0$, then by invariance $\P(uv)  = 0\leq \eee^{c_t}\P(u)\P(v)$,
so that \hUD is proved.

{\em Selective Lower Decoupling.} Assume now that the process is irreducible. This implies that
$\P_1(a) > 0$ for all $a\in \cA$. Let $\tau = \max_{a,b\in \cA} |\xi^{(a,b)}|$.
Given two words $u\in \Omega_{t}$ and $v\in \Omega_\fin$, $|v|\geq 1$, let $\xi
= \xi^{(u_t,v_1)}$. Then either $|\xi| = 0$, in which case
\begin{equation}\label{eq:puv1}
{\P(u v)}= \frac{P(u_t; v_1)}{\P_1(v_1)}{\P(u)\P(v)},
\end{equation}
or $k:=|\xi^{(u_t,v_1)}| \geq 1$, in which case
\begin{equation}\label{eq:puv2}
{\P(u\xi v)} =
\frac{P(u_t;\xi_1)\P(\xi)P(\xi_k;v_1)}{\P_1(\xi_1)\P_1(v_1)}{\P(u)\P(v)}.
\end{equation}
The factors in front of $\P(u)\P(v)$ on the right-hand sides of \eqref{eq:puv1}
and \eqref{eq:puv2} are positive and depend only on~$u_t$ and~$v_1$. We obtain a
lower bound by taking the minimum over all possible values of~$u_t$ and~$v_1$.
This implies that \hSLD holds with $\tau_t \equiv \tau$ and some~$c_t$ independent
of~$t$.

{\em Selective Symmetric Decoupling.}
Assume finally that $\widehat \P$ is another Markov process such that $\P_2 \ll
\widehat \P_2$. Then by the Markov property we have that $\P_t \ll \widehat
\P_t$ for all $t$, so $\widehat\P$ is irreducible, and one can choose the same
$\xi^{(a,b)}$ as for $\P$ (\ie we have both $\P(a\xi^{(a,b)} b) > 0$ and
$\widehat \P(a\xi^{(a,b)} b) > 0$). Considerations similar to the above imply
that \hSSD holds. \hfill\qed

Finally, for the reader's convenience, we prove some well-known properties of
the Kolmogorov--Sinai entropy that are used in the proof of
Proposition~\ref{p1.6} (see for example Section~4.3 of \cite{KH1995}).

\begin{lemma}\label{lem:propsKS} For all $\Q \in \cP_\varphi(\Omega)$, the limit
\begin{equation} \label{eq:limithKS}
h(\Q) = \lim_{t\to\infty}\frac 1 th_t(\Q)
\end{equation}
exists, is finite, and the mapping $h: \cP_\varphi(\Omega) \to [0, \infty)$ is
upper semicontinuous and affine.
\end{lemma}
\proof{} First, it follows from $\varphi$-invariance and the inequality $\log x
\leq x-1$ that
\begin{equation*}
h_{t+t'}(\Q)  - h_{t}(\Q) - h_{t'}(\Q)	 = \sum_{w\in \Omega_t}\sum_{w'\in
\Omega_{t'}} \Q(ww') \log \frac{\Q(w)\Q(w')}{\Q(ww')} \leq 0.
\end{equation*}
By subadditivity, the limit \eqref{eq:limithKS} exists, is finite, and $h(\Q) =
\inf_{t\in\nn} \frac 1 t h_t(\Q)$. Moreover, as an infimum over a family of
continuous functions, $h$ is upper semicontinuous. That $h$ is affine is an immediate
consequence of the following relation: for all $\Q^{(1)},\Q^{(2)}\in
\cP_\varphi(\Omega)$, and all $p_1 \in (0,1)$, $p_2 = 1-p_1$, we have
\begin{equation}\label{eq:preaffineQ1Q2}
	\sum_{i=1,2}p_i h_t\big(\Q^{(i)}\big) \leq 
h_t\left(\sum_{i=1,2}p_i\Q^{(i)}\right) \leq \sum_{i=1,2}p_i
h_t\big(\Q^{(i)}\big) - \sum_{i=1,2}p_i \log p_i.
\end{equation}
To complete the proof, we now establish \eqref{eq:preaffineQ1Q2}.
The first inequality follows from the concavity of $x\mapsto f(x):= -x\log x$.
Indeed, we have
\begin{equation*}
	\sum_{i=1,2}p_i h_t\big(\Q^{(i)}\big) =\sum_{w\in \Omega_t}\sum_{i=1,2}p_i
f\left(\Q^{(i)}(w)\right) \leq  \sum_{w\in \Omega_t}
f\left(\sum_{i=1,2}p_i\Q^{(i)}(w)\right)= 
h_t\left(\sum_{i=1,2}p_i\Q^{(i)}\right).
\end{equation*}
For the second inequality, we observe that
\begin{align*}
 h_t\left(\sum_{i=1,2}p_i\Q^{(i)}\right) &=-\sum_{w\in
\Omega_t}\sum_{i=1,2}p_i\Q^{(i)}(w)
\log\left(\sum_{j=1,2}p_j\Q^{(j)}(w)\right)\\
&\leq - \sum_{w\in \Omega_t}\sum_{i=1,2}p_i\Q^{(i)}(w)
\log\left(p_i\Q^{(i)}(w)\right)\\
& = - \sum_{i=1,2}p_i\sum_{w\in \Omega_t}\Q^{(i)}(w)
\log\left(\Q^{(i)}(w)\right)- \sum_{i=1,2}p_i \log p_i \sum_{w\in
\Omega_t}\Q^{(i)}(w) \\
& =  \sum_{i=1,2}p_i h_t\big(\Q^{(i)}\big) - \sum_{i=1,2}p_i \log p_i.
\end{align*}
The proof is complete.\hfill\qed

\subsection{Hidden Markov chain example}\label{ss:hiddenMarkChain}

In the section we discuss the hidden Markov chain of
Example~\ref{ex:HiddenMarkov}. For reasons of space, we only outline the main
steps of the analysis; the details are easy to fill. 

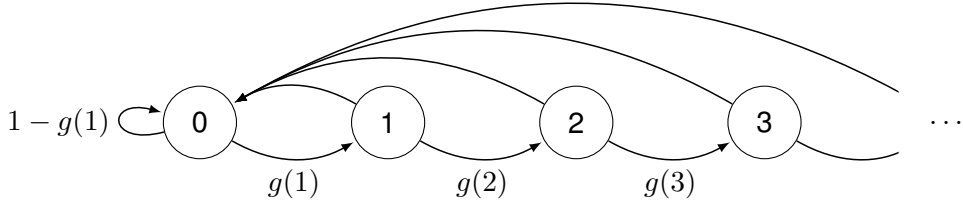
\begin{figure}[htb]
\centering
	 \begin{tikzpicture}[font=\sffamily]
        \node[state,
              draw=black,
			] (s0) {0};
        \node[state,
              right=1.5cm of s0,
              draw=black, 
			] (s1) {1};
        \node[state,
              right=1.5cm of s1,
              draw=black, 
			] (s2) {2};
        \node[state,
              right=1.5cm of s2,
              draw=black, 
			] (s3) {3};
        \node[state,
              right=1.5cm of s3,
              draw=black, 
			] (s4) {4};

        \draw[every loop,
        auto=right,
        line width=0.2mm,
        >=latex,
		]
            (s0) edge[bend right, auto=right]  node {$g(1)$} (s1)
            (s1) edge[bend right, auto=right]  node {$g(2)$} (s2)
            (s2) edge[bend right, auto=right]  node {$g(3)$} (s3)
            (s3) edge[bend right, auto=right]  node {} (s4)
            (s1) edge[bend right, auto=left] node {} (s0)
            (s2) edge[bend right, auto=left] node {} (s0)
            (s3) edge[bend right, auto=left] node {} (s0)
            (s4) edge[bend right, auto=left] node {} (s0)
            (s0) edge[loop left]             node {$1-g(1)$} (s0)
			;

\draw [fill=white,draw=white] (10.5,0.6) rectangle (9.2,-0.6) node[pos=.5]
{$\dots$};

    \end{tikzpicture}
\caption{Illustration of the Markov chain.}\label{f:hiddenmarkov}
\end{figure}

Let $(\gamma(n))_{n\in\nn_0}$ be a sequence of non-negative numbers such that
$\gamma(0) = 0$ and  $\gamma(n+1) \geq \gamma(n) + \varepsilon$ for all $n$ and
some $\varepsilon>0$.
We consider a countable Markov chain with states $0,1,2, \dots$ such that from
each state $n\in\nn_0$, we jump either to $n+1$ with probability $g(n+1) :=
\eee^{\gamma(n)-\gamma(n+1)}$ or we jump to $0$ with probability $1-g(n+1)$ (see
Figure~\ref{f:hiddenmarkov}).  

This Markov chain admits a unique invariant measure, and we denote by $\Q$ the
corresponding Markov process on $\nn_0^\nn$. Note that 
$$
\Q_1(n) = Z^{-1}\eee^{-\gamma(n)}, \quad Z=\sum_{n=0}^\infty \eee^{-\gamma(n)}.
$$
Let $\cA= \{\mathrm{a},\mathrm{b}\}$ and let $\Psi: \nn_0^\nn \to \cA^\nn$ be defined by
$\Psi (\omega_1, \omega_2, \dots) = (\psi(\omega_1), \psi(\omega_2), \dots)$,
where $\psi(0) = \mathrm{a}$ and $\psi(n) = \mathrm{b}$ for all $n\geq 1$. Our main object of
interest is the invariant probability measure on $\Omega = \cA^\nn$ defined by
$\P = \Q \circ \Psi^{-1}$. The following holds:
\begin{itemize}[leftmargin=1.2em]
	\item The measure $\P$ has full support, and for any words $u$ and $v$ we have 
	\begin{equation*}
	\P(u\mathrm{a}v) = {\P(u\mathrm{a}) \P(\mathrm{a}v)}/{\P_1(\mathrm{a})}.
	\end{equation*}
Moreover, $\P$ is reversible in the sense that $\P_t(w_1, \dots, w_t) =
\P_t(w_t, \dots, w_1)$ for all $t\in\nn$ and $w \in \Omega_t$. 
	\item Let us set
\begin{equation*}
	p_t(w) =  \frac{\P_{t+2}(\mathrm{a}w\mathrm{a})}{\P_1(\mathrm{a})}, \qquad w \in \Omega_t,
\end{equation*}
with the convention $p_0(\kappa) = \P_2(\mathrm{aa})/\P_1(\mathrm{a})=1-g(1)$, where $\kappa$ is
the empty word. Then 
$$
p_{|u|+|v|+1}(u\mathrm{a}v) = p_{|u|}(u)p_{|v|}(v)\quad \mbox{for all $u,v\in
\Omega_\fin$}.
$$
\item The quantities $\P_{t+1}(\mathrm{b}^t\mathrm{a})$, $\P_{t+1}(\mathrm{a}\mathrm{b}^t)$ and $\P_t(\mathrm{b}^t)$ are
bounded above and below by some constant (independent of $t$) times
$\eee^{-\gamma(t)}$. More generally, the quantities $\P_{t+1}(\mathrm{a}w)$,
$\P_{t+1}(w\mathrm{a})$ and $p_{t}(w)$ are bounded above and below by a constant
(independent of $t$ and $w\in \Omega_t$) times $\P_t(w)$. 

\item $\P$ satisfies \hSLD with $\tau_t \equiv 1$ and $\sup_t c_t < \infty$ (by
taking $\xi = \mathrm{a}$ in \eqref{eq:sellowerdecoup}). 
\item $\P$ satisfies \hUD with $\tau_t \equiv 0$ and $c_t = c + \sup_{n\in\nn_0}
[\gamma(n)+\gamma(t)-\gamma(n+t)]$ for some $c>0$, provided that
$(\gamma(n))_{n\in\nn_0}$ is such that $c_t = o(t)$. 
\end{itemize}

We assume now that $\widehat \P$ is constructed in the same way, with $(\widehat
\gamma(n))_{n\in \nn_0}$ satisfying the same conditions as $(\gamma(n))_{n\in \nn_0}$. We define $\widehat 
p_t$ in the same way as $p_t$. Then the following holds:
\begin{itemize}[leftmargin=1.2em]
	\item The pair $(\P, \widehat \P)$ satisfies \hSSD with $\tau_t \equiv 1$ and
$\sup_t c_t < \infty$ (by taking $\xi = \mathrm{a}$ in \eqref{eq:timereversibledec}).
\item The function $q$ defined in \eqref{eq:defqalphaentrop} can be
written as
\begin{equation*}
q(\alpha) =  \lim_{t\to\infty} \frac  1t \log \sum_{w\in \Omega_t}
\zeta_{t,\alpha}(w), \qquad \alpha \in \rr, 
\end{equation*}
where the quantity $\zeta_{t,\alpha}(w):= \eee^{(\alpha+1) \log p_t(w) - \alpha 
\log\widehat p_t(w)}$ defined for $t\in \nn_0$ and $w\in \Omega_t$ satisfies the
relation 
$$
\zeta_{|u|+|v|+1,\alpha}(u\mathrm{a}v) = \zeta_{|u|,\alpha}(u)\zeta_{|v|,\alpha}(v), \qquad u,v\in \Omega_\fin.
$$
\item We have $q(\alpha) = -\log \rho(\alpha)$, where $\rho(\alpha)$ is the
radius of convergence of the power series
\begin{equation*}
R_\alpha(x) = \sum_{t\in \nn_0} r_t(\alpha)x^t, \qquad r_t(\alpha) = 	\sum_{w\in
\Omega_t} \zeta_{t,\alpha}(w)  .
\end{equation*}
\item Let us set
\begin{equation*}
 U_\alpha(x) =\sum_{t\in \nn_0} u_t(\alpha)x^t, \qquad u_t(\alpha) =
\zeta_{t,\alpha}(\mathrm{b}^t)	,
\end{equation*}
and observe that
\begin{equation*}
u_t(\alpha) = 	(1-g(t+1))^{\alpha+1}(1-\widehat
g(t+1))^{-\alpha}\eee^{-(\alpha+1)\gamma(t) + \alpha\widehat  \gamma(t)},
\end{equation*}
where $1-g(t+1)$ and $1-\widehat g(t+1)$ are bounded from above by one and  from
below by a constant $c>0$ uniformly in~$t$.
The radius of convergence of $U_\alpha$ is 
$$
\kappa(\alpha) 	= \liminf_{t\to\infty} \exp((\alpha+1) t^{-1}\gamma(t) - \alpha
t^{-1}\widehat \gamma(t)).
$$ 

\item By sorting the words $w\in \Omega_t$ according to the position of the
first occurrence of the symbol $\mathrm{a}$ (if there is one), we get the renewal equation
$$r_t(\alpha) = \sum_{k=0}^{t-1} u_k(\alpha) r_{t-1-k}(\alpha)	 +  
u_{t}(\alpha).$$ This relation translates into the algebraic equation
$R_\alpha(x) = x  R_\alpha(x) U_\alpha(x)+ U_\alpha(x)$, so that
\begin{equation*}
 R_\alpha(x) = \frac{ U_\alpha(x)}{1-xU_\alpha(x)}.
\end{equation*}
\end{itemize}
The power series defined above have strictly positive coefficients for all
$\alpha$. They are strictly increasing functions of $x\geq 0$, and jointly lower
semicontinuous in $\alpha \in \rr$ and $x\geq 0$. As already discussed, the
quantity of interest is the radius of convergence $\rho(\alpha)$ of $R_\alpha$. 
For each fixed $\alpha$, we are in one of the following two cases: 
\begin{itemize}
	\item[(a)] There exists $x> 0$ such that $xU_\alpha(x) = 1$, and in this case
$\rho(\alpha) = x$.
\item[(b)] $x U_\alpha(x) < 1$ for all $ 0 \leq x  \leq  \kappa(\alpha)$, in
which case $\rho(\alpha) = \kappa(\alpha)$.
\end{itemize}

In both cases, we have $\rho(\alpha) = \sup\{x\geq 0: xU_\alpha(x) \leq 1\}$.
In case (a),  $\rho'(\alpha)$ can be obtained by differentiating the relation
$\rho(\alpha)U_\alpha(\rho(\alpha)) \equiv 1$, and we obtain
\begin{equation}\label{eq:deriveerho}
\rho'(\alpha) = -\frac{\sum_{n\geq 0}\partial_\alpha
u_n(\alpha)\rho^{n+1}(\alpha)}{\sum_{n\geq 0} u_n(\alpha)(n+1)\rho^{n}(\alpha)}.
\end{equation}
In case (b), we simply have $\rho'(\alpha) = \kappa'(\alpha)$. (Of course these
relations hold only if the corresponding quantities are well defined, and if
$\alpha$ is not at a transition point between the two cases.)

Different situations can occur depending on the concrete choice of $(\gamma(n))_{n\in \nn_0}$
and $(\widehat \gamma(n))_{n\in \nn_0}$. We now briefly discuss six interesting cases. We do
not give any proofs, as these examples are easily understood by substituting the
relevant values in the formulas for $\kappa(\alpha)$ and $\rho(\alpha)$ (and
their derivative). The interested reader may wish to investigate the matter
further by trying other expressions for $(\gamma(n))_{n\in \nn_0}$
and $(\widehat \gamma(n))_{n\in \nn_0}$.

\medskip{\bf Example 1.} Let $\gamma(n) = n$ and $\widehat \gamma(n) = n^2$. We
have  $\kappa(\alpha) = +\infty$ for $\alpha < 0$, $\kappa(0) = \eee$, and
$\kappa(\alpha) = 0$ for $\alpha > 0$. For $\alpha \leq 0$ we are in case (a),
and it follows from the identity $\rho(\alpha)U_\alpha(\rho_\alpha) \equiv 1$
that $\rho$, and hence $q$, are analytic on $(-\infty, 0)$. We know already that
$q(0) = 0$. When $\alpha > 0$, we have $\rho(\alpha) = \kappa(\alpha)= 0$, and
hence we are in case (b) with $q(\alpha) = + \infty$.\footnote{The fact that
$q(\alpha)$ is infinite when $\alpha>0$ also follows from the observation that
in the sum $\sum_{w\in \Omega_t}\eee^{\alpha \sigma_t(w)}\P_t(w)$, the
contribution of $w = \mathrm{b}^{t}$ grows like $\eee^{-(\alpha+1)t+\alpha t^2}$.} 
Evaluating the quantity \eqref{eq:deriveerho} in the limit $\alpha \uparrow 0$
and $\rho(\alpha) \downarrow \rho(0) = 1$ shows that $0 > \rho'(0^-) > -
\infty$. Since $q'(\alpha) =  -\frac d {d\alpha} \log \rho(\alpha) = -
\frac{\rho'(\alpha)}{\rho(\alpha)}$, we conclude that $q'(0^-)$ is finite
(numerical evaluation gives $q'(0^-) =0.3294...$).  See
Figure~\ref{sf:case1}.\footnote{A bisection method was used to find
$\rho(\alpha) = \sup\{x \geq 0: xU_\alpha(x) \leq 1\}$. After that, $q = -\log \rho$
was obtained by direct computation.}

\medskip{\bf Example 2.} By replacing $n^2$ with $\exp(2n)$ in Example 1, we
obtain the same results except that now $q'(0^-) = + \infty$. See
Figure~\ref{sf:case2}.

\medskip{\bf Example 3.}  Let  $\gamma(n) = n + c n^2$ and $\widehat \gamma(n) =
n^2$ with $c\in (0,1)$. An analysis similar to that of Example~1 shows that
$q(\alpha)$ is infinite for $\alpha > \alpha_* := c/(1 -  c)$, and finite for
$\alpha \leq \alpha_*$. The case $c=1/2$, $\alpha_* = 1$ is represented in
Figure~\ref{sf:case3}.

\medskip{\bf Example 4.}  Let  $\gamma(n) =c n^2$ and $\widehat \gamma(n) = n^2
+ n^{3/2} $ with $c\in (0,1)$. For $\alpha < \alpha_* := c/(1 -  c)$ we are in
case (a), while for $\alpha \geq  \alpha_*$ we are in case (b) with
$\kappa(\alpha) = 0$ and  $q(\alpha) = +\infty$. The function $q$ increases
continuously to $+\infty$ when $\alpha \uparrow \alpha_*$. The case $c=1/2$ is
plotted in Figure~\ref{sf:case4}.

\pgfplotstableread{plotdata.dat}{\caseonedata}

\pgfplotstableread{plotdata.dat}{\casetwodata}

\pgfplotstableread{plotdata.dat}{\casethreedata}

\pgfplotstableread{plotdata.dat}{\casefourdata}

\begin{figure}[ht]
\centering
\subfloat[Example 1]{\label{sf:case1}\begin{tikzpicture}[scale=1]
\begin{axis}[minor tick num=1,
width=7.72cm,height=5cm,
xmin=-1.2,
scaled ticks = false, tick label style ={/pgf/number format/fixed},
]
\addplot [black,thick] table [x={alpha}, y={ealpha}] {\caseonedata} ;
\addlegendentry{$q(\alpha)$} ;
\end{axis}
\end{tikzpicture}
}\hfill
\subfloat[Example 2]{\label{sf:case2}\begin{tikzpicture}[scale=1]
\begin{axis}[minor tick num=1,
width=7.72cm,height=5cm,
xmin=-1.2,
]
\addplot [black,thick] table [x={alpha}, y={ealpha}] {\casetwodata} ;
\addlegendentry{$q(\alpha)$} ;
\end{axis}
\end{tikzpicture}
}\\
\subfloat[Example 3]{\label{sf:case3}~~~\begin{tikzpicture}[scale=1]
\begin{axis}[minor tick num=1,
width=7.72cm,height=5cm,
scaled ticks = false, tick label style ={/pgf/number format/fixed},
legend pos=north west,
xmin=-1.2,
ymax=1.1e-1,
]
\addplot [black,thick] table [x={alpha}, y={ealpha}] {\casethreedata} ;
\addlegendentry{$q(\alpha)$} ;
\end{axis}
\end{tikzpicture}}\hfill 
\subfloat[Example 4]{\label{sf:case4}\begin{tikzpicture}[scale=1]
\begin{axis}[minor tick num=1,
width=7.72cm,height=5cm,
legend pos=north west,
xmin=-1.2,
ymax=20,
]
\addplot [black,thick] table [x={alpha}, y={ealpha}] {\casefourdata} ;
\addlegendentry{$q(\alpha)$} ;
\end{axis}
\end{tikzpicture}
}

\caption{Graph of $q$ for Examples 1--4. The horizontal axis represents
$\alpha$. In Examples 1--3, the line is interrupted where $q$ jumps to
$+\infty$.}
\end{figure}
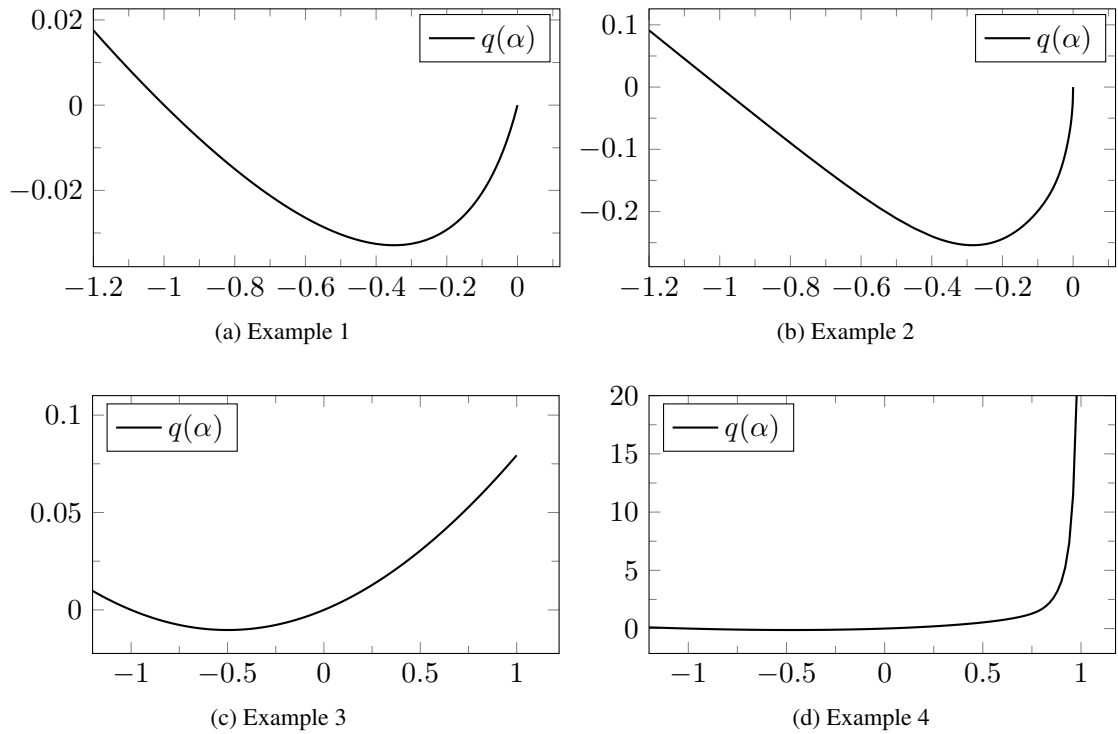

\medskip{\bf Example 5.}  Let $\gamma(n) =n$ and $\widehat \gamma(n) = 2n -
2\log(1 + n/2)$. In this case, $\kappa(\alpha) = \exp( 1-\alpha)$. One shows
that there exists $\alpha_* > 1$ (numerical evaluation gives $\alpha_* =
1.1305...$) such that we are in case~(a)  if $\alpha \leq \alpha_*$, and in
case~(b) if $\alpha > \alpha_*$. By evaluating \eqref{eq:deriveerho} as $\alpha
\uparrow \alpha_*$, $\rho(\alpha) \downarrow \rho(\alpha_*) = \kappa(\alpha_*)$
and comparing with $\kappa'(\alpha_*)$, one shows that $q$ is not differentiable
at $\alpha_*$. See Figure~\ref{f:case5}.

\begin{figure}[ht]
\centering

\pgfplotstableread{plotdata.dat}{\casefivedata}
\begin{tikzpicture}[scale=1]
\begin{axis}[minor tick num=1,
width=7.75cm,height=5cm,
legend pos=north west,
ymax=1.3,
xmin=-1.5,
xmax=1.5,
ymin=0.65,
]
\addplot [black, thick] table [x={alpha}, y={crossone}] {\casefivedata} ;
\addlegendentry{$\rho(\alpha)$} ;
\addplot [dashed,red,thick] table [x={alpha}, y={kappaalpha}] {\casefivedata};
\addlegendentry{$\kappa(\alpha)$} ;
\end{axis}
\end{tikzpicture}
\begin{tikzpicture}[scale=1]
\begin{axis}[minor tick num=1,
legend pos=north west,
width=7.75cm,height=5cm,
xmin=-1.5,
xmax=1.5,
ymax=0.4
]
\addplot [black,thick] table [x={alpha}, y={ealpha}] {\casefivedata} ;
\addlegendentry{$q(\alpha)$} ;
\end{axis}
\end{tikzpicture}

\caption{Numerical evaluation of $\rho, \kappa$ and $q$ for Example 5.}
\label{f:case5}
\end{figure}
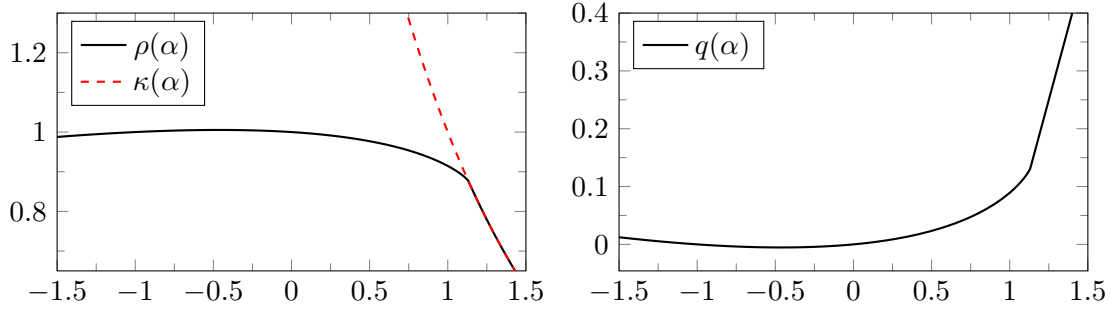
\begin{figure}[ht]
\centering

\pgfplotstableread{plotdata.dat}{\casesixdata}
~~\begin{tikzpicture}[scale=1]
\begin{axis}[minor tick num=1,
width=7.85cm,height=5cm,
ymax=7,
xmin=-1.4,
xmax=0.5,
]
\addplot [black,thick] table [x={alpha}, y={crossone}] {\casesixdata} ;
\addlegendentry{$\rho(\alpha)$} ;
\addplot [dashed,red,thick] table [x={alpha}, y={kappaalpha}] {\casesixdata};
\addlegendentry{$\kappa(\alpha)$} ;
\end{axis}
\end{tikzpicture}
\begin{tikzpicture}[scale=1]
\begin{axis}[minor tick num=1,
width=7.85cm,height=5cm,
xmin=-1.4,
xmax=0.5,
ymax=3,
]
\addplot [black,thick] table [x={alpha}, y={ealpha}] {\casesixdata} ;
\addlegendentry{$q(\alpha)$} ;
\end{axis}
\end{tikzpicture}

\caption{Numerical evaluation of $\rho, \kappa$ and $q$ for Example 6.}
\label{f:case6}
\end{figure}
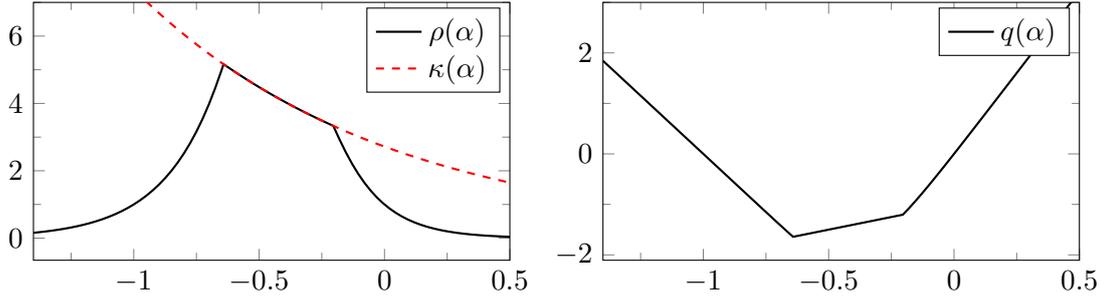

\medskip{\bf Example 6.}  Take now $\gamma(0) = 0$, $\gamma(1) = 0.01$,
$\gamma(n) = n + 5\log (1+n/5)$ for $n\geq 2$, and $\widehat \gamma(n) =10 + 2n
+ 5\log (1+n/5)$ for all $n\geq 0$. Here again, $\kappa(\alpha) = \exp(
1-\alpha)$. Explicit computations show that $\kappa(\alpha)
U_\alpha(\kappa(\alpha))<1$ on some interval $I = (-0.6418\dots, -0.2042\dots)$
and $\kappa(\alpha) U_\alpha(\kappa(\alpha)) \in (1, +\infty]$ outside of the
closure of $I$ (the numerical values in the definition of $\gamma(n)$ and
$\widehat \gamma(n)$ were chosen to ensure the existence of such an interval).
Then for $\alpha \in I$ we are in case (b), and for $\alpha \notin I$ we are in
case (a). It follows that $q$ is analytic everywhere except at the boundaries of
$I$. Explicit computations using \eqref{eq:deriveerho} show that $q$ is not
differentiable at those boundaries. See Figure~\ref{f:case6}.

\medskip
We note that in all the examples except the last, both $\P$ and $\widehat \P$
satisfy \hUD with $\tau_t \equiv 0$ and $\sup_t c_t < \infty$. In the last
example, both measures satisfy \hUD with $\sup_t c_t = \infty$ and $c_t= o(t)$.

In order to make direct comparison with the results of \cite{BJPP-2017,CJPS_phys}, we introduce the following construction.

\begin{remark} Given the pair $(\P, \widehat \P)$ constructed in this section,
one can define a new pair of measures $(\mathscr{P}, \widehat{\mathscr{P}})$ on
$\mathscr{A}^\nn$, with the product alphabet $\mathscr{A} = \cA \times \cA$,
such that $(\mathscr{P}, \widehat{\mathscr{P}})$ are related by an involution as
in Definition~\ref{def:familyinvolutions}. For any word $(u,v) \in \mathscr{A}^t
= \cA^t \times \cA^t$, define $\mathscr{P}(u,v) = \P(u) \widehat \P(v)$, and set
$\widehat{\mathscr{P}}(u,v) = \mathscr{P}(\theta_t(u,v))$ where\footnote{Since
$\P$ and $\widehat \P$ are reversible, one could as well define $\theta_t(u,v) =
((v_t, \dots, v_1),(u_t, \dots, u_1))$. This means that both cases of
Definition~\ref{def:familyinvolutions} are actually covered.} $\theta_t(u,v) =
(v,u)$. It is easy to show that the pair $(\mathscr{P}, \widehat{\mathscr{P}})$
satisfies \hSSD (with $\xi = (\mathrm{a},\mathrm{a}) \in \mathscr{A}$) and that the entropy production $\Sigma_t$
of the pair  $(\mathscr{P}, \widehat{\mathscr{P}})$ can be expressed in terms of
the entropy production $\sigma_t$ of $(\P, \widehat \P)$ by
\begin{equation*}
	\Sigma_t(u,v) : = \log \frac{ \mathscr{P}_t(u,v)}{ \widehat {\mathscr{P}}_t(u,v)} = \sigma_t(u)-\sigma_t(v), \qquad u,v \in \Omega_t .
\end{equation*}
As a consequence, we find that
\begin{equation}\label{eq:Qalphaqq}
	\mathscr{Q}(\alpha):= \lim_{t\to\infty}\frac1t\log\left\langle \eee^{	\alpha \Sigma_t}, \mathscr{P}\right\rangle = q(\alpha) + q(-\alpha-1), \qquad \alpha \in \rr.
\end{equation}
Note that
$\mathscr{Q}$ satisfies the symmetry \eqref{eq:symmetryealpha}, \ie $\mathscr{Q}(-\alpha)=\mathscr{Q}(\alpha-1)$ for all $\alpha \in \rr$.
\end{remark}
We finish with a brief comment on how, for the pairs of measures $(\mathscr{P},
\widehat{\mathscr{P}})$ constructed from the pairs $(\P, \widehat \P)$ in the
above six examples, the results of \cite{BJPP-2017,CJPS_phys} fail to
apply or to provide the global LDP in Theorem~\ref{main-2}. We remark that in
the first five cases the measures $\mathscr{P}$ and $\widehat{\mathscr{P}}$
satisfy \hUD with $\tau_t \equiv 0$ and $\sup_t c_t < \infty$, while in the last
case, both measures satisfy \hUD with $\sup_t c_t = \infty$ and $c_t= o(t)$.

\begin{itemize}
\item The results of \cite{BJPP-2017} apply to the pairs $(\mathscr{P},
\widehat{\mathscr{P}})$ of Examples 1--5 above, and give a local LDP for  in the
interval  $(\mathscr{Q}'(-1^+),\mathscr{Q}'(0^-))$. Only in Example 2 is this
interval equal to $\rr$ (see \eqref{eq:Qalphaqq}), and hence in this case
\cite{BJPP-2017} gives the full LDP. Example 6 cannot be handled by the method
of \cite{BJPP-2017}, because it does not satisfy \hUD with $\sup_t c_t < \infty$.
Note also that in Example 6, $\mathscr{Q}$ is not differentiable in $(-1,0)$,
unlike in the situation of \cite{BJPP-2017}.
\item If the sequence
$(G_t)_{t\in\nn} \subset C(\mathscr{A}^{\nn})$ defined by $G_t = \log
\mathscr{P}_t$\footnote{$\mathscr{P}_t$ is defined as the marginal of
$\mathscr{P}$ on the first $t$ coordinates of $\mathscr{A}^\nn$.} is {\em
asymptotically additive} then the results in \cite{CJPS_phys} apply, and provide the
global LDP for $t^{-1} \sigma_t$. This is the case of Examples 5 and 6 above.
Examples 1--4 clearly cannot be handled by \cite{CJPS_phys}, as under the
assumptions therein the entropic pressure $e(\alpha) = q(-\alpha)$ is finite for all $\alpha \in \rr$.
 \end{itemize}

\subsection{Weak Gibbs measures on subshifts}\label{ss:weakGibbsmeas}

LDPs for weak Gibbs measures (on shift spaces and more general dynamical
systems) have been abundantly studied; see for example
\cite{comman-2009,varandas_nonuniform_2012,varandas_weak_2015,PS-2018,CJPS_phys}
and the references therein. The weak Gibbs condition and our decoupling
assumptions are essentially incomparable (see below). We show here that given a
weak Gibbs measure supported on a subshift satisfying a suitable  {\em
specification property}, one can still construct a map $\psi_{n,t}$ satisfying
the conclusions of Proposition~\ref{prop:constructionpsi}. As our results use
\hSLD and \hSSD only through the conclusions of
Proposition~\ref{prop:constructionpsi}, they remain valid for weak Gibbs
measures.

We first introduce the notion of weak Gibbs measure on a subshift. The measure
$\P$ can be viewed as an invariant measure for the dynamical system $(\Omega^+,
\varphi)$. Recall that $\Omega^+ = \supp\, \P$ was defined in~\eqref{eq:OmegaPlus}. 
In this setup the following two conditions are natural. We again assume that 
$\tau_t = o(t)$.
\newcommand{\hWSP}{\hyperlink{hyp.WSP}{\textbf{(WSP)}}\xspace}
\newcommand{\hWGC}{\hyperlink{hyp.WGC}{\textbf{(WGC)}}\xspace}
\begin{description}
\sl
\item[\hypertarget{hyp.WSP}{Weak specification property (WSP).}]
For all
$t\in\nn$, all $u\in \Omega_t^+$, and all $v\in  \Omega_\fin^+$, there exists
$\xi \in \Omega_\fin^+$ satisfying $|\xi| \leq \tau_t$ such that $u\xi v\in
\Omega_\fin^+$.\footnote{For similar 
and weaker forms of specification properties and
 related results, see for example
 \cite{pfister_billingsley_2003,PS-2005,thompson_irregular_2012,KLO-2016,PS-2019}}
\item[\hypertarget{hyp.WGC}{Weak Gibbs condition (WGC).}] 
The measure~$\P$ is {\em weak
Gibbs} with respect to some potential $f\in C(\Omega^+)$; \ie there exists a
real number $p$ (called {\em pressure}) and a real sequence $(d_t)_{t\in\nn}$
such that $d_t = o(t)$, and for all $\omega \in \Omega^+$ and all $t\in\nn$, 
\begin{equation*}
 \eee^{-d_t + S_tf(\omega)- tp} \leq \P_t(\omega_1, \dots, \omega_t)\leq 
\eee^{d_t + S_tf(\omega)- tp}.
\end{equation*}
\end{description}

Measures satisfying \hWGC with $\sup_{t} d_t < \infty$ are called {\em Gibbs
measures} (see Example~\ref{ex:WGSS}). Without loss of generality, we shall
assume that $p = 0$ (by replacing $f$ with $f-p$ if necessary). 

Note that \hWSP is a condition on the structure of the set $\Omega^+$, while \hWGC
is a condition\footnote{From the point of view of dynamical systems, one is
first given a subshift $\Omega^+$ satisfying \hWSP, and then one introduces (weak)
Gibbs measures on it.}  on~$\P$. Once the set $\Omega^+$ is fixed, \hWGC implies a
strong lower bound on the probability of the ``allowed'' words: 
\begin{equation}\label{eq:logPNboundedwg}
\P_t(w) \geq \eee^{-Ct}, \qquad t\in\nn, \quad w\in \Omega_{t}^+,
\end{equation}
where $C = \|f\| + \sup_{t} d_t/t$.
Our decoupling assumptions are different in philosophy, as they are formulated
at the level of measures only.
 They compare to \hWSP and \hWGC as follows.
\begin{itemize}
\item As mentioned in Example~\ref{ex:WGSS}, if \hWSP holds and $\P$ is a Gibbs
measure (\ie \hWGC holds with $\sup_{t} d_t < \infty$), then the \hUD and \hSLD
assumptions are satisfied. On the contrary, if $\sup_{t} d_t = \infty$,
then \hWSP and \hWGC do not imply any of our decoupling assumptions in general. 
\item \hSLD implies \hWSP.
\item Even put together, \hSLD and \hUD do not imply \hWGC in general, as
\eqref{eq:logPNboundedwg} may fail. Indeed, \hSLD ensures that there is one $\xi
\in \Omega_\fin$, $|\xi|\leq  \tau_t$ such that $\IP(u \xi v) \geq \eee^{- c_t}
\IP(u) \IP(v)$, and says nothing about  $\IP(u \xi' v)$ for $\xi' \neq \xi$. See
Example~\ref{ex:HiddenMarkov} and Appendix~\ref{ss:hiddenMarkChain}.
\end{itemize}

\begin{remark}
It has been shown very recently \cite{PS-2019} that under quite general conditions
(and in particular, without any finite-type assumption on the subshift),
equilibrium measures for absolutely summable interactions satisfy \hWGC.
\end{remark}

We now establish an analogue of Proposition~\ref{prop:constructionpsi} for \hWSP
and \hWGC.
Here, $n$, $t$, $t'$, $N$, $\P^{(n)}$ and $\Lambda_{t'}$ are as in
Section~\ref{subs:constrPsint}. 

\begin{proposition}\label{prop:weakGibbsdec}
Assume \hWSP and \hWGC. Then there exists a map $\psi_{n,t} : \Omega_{t'} \to
\Omega_t$ such that the following holds.
\begin{enumerate}
\item We have \begin{equation}\label{eq:PndecouplsetsWG}
\P_{t'}^{(n)}\circ \psi_{n,t}^{-1} \leq  \eee^{g(n,t)} \, \P_{t}, \quad
\text{with } \lim_{n\to\infty} \limsup_{t\to\infty} \frac 1 t g(n,t)  = 0.	
\end{equation}
\item Let $\widehat \P \in \cP_\varphi(\Omega)$, and assume one of the
following: (a) $\widehat \P = \Theta \P$ with $\Theta$ as in
Definition~\ref{def:familyinvolutions} and  $\theta_t(\Omega_t^+) =
\Omega_t^+$; or (b) there exists $\widehat d_t = o(t)$ and  $\widehat f \in
C(\Omega)$ such that for all $\omega\in \Omega^+$,\footnote{Note that no
requirement is made for $\omega \notin \Omega^+$.}
\begin{equation}\label{eq:weakGibbsshiftNrerev}
 \eee^{-\widehat d_t + S_t \widehat f(\omega)} \leq \widehat\P_t(\omega_1,
\dots, \omega_t)\leq  \eee^{\widehat d_t + S_t\widehat f(\omega)}.
\end{equation}  	
Then $\psi_{n,t}$ can be chosen so that, in addition to
\eqref{eq:PndecouplsetsWG},
\begin{equation}\label{eq:sigmapsicompatibleWG}
\lim_{n\to\infty }\limsup_{t\to\infty}\frac 1 t \sup_{w\in
\Lambda_{t'}}\left|\sigma_{t}(\psi_{n,t}(w)) - \sum_{k=0}^{N-1}
\sigma_n(w_{[kn+1, (k+1)n]}) \right|  = 0,
\end{equation}
and there exists $c>0$ such that
\begin{equation}\label{eq:rho_tbddWG}
|\sigma_t(w)| \leq 	ct, \qquad t\in\nn, \quad w\in \Omega_{t}^+.
\end{equation}

\end{enumerate}

\end{proposition}
\proof{}
We first prove 1. For each $w\in \Omega_{t'}$, we write $w = w^{1}w^{2} \dots
w^{N}$ with $w^i \in \Omega_n$. Setting
\begin{equation}\label{eq:DefPsiWG}
		\psi_{n,t}(w) =b w^1 \xi^1 w^2 \xi^2 \dots  w^{N-1}\xi^{N-1} w^N \in 
\Omega_{t}
\end{equation}
for some suitable $\xi^i \in \Omega_n$ satisfying $|\xi^i| \leq \tau_n$ for all
$i$, we shall prove that
\begin{equation}\label{eq:toproveWG}
	\P(\psi_{n,t}(w))\geq \eee^{-g_1(n,t)}\P^{(n)}(w) , \quad \text{with }
\lim_{n\to\infty} \limsup_{t\to\infty} \frac 1 t g_1(n,t)  = 0.
\end{equation}
Then, the conclusion of Part 1 follows from the same combinatorial argument as
in Proposition~\ref{prop:constructionpsi} (see the discussion after
\eqref{eq:Ppsintprop31}).
In order to prove \eqref{eq:toproveWG}, we assume that $w^i \in \Omega_n^+$ for
all $i$ (equivalently, that $w\in \Lambda_{t'}$), as the result is trivial
otherwise. By following the same strategy as in
Proposition~\ref{prop:constructionpsi}, using \hWSP instead of \hSLD, we can choose 
$\xi_1, \dots, \xi^{N-1}$ and $b$ such that $\psi_{n,t}(w) \in \Omega_{t}^+$. 
Next,  let $\omega \in \Omega^+$ be such that $\omega_{[1,t]} = \psi_{n,t}(w)$.
By \hWGC,
\begin{equation}\label{eq:boundPtpsiWG}
	\P_t(\psi_{n,t}(w))  \geq  \eee^{S_t f(\omega) - d_t}
\geq \eee^{-g_1(n,t)}\prod_{i=1}^{N}\P_n(w^i) ,
\end{equation}
where $g_1(n,t) = 	 d_t + N d_n + (t-t')\|f\|$. The relations $N \leq t/n$ and
\eqref{eq:boundintgpart} imply \eqref{eq:toproveWG}, which completes the proof
of Part 1.

We now prove Part 2. By combining \eqref{eq:boundPtpsiWG} with the corresponding
upper bound, we obtain
\begin{equation}\label{eq:sigmapsicompatibleWG2}
\lim_{n\to\infty }\limsup_{t\to\infty}\frac 1 t \sup_{w\in
\Lambda_{t'}}\left|\log\P_{t}(\psi_{n,t}(w)) -  \log\P^{(n)}_{t'}(w) \right|  =
0.
\end{equation}
 Assume first that  $\widehat \P$ satisfies (b). By
\eqref{eq:weakGibbsshiftNrerev}, the relation \eqref{eq:sigmapsicompatibleWG2}
also holds with $\P$ replaced with $\widehat \P$, and
\eqref{eq:sigmapsicompatibleWG} immediately follows. Moreover, by \hWGC and
\eqref{eq:weakGibbsshiftNrerev}, we obtain \eqref{eq:rho_tbddWG} with $c = \|f -
\widehat f\| + \sup_{t}(d_t + \widehat d_t)/t$.

Assume now that $\widehat \P$ satisfies (a) and let $w \in\Lambda_{t'}$. Then
$\psi_{n,t}(w) \in \Omega_{t}^+$ and $\theta_t (\psi_{n,t}(w))\in
\Omega_{t}^+$. Hence there exists $\widehat \omega \in \Omega^+$ such that
$\widehat \omega_{[1,t]} = \theta_t (\psi_{n,t}(w))$. Since $\widehat
\P_t(\psi_{n,t}(w)) = \P_t(\theta_t (\psi_{n,t}(w)))$, we obtain by \hWGC that
\begin{equation*}
 \eee^{-d_t + S_t  f(\widehat \omega)} \leq \widehat\P_t(\omega_1, \dots,
\omega_t)\leq  \eee^{d_t + S_t f(\widehat \omega)}.
\end{equation*}  
This and \hWGC imply \eqref{eq:rho_tbddWG} with $c = 2\|f\| + 2\sup_{t}d_t/t$.
Using the notation \eqref{eq:DefPsiWG}, and introducing $\widehat b =
\theta_{|b|}(b)$, $\widehat \xi^i = \theta_{|\xi^i|}(\xi^i)$, $\widehat w^i =
\theta_{n}(w^i)$, we have either $\theta_t (\psi_{n,t}(w)) = \widehat b \widehat
w^1 \widehat \xi^1 \dots\widehat\xi^{N-1} \widehat w^N$ or $\theta_t
(\psi_{n,t}(w)) = \widehat w^N  \widehat\xi^{N-1}\dots  \widehat \xi^1\widehat
w^1\widehat b $  (see Definition~\ref{def:familyinvolutions}).
Computations similar to \eqref{eq:boundPtpsiWG} show that
\eqref{eq:sigmapsicompatibleWG2} also holds with $\P$ replaced with $\widehat
\P$, so that \eqref{eq:sigmapsicompatibleWG} again follows. This completes the proof.\hfill\qed 	

By using Proposition~\ref{prop:weakGibbsdec} instead
Proposition~\ref{prop:constructionpsi}, our results apply as follows.
\begin{itemize}
\item All the conclusions of Theorem~\ref{thm:summarythmlev1} are valid under
\hWSP and \hWGC.
\item All the conclusions of Theorem~\ref{main-2} are valid under \hWSP, \hWGC, and
the assumptions in Part 2 of Proposition~\ref{prop:weakGibbsdec}. The finiteness
of $q(\alpha)$  for all $\alpha\in \rr$ follows from \eqref{eq:rho_tbddWG}.
\item All the conclusions of Theorem~\ref{t1.9} are valid under \hWSP and \hWGC. The
estimates requiring \hUD in Proposition~\ref{p1.6} can easily be adapted by using
the following consequence of \hWGC: for all $w^1, w^2, \dots, w^{N} \in \Omega_n$
and all $\xi^1, \xi^2, \dots, \xi^{N-1} \in \Omega_{\tau_n}$, we have
\begin{equation*}
 \P_{Nn + (N-1)\tau_n}(w^1 \xi^1 w^2 \xi^2 \cdots w^{N-1} \xi^{N-1}w^n) \leq
\eee^{h(n,t)}	\prod_{i=1}^N \P_n(w^i),
\end{equation*}
where $h(n,t): = d_{Nn + (N-1)\tau_n} + Nd_n + (N-1)\tau_n\|f\|$ satisfies  
\[\lim_{n\to\infty} \limsup_{t\to\infty} t^{-1} h(n,t) = 0.\]
\end{itemize}

\addcontentsline{toc}{section}{References}

\newcommand{\etalchar}[1]{$^{#1}$}
\def\polhk#1{\setbox0=\hbox{#1}{\ooalign{\hidewidth
  \lower1.5ex\hbox{`}\hidewidth\crcr\unhbox0}}} \def\pre{{Phys. Rev. E\
  }}\def\polhk#1{\setbox0=\hbox{#1}{\ooalign{\hidewidth
  \lower1.5ex\hbox{`}\hidewidth\crcr\unhbox0}}}
\providecommand{\bysame}{\leavevmode \hbox to3em{\hrulefill}\thinspace}
\providecommand{\og}{``}
\providecommand{\fg}{''}
\providecommand{\smfandname}{and}
\providecommand{\smfedsname}{eds.}
\providecommand{\smfedname}{ed.}
\providecommand{\smfmastersthesisname}{Master Thesis}
\providecommand{\smfphdthesisname}{Thesis}

\end{document}